\documentclass[12pt,twoside]{book}

\newcommand{\essaytitle}{Emergent Features in $\mathbf{U(N) \times U(\tilde{N})}$ Bi-adjoint Cubic Theory
}
\newcommand{\shortessaytitle}{On the $U(N)\times U(\tilde N) $ Bi-adjoint $\phi^3$ Theory} % Make this shorter if you see it spilling over within the header
\newcommand{\yourname}{Lauren Smyth}
\newcommand{\yoursupervisor}{Dr Freddy Cachazo}

\usepackage[utf8]{inputenc}
\usepackage[T1]{fontenc}
\usepackage{fancyhdr}
\usepackage{lipsum}
\usepackage{graphicx}
\usepackage{titlesec}
\usepackage{appendix}
\usepackage[sectionbib]{chapterbib}
\usepackage[breakwords]{truncate}
\usepackage{lastpage}
\usepackage{amsmath}
\usepackage{mathrsfs}
\usepackage{amssymb}
\usepackage[font={small}]{caption}
\usepackage{cite} 
\usepackage{physics} 
\usepackage{graphicx,color}
\usepackage[dvipsnames]{xcolor}
\usepackage{hyperref}
\definecolor{darkbyzantium}{rgb}{0.36, 0.22, 0.33}
\definecolor{britishracinggreen}{rgb}{0.0, 0.26, 0.15}
\hypersetup{
	colorlinks=true,         
	linkcolor=black,          
	citecolor=darkbyzantium,        
	urlcolor=darkbyzantium 
}

\renewcommand*\thesection{\arabic{section}}

  \usepackage[nottoc,notlof,notlot]{tocbibind}

\titleformat{\chapter}[display]
  {\bfseries\large}{}{-10ex}
  {\titlerule\vspace{2ex}\filright\Huge}
  [\vspace{1ex}\titlerule]

\usepackage{amsthm}
\usepackage{subcaption} 
\usepackage{float}  % in preamble
\usepackage{booktabs}
\usepackage[dvipsnames]{xcolor}
\usepackage{epsfig}
\usepackage{dcolumn}

\usepackage{tikz}
\usetikzlibrary{calc}
\usepackage{ytableau}
\usetikzlibrary{shapes, positioning}
\usetikzlibrary{shapes.geometric, arrows}
\usetikzlibrary{decorations.pathmorphing}
\usepackage{upgreek}
\usepackage{amsfonts}
\usepackage{setspace}
\usepackage{comment}
\usepackage{enumitem}
\usetikzlibrary{decorations.markings}
\tikzset{
  gluon/.style={decorate, decoration={coil, amplitude=4pt, segment length=5pt}, draw=black},
  fermion/.style={->, thick},
  vertex/.style={circle, fill=black, inner sep=1.5pt},
}
\usepackage{array,multirow,bigdelim,arydshln}
\usepackage{xparse}
\theoremstyle{plain}  % This will format the theorem as normal
\begin{document}
%%%%%%%%%%%%%%%%%%%%%%%%%%%%%%
%% TITLE PAGE - DO NOT EDIT %%
%%%%%%%%%%%%%%%%%%%%%%%%%%%%%%
% \setcounter{page}{1} % sets the page number for the book compilation (ignore this)
%%%%%%%%%%%%%%%%%%%%%%%%%%%
%% DO NOT EDIT THIS FILE %%
%%%%%%%%%%%%%%%%%%%%%%%%%%%

\setlength{\headheight}{15pt} 

% Title page
\begin{center}
\includegraphics[height=0.35\textwidth]{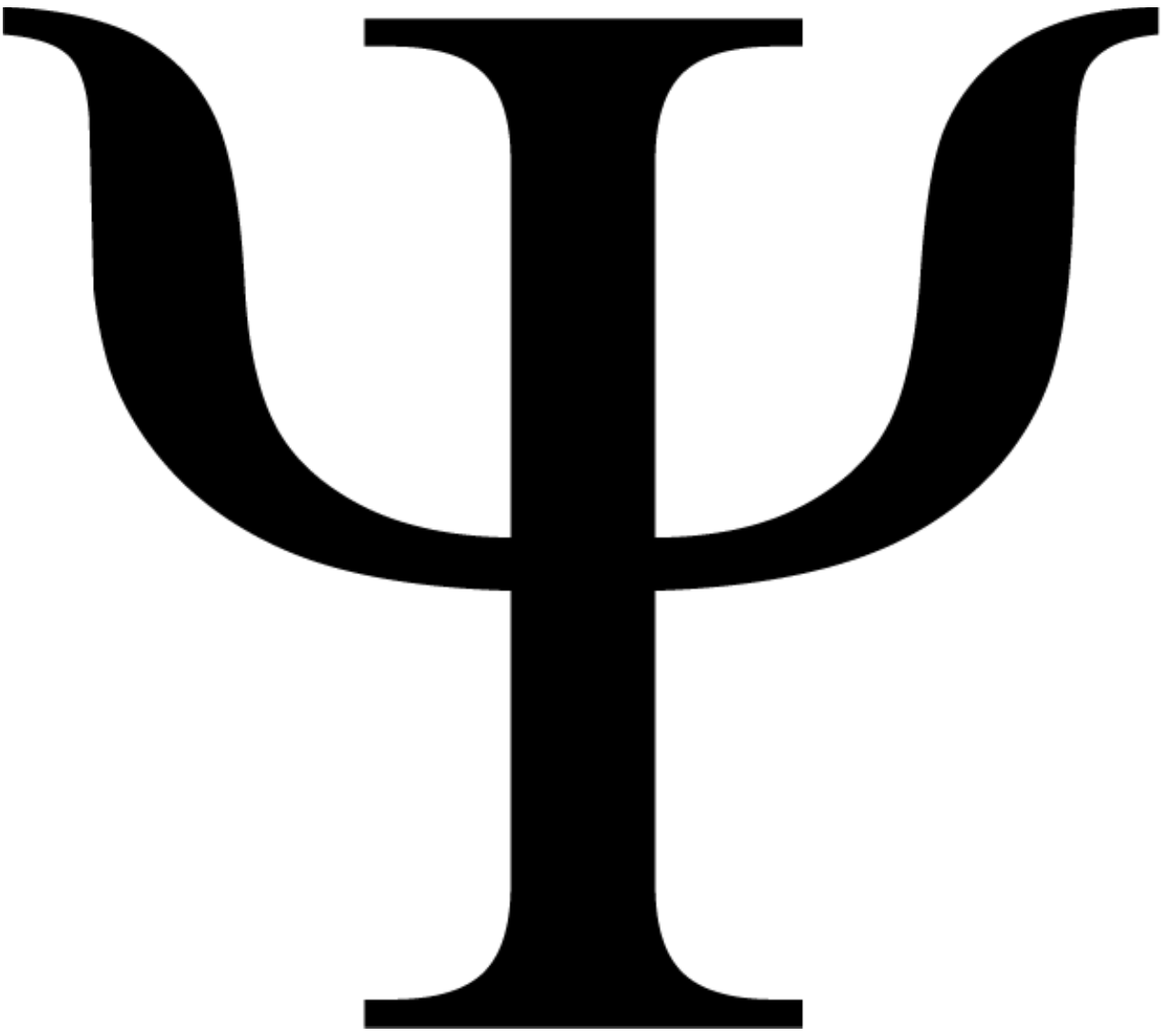}\\[2cm]
\textbf{\huge\essaytitle}\\[3cm]
\textbf{\Large\yourname}\\
\vspace*{\fill}
 \large{\textbf{An essay submitted}\\ 
  \textbf{for partial fulfillment of}\\ 
  \textbf{Perimeter Scholars International}\\[2cm]
  \textbf{June, 2025}}
\end{center}

\pagestyle{empty} % leaves a the next page black

\frontmatter % makes Roman page numbers
\pagestyle{fancy} % undoes the empty page style

% Table of contents
\tableofcontents
\mainmatter % makes Arabic page numbers

% % The heading on the front page
\chapter*{\essaytitle} 

\vspace{-0.5cm}
\centerline{\textbf{\yourname}} 
\vspace{0.3cm}
\centerline{Supervisor: \yoursupervisor} 
\vspace{0.5cm}

%%%%%%%%%%%%%%%%%%%%%%%%%%%%%%%%
%% ESSAY ABSTRACT - EDIT THIS %%
%%%%%%%%%%%%%%%%%%%%%%%%%%%%%%%%
\begin{quote}

This work investigates the role of the $U(N)\times U(\tilde{N})$ global symmetry in tree-level scattering amplitudes of the bi-adjoint $\phi^3$ theory from three perspectives: combinatorics, correlation functions, and a massive extension of the CHY formalism. We derive a planar scattering potential whose extrema reproduce Dolan and Goddard’s massive scattering equations, providing physical intuition of the construction. This potential enables the counting of kinematic invariants via maximally symmetric Ferrers shapes, and it is expressed in terms of conformally invariant cross-ratios. We find that the $U(1)$ decoupling identity provides a physical interpretation of two different Catalan recursion relations, and also reveals an interplay between Catalan and Narayana numbers in the $U(1)$ splitting. Finally, we construct correlation functions for a fixed particle ordering using the CHY formalism, offering new insights into the dynamics of such amplitude structures. We derive a closed form expression of the reduced number of solutions for this set-up, as well as an off-shell scattering potential.
\end{quote}

%%%%%%%%%%%%%%%%%%%%%%%%%%%%%%%%%%%%%%%%%%%%
%% YOUR ESSAY CHAPTERS - EDIT THESE FILES %%
%%%%%%%%%%%%%%%%%%%%%%%%%%%%%%%%%%%%%%%%%%%

\section*{Statement of original research}
Chapter \ref{sec:introduction} provides a literature review, while Chapters \ref{sec: section2} - \ref{sec: section4} include original research. Chapter \ref{sec: conclusions} summarises the findings and outlines possible future directions.

\section{Introduction}\label{sec:introduction}

The combinatorial explosion of Feynman diagrams at tree-level for an increasing number of external particles implies that scattering amplitudes are inherently as intricate as their computations suggest. However, recent developments into the study of these tree-level scattering amplitudes have revealed deep connections with rich mathematical structures and uncovered remarkably simple patterns hidden beneath the apparent complexity.

In 1986, Parke and Taylor made a striking discovery: tree-level gluon amplitudes simplify dramatically in the spinor-helicity formalism, yielding compact expressions for maximally-helicity-violating (MHV) amplitudes with any number of external gluons~\cite{Parke:1986gb}. This revealed a hidden simplicity, even for six gluons in next--to--MHV (NMHV) scattering, where traditional Feynman diagram methods require over 200 terms. Building on insights from twistor theory, Nair interpreted MHV amplitudes as current correlators in a two-dimensional Wess--Zumino--Witten model on $\mathbb{CP}^1$~\cite{Nair:1988bq}. This perspective inspired Witten, who in 2003 proposed a twistor string theory in $\mathbb{CP}^{3|4}$ that geometrically encodes $\mathcal{N}=4$ super--Yang--Mills amplitudes~\cite{Witten_2004}. Building on this, Roiban, Spradlin, and Volovich (RSV) showed that tree-level amplitudes can be computed by integrating over the moduli space of holomorphic maps from $n$-punctured spheres into connected curves in twistor space -- a formulation now known as the \emph{Witten--RSV formula}~\cite{Roiban_2004,Witten_2004,Cachazo_2018}.

A further leap came with the Britto--Cachazo--Feng--Witten (BCFW) recursion relations~\cite{Britto_2005}, which use analytic continuation to reconstruct all tree-level amplitudes from lower-point on-shell data, making factorization properties manifest. More recently, the study of positive geometry has unified many of these ideas \cite{arkanihamed2014scatteringamplitudespositivegrassmannian, Arkani-Hamed:2012zlh}: in planar $\mathcal{N}=4$ SYM, all tree-level amplitudes are encoded in the volume of the amplituhedron, a region in the positive Grassmannian defined in momentum-twistor space~\cite{Arkani-Hamed:2013jha}. Similarly, the associahedron -- known in mathematics since the $1960$'s -- emerges in the bi-adjoint $\phi^3$ theory, whose scattering form captures tree-level kinematic structure~\cite{Arkani_Hamed_2014, Arkani_Hamed_2018}. 

At the algebraic level, colour-ordered amplitudes satisfy relations such as Kleiss--Kuijf~\cite{KLEISS1989616} and Bern--Carrasco--Johansson (BCJ)~\cite{Bern_2008}, reflecting a colour-kinematics duality that reduces the independent basis of amplitudes and enables the double-copy construction of gravity amplitudes. In parallel, the Cachazo--He--Yuan (CHY) formalism expresses tree-level amplitudes as contour integrals over the moduli space of marked points on $\mathbb{CP}^1$, localised by the scattering equations that link external kinematics to puncture locations on $\mathbb{CP}^1$~\cite{Cachazo_2013,Cachazo_2014_scalars,Cachazo_2015,Cachazo_2014,Dolan_2014, Cachazo_2014_short}.

This essay aims to explore the role of global $U(N)\times U(\tilde{N})$ symmetry in bi-adjoint $\phi^3$ tree-level amplitudes through three complementary lenses: combinatorics, correlation functions, and a massive extension of the CHY formalism. By focusing on this relatively simple yet structurally rich theory, we seek to uncover how such a symmetry governs the organisation of scattering data and connects with deeper mathematical patterns. The broader goal of such research is to better understand how modern amplitude techniques—-particularly those rooted in geometry and symmetry—-simplify and organise tree-level scattering processes beyond traditional Feynman diagrammatics.

This introductory section concludes with an overview of the CHY formalism and the bi-adjoint $\phi^3$ scalar theory. Section~\ref{sec: section2} provides a physical explanation of how the massive extension of the CHY formulation, introduced by Dolan and Goddard in 2014~\cite{Dolan_2014}, reproduces tree-level amplitudes in the bi-adjoint $\phi^3$ theory. This includes a derivation of the scattering potential underlying the relevant scattering equations, a discussion on the counting of independent kinematic invariants, and a demonstration of the $SL(2,\mathbb{C})$ invariance of the potential via cross-ratios. Section~\ref{sec: section3} explores the physical interpretation of Catalan recursion relations as building and factorising diagrams using the $U(1)$ decoupling identity, uncovering connections to the Narayana numbers and diagram factorisation. Finally, Section~\ref{sec: section4} examines the relationship between CHY amplitudes and correlation functions for a specific colour ordering, including counting the new number of solutions to the corresponding scattering equations and deriving an off-shell scattering potential. 
\subsection{Introduction to CHY}\label{subsec: introduction to CHY}
As mentioned previously, traditional methods for obtaining amplitudes through perturbative Feynman diagrams become computationally formidable as the number of particles in a scattering process increases. 

The idea of the Cachazo-He-Yuan formulation (CHY) of scattering amplitudes \cite{Cachazo_2013,Cachazo_2014,Cachazo_2014_short,Cachazo_2014_scalars,Cachazo_2015,Dolan_2014} is to bypass this perturbative and enumerative difficulty in accessing these amplitudes by providing a direct link from the kinematic data of particle interactions to the scattering amplitudes themselves. This is achieved for massless particles at tree-level by a connection between the space of kinematic invariants $s_{ab}=(k_a+k_b)^2$ (the propagators of diagrams) for an $n-$particle scattering process and the moduli space of Riemann spheres with $n-$punctures, ${\cal M}_{0,n}$, via the \textit{scattering equations} \cite{Cachazo_2014}:
\begin{equation}
    \sum_{\substack{b\neq a \\ b=1}}^n \frac{s_{ab}}{x_a-x_b} = 0,
    \label{eq: scattering equations for massless particles}
\end{equation}
where $n$ denotes the particle number and $x_a$'s are the punctures on the Riemann sphere $\mathbb{CP}^1$, three of which will be fixed as a result of the $\mathrm{SL}(2,\mathbb{C})$ invariance. This $\mathrm{SL}(2,\mathbb{C})$-invariant system of $n-3$ polynomial equations, with $(n-3)!$ solutions~\cite{Cachazo_2017, Cachazo_2014, Dolan_2014_poly}, forms the backbone of the tree-level S-matrix for massless particles in any dimension~\cite{Cachazo_2014_short, Dolan_2014}. Notably, the number of solutions matches the number of independent colour-ordered amplitudes after imposing the BCJ relations~\cite{Bern_2008}, indicating that the scattering equations encode the minimal data required to reconstruct the full tree-level S-matrix.

Defined in Ref. \cite{Cachazo_2014},  Eq.~\eqref{eq: scattering equations for massless particles} exhibits the remarkable property of \emph{Kawai-Lewellen-Tye (KLT) orthogonality}, crucial for constructing gravity amplitudes from the Witten-RSV gauge theory formula \cite{Witten_2004, Roiban_2004}. Originating from string theory, the scattering equations are linked to KLT orthogonality and closed string amplitudes in the high-energy limit \cite{Cachazo_2014}, with the Koba-Nielson factor~\cite{KOBA1969517} generating the scattering equations\cite{Jepsen2015}. Additionally, algebraic relations from string theory disk amplitude calculations \cite{mafra2011completenpointsuperstringdisk, Broedel_2013, STIEBERGER2016104, Barreiro_2014} have contributed to the formalism.

Their universal role becomes clear in the CHY formulation, where amplitudes\footnote{In theories with ``colour'' symmetry (see Section \ref{subsec:22}), these would actually be \emph{partial amplitudes}—-colour-stripped amplitudes with fixed external ordering, prior to any colour or kinematic dressing.} across a wide range of theories are expressed as
\begin{equation}
\label{eq:genCHY}
A^{\rm massless}_{n} = \frac{1}{\mathrm{Vol}(\mathrm{SL}(2,\mathbb{C}))} \int \prod_{a=1}^n dx_a \prod_{a=1}^n \delta\left( \sum_{\substack{b=1 \\ b\neq a}}^n \frac{s_{ab}}{x_a - x_b} \right)\, \mathcal{I}_L\,\mathcal{I}_R\,.
\end{equation}
For a guide on how to evaluate a CHY integral to obtain a tree-level, scalar amplitude see Appendix \ref{sec: calculating a 4point chy amplitude}.

The division by $\mathrm{Vol}(\mathrm{SL}(2,\mathbb{C}))$ removes redundancy from gauge transformations on the moduli space, ensuring the amplitude is properly normalized and $\mathrm{SL}(2,\mathbb{C})$-invariant. For more information on gauge-fixing and the geometry of $\mathbb{CP}^1$ in the CHY formalism, see Appendix \ref{subsec: cp1 and chy}.

After integrating over the punctures on $\mathbb{CP}^1$, Eq.~\eqref{eq:genCHY} demonstrates that the scattering equations-imposed by the delta functions-play a central role in constructing the amplitude by localising the integral. Although the formula is written in terms of delta functions, they are meant to be interpreted as multi--dimensional complex residues, so all solutions, complex and real, must be used. However, in Ref.~\cite{Cachazo_2017}, a region of the kinematic space where all solutions are real was discovered and in it the delta functions recover their usual meaning. These can be accessed via the \emph{positive kinematic region} $\mathcal{K}^+_n$, defined by the positivity of all Mandelstam variables $s_{ij}$ in the chosen basis. This region has dimension $\frac{n(n-3)}{2}$, corresponding to the number of independent kinematic invariants after accounting for momentum conservation and on-shell conditions.

In $\mathcal{K}^+_n$, the scattering equations are equivalent to finding the equilibrium configuration of $n-3$ mutually repelling particles on a finite real interval, under the ``scattering potential''
\begin{equation}
\label{eq: potential of the scattering equations}
\mathcal{S}(x) = \sum_{1 \leq a < b \leq n} s_{ab} \log|x_a - x_b|,
\end{equation}
where the scattering equations arise by extremising $\mathcal{S}(x)$. Imposing $SL(2,\mathbb{C})$ invariance on Eq.~\eqref{eq: potential of the scattering equations} leads to momentum conservation (for a derivation, see Appendix \ref{subsec: invariance of the scattering potential implies momentum conservation}).

The integrands $\mathcal{I}_L$ and $\mathcal{I}_R$ in Eq.~\eqref{eq:genCHY} encode the specific theory under consideration. They are built from $\mathrm{SL}(2,\mathbb{C})$-covariant factors $1/(x_a - x_b)$, such as the Parke-–Taylor factors\footnote{Named after the Parke Taylor formula for MHV gluon amplitudes derived in 1986~\cite{Parke:1986gb}.}
\begin{equation}
\label{eq: parke taylor factors definition}
\text{PT}(a(1)a(2)\cdots a(n)) = \frac{1}{(x_{a(1)} - x_{a(2)})(x_{a(2)} - x_{a(3)})\cdots(x_{a(n)} - x_{a(1)})},
\end{equation}
and polynomials in $k_a \cdot k_b$, $k_a \cdot \epsilon_b$, and $\epsilon_a \cdot \epsilon_b$, typically constructed from \emph{reduced Pfaffians} $\mathrm{Pf}'\Psi$ of matrices encoding the kinematic data. For more details on such mathematical objects see Appendix \ref{sec: chy building blocks}.

Comparing these building blocks across Yang–-Mills theory and Einstein gravity reveals the double-copy structure: while Yang–-Mills amplitudes involve a single factor of $\mathrm{Pf}'\Psi$ of such a matrix $\Psi$, graviton amplitudes involve $(\mathrm{Pf}'\Psi)^2$~\cite{Cachazo_2014_short, adamo2022snowmasswhitepaperdouble}.

With the CHY formalism introduced, we now turn to the bi-adjoint $\phi^3$ theory, the scattering theory which underpins this work.

\subsection{\texorpdfstring{The Bi-Adjoint Scalar $\phi^3$ Theory}{The Bi-Adjoint phi3 Theory and Colour Ordering}}\label{subsec:22}

In the style of Ref.~\cite{Cachazo_2014_scalars}, the \emph{colour-dressed}\footnote{By \emph{colour-dressed}, we mean the expression includes both the part containing the colour symmetry of the theory as well as the kinematics part (see Appendix~\ref{sec: colour ordered amplitudes}).} integrands $\tilde{\mathcal{I}}_L$ and $\tilde{\mathcal{I}}_R$ for a CHY style amplitude in Yang-Mills and Einstein gravity can be written as~\cite{Cachazo_2014_short}
\begin{equation}
\label{eq:genintegrands}
\tilde{\mathcal{I}}_L\tilde{\mathcal{I}}_R = 
\left(\sum_{\rho \in S_n / \mathbb{Z}_n} 
\frac{
\mathrm{Tr} \big( T^{a_{\rho(1)}} \cdots T^{a_{\rho(n)}} \big) }
{(x_{\rho(1)} - x_{\rho(2)}) \cdots (x_{\rho(n)} - x_{\rho(1)})}
\right)^{2 - \mathbf{s}}
\left( \mathrm{Pf}'\Psi \right)^{\mathbf{s}},
\end{equation}
where $\mathbf{s}$ denotes the spin of the external particles, $\Psi$ denotes a $2n\times 2n$ antisymmetric matrix encoding the polarisation and kinematic data, and $T^{a_{\rho(j)}}$ denote the $U(N)$ colour group generators. The sum is over the set $S_n/\mathbb{Z}_n$, modding out the subset of cyclic permutations $\mathbb{Z}_n$ which preserve the trace to prevent over counting, thus keeping only the cyclically inequivalent orderings\cite{https://doi.org/10.5170/cern-2014-008.31,Elvang:2013cua, peskin2019concepts}. For $\mathbf{s}=1,2$, Eq.~\eqref{eq:genintegrands} yields the integrands for Yang-Mills and Einstein gravity, respectively, with the former exhibiting a clear $U(N)$ global symmetry structure. But what about the case of $\mathbf{s}=0$?

For $\mathbf{s}=0$, the integrand describes amplitudes for the \emph{bi-adjoint $\phi^3$ scalar theory}, where each external particle carries two ``colour'' indices, each transforming under the adjoint representation. The term ``colour'' is used loosely here: borrowed from gauge theory, it actually refers to ``flavour'' in this context, as the theory admits a global $U(N) \times U(\tilde{N})$ symmetry. This is reflected in the presence of two independent Parke--Taylor factors with particle orderings governed by Eq.~\eqref{eq:genintegrands} when $\mathbf{s}=0$. The theory can be interpreted as the ``zeroth copy'' of Yang-Mills theory, and serves as a useful laboratory for studying colour-kinematics duality and the double copy construction~\cite{Brown_2018}.

\begin{figure}[hbt!]
\begin{center}
  \begin{tikzpicture}[scale=1, thick]

    % Draw the big outer circle
    \draw[black, ultra thick] (0,0) circle(2);

    % Boundary points (1, 5, 4, 3, 2)
    \foreach \i/\angle/\label in {
        1/72/$4$,
        2/144/$3$,
        3/216/$2$,
        4/288/$1$,
        5/360/$5$
    } {
        \coordinate (\i) at (\angle:2);
        \fill[black] (\i) circle(0.08); % <--- add this line to draw a black dot at each boundary point
        \node at (\angle:2.3) {\small \label};
    }

    % Draw red connections between the boundary points
    \draw[red, thick] (1) -- (2) -- (5) -- (3) -- (4) -- (1);

    % Internal points
    \coordinate (M1) at (0, -1.25);
    \coordinate (M2) at (1, 0.0);
    \coordinate (M3) at (0.0, 1.15);

    % Fill internal points with red dots
    \fill[red] (M1) circle(0.08);
    \fill[red] (M2) circle(0.08);
    \fill[red] (M3) circle(0.08);

    % Black internal lines
    \draw[black, thick] (1) -- (M3) -- (M2) -- (M1) -- (3);
    \draw[black, thick] (5) -- (M2);
    \draw[black, thick] (2) -- (M3);
    \draw[black, thick] (4) -- (M1);

  \end{tikzpicture}
\end{center}
    \caption{Pictorial representation of a Feynman diagram (shown within the circle in black) in the bi-adjoint $\phi^3$ theory. This amplitude is under the ordering $m(12345|12534)$.}
    \label{fig:biadjoint amplitude example}
\end{figure}
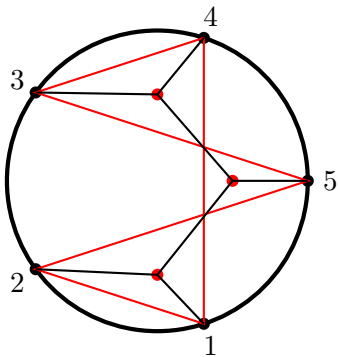

Amplitudes in theories of colour can be expanded in a trace basis, separating colour and kinematic data~\cite{peskin2019concepts,Thomson_2013}. For a review, see Appendix \ref{sec: colour ordered amplitudes}. Similarly, the full colour-dressed bi-adjoint amplitude involves a double trace structure, reflecting the fact that the massless scalar fields \(\phi^{a\tilde{a}}\) transform under the adjoint of \(U(N) \times U(\tilde{N})\). Explicitly, the full amplitude is
\vspace{-1mm}
\begin{equation}
\mathcal{A}_n^{\phi^3}(\{k_i, a_i, \tilde{a}_i\}) = \sum_{\alpha, \beta \in S_n / \mathbb{Z}_n} 
\mathrm{Tr} \big( T^{a_{\alpha(1)}} \cdots T^{a_{\alpha(n)}} \big) 
\mathrm{Tr} \big( \tilde{T}^{\tilde{a}_{\beta(1)}} \cdots \tilde{T}^{\tilde{a}_{\beta(n)}} \big)
\, m_n(\alpha|\beta),
\label{eq: biadjoint phi3 colour ordered formal expression}
\end{equation}
where \(m_n(\alpha|\beta)\) is the double-partial amplitude corresponding to the pair of orderings \(\alpha\) and \(\beta\). The main computational task is to evaluate the colour-ordered partial amplitudes, which are simplified by only receiving contributions from diagrams with a specific colour structure-encoded by the Parke–Taylor factors-thereby limiting the number of poles\cite{https://doi.org/10.5170/cern-2014-008.31}. Fig.~\ref{fig:biadjoint amplitude example} illustrates the double-colour structure of a sample diagram, which only has poles $s_{12}$ and $s_{34}$. 
Although this construction for tree-level amplitudes in the theory are thoroughly established, extending them to the massive case is highly non-trivial. In Section \ref{sec: section2}, we endeavour to provide a physical motivation to an extension to this theory for particles under a planar ordering.

\section{A Massive Planar Scattering Potential: Combinatorial Structures \& Cross-Ratios}\label{sec: section2}
In 2014, Dolan and Goddard introduced a massive extension of the scattering equations that successfully reproduced tree-level amplitudes in a $\phi^3$ theory for scalar particles of mass $m$~\cite{Dolan_2014}. Their modification introduced mass-dependent terms into Eq.~\eqref{eq: scattering equations for massless particles}:
\begin{equation}
    \sum_{\substack{b \neq a \\ b \in A}} \frac{k_a \cdot k_b}{x_a - x_b} + \frac{m^2}{2(x_a - x_{a-1})} + \frac{m^2}{2(x_a - x_{a+1})} = 0,
    \label{eq:massive_dolan_goddard}
\end{equation}
where $A = \{1, 2, \dots, n\}$. Crucially, this construction relies on the \emph{cyclic ordering} of the massive particles - a non-trivial and physically meaningful input.

The success of this approach lies in its ability to correctly reproduce the desired amplitudes. However, a natural question arises: \emph{why} does this work, and physically, \emph{how} does the cyclicity arise? In what follows, we argue that this cyclic ordering can be understood as emerging naturally from the structure of \emph{planar} Feynman diagrams in massive $\phi^3$ theory.

\subsection{Planar Kinematics}\label{subsec: Planar Feynman Diagrams and Planar Kinematic Invariants}
To exploit the structure of the positive kinematic region $\mathcal{K}^+_n$ described in Section~\ref{subsec: introduction to CHY}, we seek a planar basis for the kinematic data. In this basis in $\mathcal{K}^+_n$, the scattering potential is naturally expressed, and the connection to planar Feynman diagrams, as suggested by the ordering in Eq.~\eqref{eq:massive_dolan_goddard}, becomes manifest.

\begin{figure}[hbt!]
\begin{center}
\begin{tikzpicture}[scale=1]

    % First Circle - Planar Graph (Green)
    \draw (0,0) circle (2);

    % Outer Black Nodes (smaller)
    \foreach \i in {1,2,3,4,5,6} {
        \node[fill=black, circle, minimum size=3pt, inner sep=1.8pt] (A\i) at ($(0,0) + (90-60*\i:2)$) {};
        \node at ($(0,0) + (90-60*\i:2.6)$) {\textbf{\large \i}};
    }

    % Internal Green Nodes (smaller)
    \node[fill=green, circle, minimum size=3pt, inner sep=1.8pt] (G1) at (-0.9, 0.1) {};
    \node[fill=green, circle, minimum size=3pt, inner sep=1.8pt] (G2) at (0.3,0.8) {};
    \node[fill=green, circle, minimum size=3pt, inner sep=1.8pt] (G3) at (0.0,0.1) {};
    \node[fill=green, circle, minimum size=3pt, inner sep=1.8pt] (G4) at (0.3,-0.7) {};

    % Edges (Green)
    \draw[green, thick] (A1) -- (G2);
    \draw[green, thick] (A2) -- (G4);
    \draw[green, thick] (A3) -- (G4);
    \draw[green, thick] (A4) -- (G1);
    \draw[green, thick] (A5) -- (G1);
    \draw[green, thick] (A6) -- (G2);
    \draw[green, thick] (G1) -- (G3) -- (G2);
    \draw[green, thick] (G3) -- (G4);

    % Second Circle - Non-Planar Graph (Red)
    \begin{scope}[xshift=8cm]

        \draw (0,0) circle (2);

        % Outer Black Nodes (smaller)
        \foreach \i in {1,2,3,4,5,6} {
            \node[fill=black, circle, minimum size=3pt, inner sep=1.8pt] (B\i) at ($(0,0) + (90-60*\i:2)$) {};
            \node at ($(0,0) + (90-60*\i:2.6)$) {\textbf{\large \i}};
        }

        % Internal Red Nodes (smaller)
        \node[fill=red, circle, minimum size=3pt, inner sep=1.8pt] (R1) at (-0.7, -0.2) {};
        \node[fill=red, circle, minimum size=3pt, inner sep=1.8pt] (R2) at (0.3,0.8) {};
        \node[fill=red, circle, minimum size=3pt, inner sep=1.8pt] (R3) at (0.0,0.0) {};
        \node[fill=red, circle, minimum size=3pt, inner sep=1.8pt] (R4) at (0.0,-0.7) {};

        % Edges (Red)
        \draw[red, thick] (B1) -- (R2);
        \draw[red, thick] (B2) -- (R1);
        \draw[red, thick] (B3) -- (R4);
        \draw[red, thick] (B4) -- (R4);
        \draw[red, thick] (B5) -- (R1);
        \draw[red, thick] (B6) -- (R2);
        \draw[red, thick] (R2) -- (R3);
        \draw[red, thick] (R1) -- (R3);
        \draw[red, thick] (R3) to[out=30,in=30] (R4);

    \end{scope}

    % Check and Cross Marks
    %\node at (3.5, 0) {\Huge \textcolor{green}{$\checkmark$}};
    %\node at (11.5, 0) {\Huge \textbf{\textcolor{red}{X}}};

\end{tikzpicture}
\end{center}
\caption{Planar (left) and non-planar (right) graphs for $ n = 6 $ particles. The planar graph has poles at $ s_{16}, s_{23}, s_{45} $, while the non-planar one has poles at $ s_{16}, s_{34}, s_{25} $, with $ s_{25} $ being a non-planar invariant.}
\label{fig:planar_graph_comparison}
\end{figure}
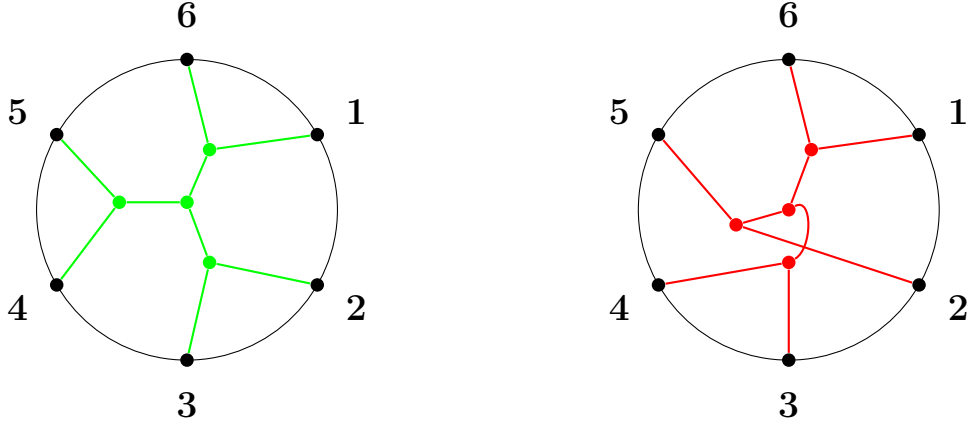

To make contact with the diagrammatic structure of planar amplitudes (Fig. \ref{fig:planar_graph_comparison}), we express Eq.~\eqref{eq: potential of the scattering equations} in terms of the $\frac{n(n-3)}{2}$ independent \emph{planar} kinematic invariants\footnote{Remarkably, the number of planar and non-planar kinematic invariants coincides once momentum conservation is imposed. For details on constructing a general kinematic invariant basis, see Ref.~\cite{Cachazo_2017}.}. This choice induces a planar ordering on the external particles, aligning the kinematic input with the structure of planar Feynman diagrams. It is essential to verify that $SL(2,\mathbb{C})$ invariance is maintained under this reparameterisation, so that the resulting potential under the deformation $s_{ij} \to s_{ij} - m^2$ remains $SL(2,\mathbb{C})$ invariant.

To test whether Eq.~\eqref{eq:massive_dolan_goddard} arises from planar Feynman diagrams, we begin by examining the case $n=6$, since it is in this setting that planar kinematic invariants of length greater than two first appear in the potential. By the \emph{length} of a kinematic invariant, we mean the number of momentum labels it includes. For example, 
\begin{equation}
    \label{eq: def length of kinematic invariant}
    s_{i,i+1, \ldots, i+k-1} = (k_i + \cdots + k_{i+k-1})^2
\end{equation}
has length $k$. 

We first identify a method for expressing non-planar kinematic invariants (i.e., those not of the planar form $s_{i,i+1,\ldots,i+k-1}$ for\footnote{Momentum conservation imposes this restriction on $k$, as kinematic invariants must have length between 2 and $n-2$.} $2\leq k \leq n-2$) in terms of planar ones. These two sets are related via momentum conservation and the identity
\begin{equation}
    s_{i,i+1,i+2} = s_{i,i+1} + s_{i+1,i+2} + s_{i,i+2}, 
    \label{eq: relating kinematic invariants}
\end{equation}
in which clearly $s_{i,i+2}$ is not a planar kinematic invariant. We will exploit this fact in particular to obtain our basis. For example,
\begin{equation}
    s_{26} = s_{612} - s_{12} - s_{61}.
\label{eq: planar basis example 1}
\end{equation}
However, we must remember that the planar basis for $n=6$ consists of nine\footnote{This comes from $\frac{6(6-3)}{2}=9$. } independent elements, dictated by the momentum conservation condition
\begin{equation}
    s_{i,i+1} + s_{i+1,i+2} + s_{i,i+2} = s_{i+3,i+4} + s_{i+4,i+5} + s_{i+3,i+5} \quad \text{mod}(6).
    \label{eq: momentum conservation for n=6}
\end{equation}
Given Eq.~\eqref{eq: momentum conservation for n=6}, we can select three planar kinematic invariants of length three --- for example, $s_{123}, s_{234}, s_{345}$ --- and discard the rest using momentum conservation. Substituting this choice into Eq.~\eqref{eq: planar basis example 1} further simplifies the non-planar kinematic invariants, yielding
\begin{equation}
    s_{26} = s_{345} - s_{12} - s_{61}
\end{equation}
Substituting our planar basis for $n=6$,
\begin{align}
    s_{i,i+1} &\quad i \in \{1, \dots, 6\} \mod(6) \nonumber \\
    s_{i,i+1,i+2} &\quad i \in \{1, \dots, 3\} \mod(6), \nonumber
    \label{eq: planar basis for n=6}
\end{align}
 into Eq.~\eqref{eq: potential of the scattering equations}, we obtain the following expression for the scattering potential: 
 \small{
\begin{equation}
     \mathcal{S}(x) = \sum_{\substack{i=1\\ \text{mod}(6)}}^6 s_{i,i+1} \log{\left(\frac{|x_{i,i+1}||x_{i+2,i-1}|}{|x_{i,i+2}||x_{i+1,i-1}|}\right)} + \sum_{\substack{i=1\\ \text{mod}(6)}}^3 s_{i,i+1,i+2} \log{\left(\frac{|x_{i,i+2}||x_{i+3,i+5}|}{|x_{i,i+3}||x_{i+2,i+5}|}\right)},\label{eq: n=6 massless scattering potential}
 \end{equation}}
with $x_{ij}\equiv x_i-x_j$.

This expression is composed entirely of \textit{cross ratios}-—projective invariants that quantify the relative spacing of four distinct, ordered points on a projective line, which remain unchanged under projective transformations~\cite{shafarevich2012linear}. On $\mathbb{CP}^1$, it uniquely characterises the configuration of four punctures up to Möbius transformations, making the structure manifestly $SL(2,\mathbb{C})$-invariant \cite{ahlfors1979complex}.

As hinted at previously, to introduce mass in an $SL(2,\mathbb{C})$-invariant manner, we can simply shift the kinematic invariants as
\begin{align*}
     s_{i,i+1} &\to s_{i,i+1} - m^2, \\
     s_{i,i+1,i+2} &\to s_{i,i+1,i+2} - m^2,
\end{align*}
to mimic the effect of massive propagators. This yields the massive scattering potential for $n=6$ particles:
\begin{align}
        \mathcal{S}(x) = &\sum_{\substack{i=1\\ \text{mod}(6)}}^6 (s_{i,i+1}-m^2) \log{\left(\frac{|x_{i,i+1}||x_{i+2,i-1}|}{|x_{i,i+2}||x_{i+1,i-1}|}\right)} \nonumber \\
        &+ \sum_{\substack{i=1\\ \text{mod}(6)}}^3 (s_{i,i+1,i+2}-m^2) \log{\left(\frac{|x_{i,i+2}||x_{i+3,i+5}|}{|x_{i,i+3}||x_{i+2,i+5}|}\right)}.
        \label{eq: scattering potential massive n 6}
\end{align}
Eq.~\eqref{eq: scattering potential massive n 6} reproduces Eq.~\eqref{eq:massive_dolan_goddard} for $n=6$ upon imposing the critical point condition
\begin{equation*}
    \frac{\partial \mathcal{S}}{\partial x_a} = 0.
\end{equation*}

The next section generalises Eq.~\eqref{eq: scattering potential massive n 6} to $n$ particles using a specific counting method for planar kinematic invariants. We then examine the structure of the resulting formula, highlighting its connection to particular cross ratios.

\subsection{\texorpdfstring{Maximally Symmetric Ferrers Shapes \& Cross Ratios in $SL(2,\mathbb{C})$-Invariant Scattering}{Maximally Symmetric Ferrers Shapes \& Cross Ratios in SL(2,C)-Invariant Scattering}}\label{subsec: Counting Planar Mandelstam Invariants via Maximally Symmetric Ferrers Shapes.}

The planar scattering potential for $n-$particles is given by
\begin{equation}
    \mathcal{S}(\{x_i\}) = \sum_{k=2}^{n-2} \sum_{i=1}^{k} \left( s_{i,\ldots,i+n-k-1} - m^2 \right) 
\log \left( \frac{|x_{i,i+n-k-1}|\, |x_{i+n-k,i-1}|}{|x_{i,i+n-k}|\, |x_{i+n-k-1,i-1}|} \right)
\label{eq: scattering potential - general - planar ordered particles - planar kinematic invariants}
\end{equation}
This formula was motivated by a recurring structure in the potential when all kinematic invariants--—across the full range of possible lengths--were included in the basis. Instead of truncating to favour shorter invariants through momentum conservation---which imposes a maximum length of $\frac{n}{2}$ for even $n$ and $\frac{n-1}{2}$ for odd $n$---we retain all invariants of all possible lengths, and in doing so, sidestepping the need to distinguish between even and odd particle numbers. The form of Eq.~\eqref{eq: scattering potential - general - planar ordered particles - planar kinematic invariants} then provides a direct route to counting planar kinematic invariants via a double sum structure:
\begin{equation}
    \sum_{k=2}^{n-2} \
\sum_{i=1}^{k}s_{i,i+1,\cdots, i+n-k-1}.
    \label{eq: counting structure for planar kinematic invariants}
\end{equation}
 Does this double sum correctly count the number of planar kinematic invariants? Let's think of $s_{i,i+1,\cdots, i+n-k-1}$ as objects rather than Mandelstam invariants for our counting exercise. The inner sum runs over $i$, with $n$ and $k$ fixed, counting $k$ objects:
\begin{equation*}
    \left(\sum_{k=2}^{n-2} \
\sum_{i=1}^{k}s_{i,i+1,\cdots, i+n-k-1} \right) = \sum_{k=2}^{n-2} k.
\end{equation*}
This can be calculated using the integer summation formula, often attributed to Gauss~\cite{rosen2011discrete},
\begin{equation*}
    \sum_{k=1}^m k = \frac{m(m+1)}{2},
\end{equation*}
but we must be careful because the sum we want to calculate begins at $k=2$, not at $k=1$. Proceeding we find 
\begin{equation*}
    \sum_{k=2}^{n-2} k = \frac{(n-2)(n-1)}{2}-1 = \frac{n(n-3)}{2},
\end{equation*}
giving us the correct number for the basis of kinematic invariants. Great, this works, but how does the counting work in the double sum? How the sum counts the planar kinematic invariants is through the following partitioning: 
\begin{enumerate}
    \item The double sum considers kinematic invariants of all possible lengths, from the minimal length of $2$ to the maximal length of $n-2$.
    \item For all possible lengths, list the planar kinematic invariants until the last index of the length is $n-1$,\footnote{This pivot for counting is entirely for our convenience, but one can devise a similar listing rule.} then stop and move to the next length until you reach invariants of length $n-2$. 
\end{enumerate}
\begin{figure}[hbt!]
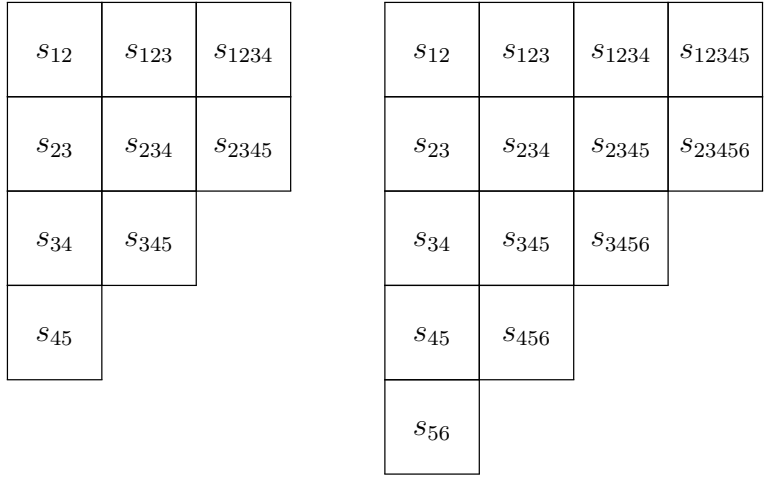

    \centering
    \ytableausetup{boxsize=3em} % Set larger box size
    \newcommand{\inv}[1]{\scriptstyle #1} % Shrink text inside boxes
    \begin{tabular}{c @{\hspace{3em}} c}% Create two columns (left and right)
        % n=6 Ferrers Diagram (Left Side)
        \begin{ytableau}
            s_{12} & s_{123} & s_{1234} \\
            s_{23} & s_{234} & s_{2345} \\
            s_{34} & s_{345} \\
            s_{45} 
        \end{ytableau}
        &
        % n=7 Ferrers Diagram (Right Side)
        \begin{ytableau}
            s_{12} & s_{123} & s_{1234} & s_{12345} \\
            s_{23} & s_{234} & s_{2345} & s_{23456} \\
            s_{34} & s_{345} & s_{3456} \\
            s_{45} & s_{456} \\
            s_{56}
        \end{ytableau}
    \end{tabular}  
    \caption{Counting planar kinematic invariants visualised as a staircase Ferrer's diagram. Examples in this figure include $n=6$ (left) and $n=7$  (right) particles, with the planar kinematic invariants ordered in the blocks.}
    \label{fig:ferrers}
\end{figure}

Following these steps, one realises that there are $n-2$ elements for invariants of length $2$, $n-3$ elements for invariants of length $3$, and so on down to $2$ elements for invariants of length $n-2$. Partitioning our objects in such a way can be represented as a \emph{Ferrer's shape} of size $n(n-3)/2$ of partition $(2,3,....,n-2)$, such as the cases for $n=6$ and $n=7$ as shown in Fig. \ref{fig:ferrers}. A Ferrer's shape of a partition $(a_1, a_2,...,a_k)$ with size $n$ is a configuration of $n$ square boxes arranged such that the $i ^{\mathrm{th}}$ row contains $a_i$ boxes, with all rows aligned vertically \cite{stanley_ec1, bona2016walk}. They provide a clear visual representation of how objects can be partitioned, given the bijection between partitions of $n$ and Ferrer's shapes \cite{bona2016walk}.

By inspection, the number of blocks in such a Ferrer's shape for $n$ particle dynamics can be obtained by adding $n-2$ to the number of blocks for $n-1$ particle dynamics.
A notable feature of its symmetric structure is that the conjugate of this Ferrer's shape is partitioned as $(n-2, n-3, \dots, 3, 2)$, and also for $k$ number of blocks in a column, it contains kinematic invariants of length $n-k$. We have established that the outer sum counts the length of the kinematic invariant being assigned to a block within a column of $k$ blocks, while the inner sum allocates all such invariants to each of these $k$ blocks. A more thorough examination of this way of counting the rank of the basis of kinematic invariants is given in Appendix \ref{appendix: Combinatorial Connection to the Polynomial Form of the Scattering Equations}, including a connection to the polynomial form of the scattering equations. 

\begin{figure}
    \centering
\begin{tikzpicture}[>=stealth] % Use 'stealth' arrows
    % Draw the circle
    \draw[thick, blue] (0,0) circle (3cm);
    
    % Define the six points
    \foreach \i in {0,60,...,300} {
        \coordinate (P\i) at (\i:3cm);
    }
    
    % Draw arrows between the vertices anticlockwise
    \foreach \i/\j/\k in {180/240/4, 240/300/5, 300/0/6} {
        \draw[ultra thick, black, <-] (P\i) -- (P\j);
        % Position the label near the middle of each line
        \path (P\i) -- (P\j) node[midway, above=6pt, black] {$k_{\k}$};
    }
    \foreach \i/\j/\k in {0/60/1, 60/120/2, 120/180/3} {
        \draw[ultra thick, black, <-] (P\i) -- (P\j);
        % Position the label near the middle of each line
        \path (P\i) -- (P\j) node[midway, below=5pt, black] {$k_{\k}$};
    }
    
    % Draw the vertices as small dots
    \foreach \i/\n in {0/1, 60/2, 120/3, 180/4, 240/5, 300/6} {
        \fill (P\i) circle (2.8pt);
        % Label each point
        \node at ($(P\i) + (0.1,0.5)$) {$\n$};
    }
\end{tikzpicture}
\caption{Momentum conservation implies that the momenta $k_i$ form a closed loop — each vector connecting to the next without any leftover momentum.}
    \label{fig: momentum conservation label}
\end{figure}
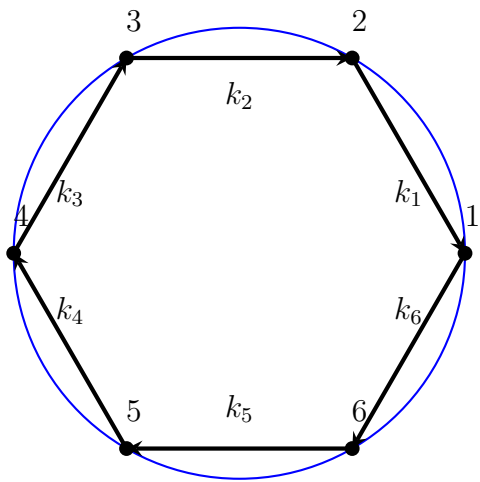

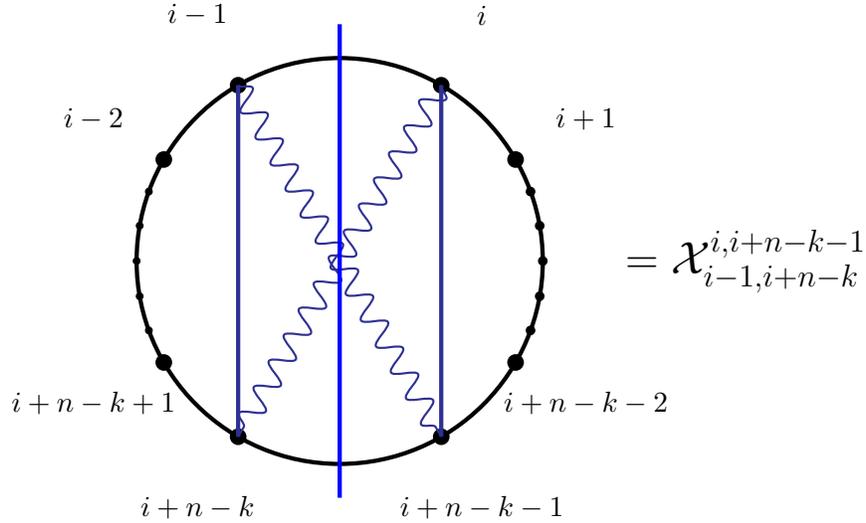
\begin{figure}[hbt!]
    \centering
    \begin{tikzpicture}[scale=0.9]
    % Draw the circle
    \draw[ultra thick, black] (0,0) circle (3);
    
    % Define points on the circle
    \foreach \angle/\label in {120/{i-1},150/{i-2}, 60/i, 30/i+1, 240/{i+n-k}, 210/{i+n-k+1} ,300/{i+n-k-1}, 330/{i+n-k-2}} {
        \node[black] (\label) at (\angle:4.2) {\small $\label$};
        \fill[black] (\angle:3) circle (3.5pt);
    }
    % Draw vertical pink lines (from i to i+n-k-1 and i-1 to i+n-k)
    \draw[Blue, ultra thick] (60:3) -- (300:3);  % i to i+n-k-1
    \draw[Blue, ultra thick] (120:3) -- (240:3); % i-1 to i+n-k

   \foreach \angle in {20, 10, 0, 350,340} {
    \filldraw (\angle:3) circle (1.8pt);
}

% Small circles between i-2 and i+n-k+1
\foreach \angle in {200, 190, 180, 170,160} {
    \filldraw (\angle:3) circle (0.05);
}

    % Wavy lines (propagators) with increased amplitude
    \draw[decorate, decoration={snake, amplitude=1.5mm, segment length=4mm}, Blue, thick] (120:3) -- (300:3);
    \draw[decorate, decoration={snake, amplitude=1.5mm, segment length=4mm}, Blue, thick] (60:3) -- (240:3);

    % Symmetry axis (adjusted for better alignment)
    \draw[blue, ultra thick] (0.0,-3.5) -- (0.0,3.5);
    
     %Equation on the right with refined spacing 
     \node at (6,0.0) {\Large $= \mathcal{X}_{i-1, i+n-k}^{i,i+n-k-1}$};
\end{tikzpicture}
    \caption{Cross ratios of Eq.~\eqref{eq: scattering potential - general - planar ordered particles - planar kinematic invariants} visualised as chords on a planar Feynman diagram. The cross ratio is visualised by drawing a line between the particles to represent the differences in the punctures present in the numerator, while the wavy lines correspond to the puncture differences in the denominator of the logarithm's argument, with the $ \mathcal{X}_{i-1, i+n-k}^{i,i+n-k-1}$ notation encoding this structure.}
    \label{fig: cross ratio planar feynman diagram}
\end{figure}

The argument of the logarithm in Eq.~\eqref{eq: scattering potential - general - planar ordered particles - planar kinematic invariants} highlights a particular cross ratio emerging. To explore this structure further, it is best first to geometrize momentum conservation using chord on a circle. Fig. \ref{fig: momentum conservation label} shows a simplistic way of thinking about the momentum vectors in a Feynman diagram.
A planar kinematic invariant can be described as any chord of the circle upon which we are drawing our planar Feynman diagram. So, the blue chord bisecting the Feynman diagram of Fig. \ref{fig: cross ratio planar feynman diagram} visualises how $s_{i,i+1,\ldots,i+n-k-1} = s_{i+n-k, i+n-k+1,\ldots,i-1}$ via momentum conservation.

This allows us to replace the planar kinematic invariants in Eq.~\eqref{eq: scattering potential - general - planar ordered particles - planar kinematic invariants} by a simple notation encoding both the cross ratio structure and momentum conservation, given by $\mathcal{X}_{i-1, i+n-k}^{i,i+n-k-1}$, as shown in Fig. \ref{fig: cross ratio planar feynman diagram}. Thus, Eq.~\eqref{eq: scattering potential - general - planar ordered particles - planar kinematic invariants}, can be equivalently expressed by
\vspace{-1mm}
\begin{equation}
    \mathcal{S}(\{x_i\}) = \sum_{k=2}^{n-2} \
\sum_{i=1}^{k}\left(\mathcal{X}^{i,i+n-k-1}_{i-1,i+n-k} - m^2 \right) 
\log \left( \frac{|x_{i,i+n-k-1}| \cdot |x_{i+n-k,i-1}|}{|x_{i,i+n-k}| \cdot |x_{i+n-k-1,i-1}|} \right).
\label{eq: scattering potential - general - planar ordered particles - chords}
\end{equation}
This offers a physical interpretation of Dolan and Goddard’s massive extension of the scattering equations, as well as insight into the combinatorial structures underlying colour-ordered Feynman diagrams. The next section returns to the massless case to explore these structures further, revealing how recursive number sequences arise through the construction and factorisation of diagrams.

\section{Catalan Geometry: Building \& Breaking Diagrams}\label{sec: section3}
While the Fibonacci sequence is likely more widely recognized, another sequence of natural numbers—--the \textit{Catalan numbers} $C_n$—--plays a powerful role in combinatorics and carries deep physical significance as well.
\begin{figure}[hbt!]
    \centering
    \begin{equation}
        1,1,2,5,14,42,132,429,1430,4862,...
    \end{equation}
    \caption{The first ten Catalan numbers, starting from $C_0=1$.}
    \label{fig: catalan number sequence}
\end{figure}
These famous numbers have several definitions, some of which highlight the deep combinatorial nature of the number sequence, such as 
\begin{equation}
    C_n = \frac{1}{n+1}\binom{2n}{n},
    \label{eq: combinatorial definition of the Catalan numbers}
\end{equation}
which expresses $C_n$ in terms of the central binomial coefficients\footnote{They are called central given that they make up the centre of the even-numbered rows in \textit{Pascal's triangle} \cite{oeisA000984}.}. Others are more recursive, demonstrating how to count the number of Dyck words of length $2n$ or how convex polygons with $n-2$ sides can be triangulated \cite{stanley_ec1}.

For counting amplitudes, $C_n$ play an important role. The number of planar Feynman diagrams with $n$ external particles in the bi-adjoint $\phi^3$ theory is given by the $(n-2)^{\text{th}}$ Catalan number $C_{n-2}$ \cite{Cachazo_2014_scalars}. This combinatorial fact provides a natural gateway into understanding the physical significance of the various recursive definitions of the Catalan numbers. These recursive structures reveal how amplitudes can be systematically constructed and naturally decomposed into physical and combinatorial substructures.

In this section, we explore how two recursion relations illuminate the role of the $U(1)$ decoupling relation in organising the amplitude's structure, and how the refined splitting of diagrams relates closely to the Narayana numbers---a refinement of the Catalan sequence that captures the distribution of specific diagram topologies.
\subsection{Diagram Splitting \& A Catalan Recursion Relation}
One recursion relation that generates the Catalan numbers is \textit{Segner's recurrence} \cite{Segner1758},
\begin{equation}
C_n = \sum_{k=1}^{n} C_{k-1} C_{n-k}, \quad C_0=1,
\label{eq:catalan recursion relation}
\end{equation}
which bijectively maps to the recursion
\begin{equation}
P_n = \sum_{k=2}^{n-1} P_{k} P_{n-k+1}, \quad P_2=1,
\label{eq:planar_feynman recursion relation}
\end{equation}
where $ P_n $ counts the number of planar Feynman diagrams with $ n $ external particles. This mirrors the recursive structure of building tree diagrams by successively inserting cubic vertices. But what is the physical intuition behind counting this deconstruction of diagrams? 

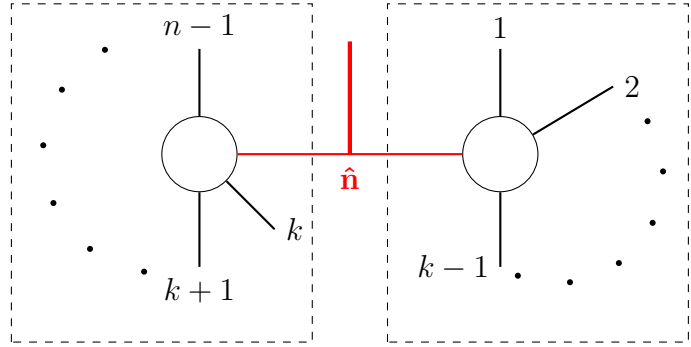
\begin{figure}[h]
\begin{center}
\begin{tikzpicture}

    % Define the positions of the two main vertices
    \node[draw, circle, minimum size=10mm] (VL) at (-2,-1) {}; % Left subdiagram
    \node[draw, circle, minimum size=10mm] (VR) at (2,-1) {};  % Right subdiagram

    % External legs
    \draw[thick] (VL) -- (-2,0.4) node[above] {$ n-1 $};
    \draw[thick] (VL) -- (-1,-2) node[right] {$ k $};
    \draw[thick] (VL) -- (-2,-2.5) node[below] {$ k+1 $};
    
    \draw[thick] (VR) -- (2,-2.5) node[left] {$ k-1 $};
    \draw[thick] (VR) -- (2,0.4) node[above] {$ 1 $};
    \draw[thick] (VR) -- (3.5,-0.1) node[right] {$ 2 $};
    
    % Horizontal propagator
    \draw[thick, red] (VL) -- (VR);
    
    % Vertical line in the middle
    \draw[ultra thick, red] (0,0.5) -- (0.0,-1.0) node[below] {$\mathbf{\hat{n}}$};
    
    % Enclosing shapes for subdiagrams
    \draw[dashed] (-4.5,-3.5) rectangle (-0.5,1);
    \draw[dashed] (0.5,-3.5) rectangle (4.5,1);
    
     \pgfmathsetmacro{\xstart}{3.5}
    \pgfmathsetmacro{\ystart}{-0.1}
    \pgfmathsetmacro{\xend}{2}
    \pgfmathsetmacro{\yend}{-2.5}
    \pgfmathsetmacro{\cx}{(\xstart + \xend)/2}
    \pgfmathsetmacro{\cy}{(\ystart + \yend)/2}
    \pgfmathsetmacro{\r}{sqrt((\xend - \cx)^2 + (\yend - \cy)^2)}

    % Place 6 evenly spaced dots along semicircle (180 to 0 degrees)
    \foreach \i in {1,...,6} {
        \pgfmathsetmacro{\theta}{-300 -(\i*200/7)}
        \pgfmathsetmacro{\x}{\cx + \r*cos(\theta)}
        \pgfmathsetmacro{\y}{\cy + \r*sin(\theta)}
        \fill (\x,\y) circle[radius=0.04];
    }
    % Left arc: from n-1 to k+1
    \pgfmathsetmacro{\xstartL}{-2.0}
    \pgfmathsetmacro{\ystartL}{0.5}
    \pgfmathsetmacro{\xendL}{-2}
    \pgfmathsetmacro{\yendL}{-2.5}
    \pgfmathsetmacro{\cxL}{-2.5}  % move center to the left
    \pgfmathsetmacro{\cyL}{(\ystartL + \yendL)/2}
    \pgfmathsetmacro{\rL}{sqrt((\xstartL - \cxL)^2 + (\ystartL - \cyL)^2)}

    \foreach \i in {1,...,6} {
        \pgfmathsetmacro{\theta}{90 + (\i*200/7)}
        \pgfmathsetmacro{\x}{\cxL + \rL*cos(\theta)}
        \pgfmathsetmacro{\y}{\cyL + \rL*sin(\theta)}
        \fill (\x,\y) circle[radius=0.04];
    }
\end{tikzpicture}
\end{center}
\caption{Diagram Factorisation as suggested by Eq.~\eqref{eq:planar_feynman recursion relation}.}
\label{fig: planar feynman recursion relation factorisation}
\end{figure}

The structure of Eq.~\eqref{eq:planar_feynman recursion relation} implies that the planar diagrams split into two sub-diagrams of various sizes. This is visualised in Fig. \ref{fig: planar feynman recursion relation factorisation}. Inferred from the figure is the special nature of the $n^{th}$ particle, which for a given $k\in \{2,...,n-1\}$, splits the $n$-particle diagram into two smaller diagrams - one with particles $1$ through to $k-1$ and the other with particles $k$ through to $n-1$. 
Remarkably, this splitting property is nothing but a fundamental relation among colour-ordered amplitudes in disguise: \emph{the $U(1)$ decoupling identity}. For an $n$-particle amplitude, it states that the sum over all $(n - 1)$ cyclic permutations of external legs —-- where a fixed leg remains in place while the others cycle --- must vanish \cite{Elvang:2013cua,Mangano:1990by}. Cycling through the $n^{th}$ particle just like for our relation, simply gives the sum
\begin{equation}
    \sum_{k=1}^{n-1}A_{n}(1,2,3,...,k,n,k+1,...,n-1)=0.
    \label{eq: the U(1) decoupling relation}
\end{equation}
This relation, also known as the \textit{photon decoupling identity}, arises when one of the generators $ T_a $ in Eq.~\eqref{eq: biadjoint phi3 colour ordered formal expression} is proportional to the identity~\cite{Dixon:1996wi}, reflecting the fact that a photon does not self-interact.

\begin{figure}[hbt!]
\begin{center}
\begin{tikzpicture}[scale = 0.9]

    % First Circle - Good Graph (Green, Planar)
    \draw (0,0) circle (2);
    
    % Nodes for the first diagram
    \foreach \i in {1,2,3,4} {
        \node[fill=black, circle, inner sep=1.8pt] (A\i) at ($(0,0) + (45-90*\i:2)$) {};
        \node at ($(0,0) + (-45+90*\i:2.6)$) {\textbf{\large \i}};
    }

    % Edges
    \draw[Red, thick] (A1) -- (A2);
    \draw[Red, thick] (A2) -- (A3);
    \draw[Red, thick] (A3) -- (A4);
    \draw[Red, thick] (A4) -- (A1);
    \draw[black, thick] (A1) -- (0,0);
    \draw[black, thick] (A2) -- (0,0);
    \draw[black, thick] (A3) -- (0,0);
    \draw[black, thick] (A4) -- (0,0);

    % Mark center
    \filldraw[fill=Red, draw=black] (0,0) circle (0.7);

    % Second diagram - RED GRAPH 1
    \begin{scope}[xshift=5cm]
        \draw (0,0) circle (2);

        \foreach \i in {1,2,3,4} {
            \node[fill=black, circle, inner sep=1.8pt] (B\i) at ($(0,0) + (45-90*\i:2)$) {};
            \node at ($(0,0) + (-45+90*\i:2.6)$) {\textbf{\large \i}};
        }

        \draw[Red, thick] (B1) -- (B2);
        \draw[Red, thick] (B3) -- (B4);
        \draw[Red, thick] (B4) -- (B2);
        \draw[Red, thick] (B1) -- (B3);
        \draw[black, thick] (B3) -- (0,0.75) ;
        \draw[black, thick] (B4) -- (0,0.75) ;
        \draw[black, thick] (0,0.75) -- (0,-0.75) ;
        \draw[black, thick] (B1) -- (0,-0.75) ;
        \draw[black, thick] (B2) -- (0,-0.75) ;

         \node[circle, fill=black, inner sep=1.8pt] at (0,0.75) {};
         \node[circle, fill=black, inner sep=1.8pt] at (0,-0.75) {};
    \end{scope}

    % Third diagram - RED GRAPH 2
    \begin{scope}[xshift=10cm]
        \draw (0,0) circle (2);

        \foreach \i in {1,2,3,4} {
            \node[fill=black, circle, inner sep=1.8pt] (C\i) at ($(0,0) + (45-90*\i:2)$) {};
            \node at ($(0,0) + (-45+90*\i:2.6)$) {\textbf{\large \i}};
        }
        \node[circle, fill=black, inner sep=1.8pt] at (0.75,0.0) {};
        \node[circle, fill=black, inner sep=1.8pt] at (-0.75,0.0) {};
        
        \draw[Red, thick] (C1) -- (C4);
        \draw[Red, thick] (C2) -- (C3);
        \draw[Red, thick] (C4) -- (C2);
        \draw[Red, thick] (C1) -- (C3);
        \draw[black, thick] (C1) -- (0.75,0.0);
        \draw[black, thick] (C4) -- (0.75,0.0);
        \draw[black, thick] (-0.75,0.0) -- (0.75,0.0);
        \draw[black, thick] (C2) -- (-0.75,0.0);
        \draw[black, thick] (C3) -- (-0.75,0.0);
        
    \end{scope}

    \node at (2.5, 0) {\Huge $\mathbf{=}$}; 
    \node at (7.5, 0) {\Huge $\mathbf{+}$};

\end{tikzpicture}

\end{center}
    \caption{Pictorial representation of the Catalan recursion relation and the $U(1)$ decoupling relation for $n=4$ particles in the bi-adjoint theory.}
    \label{fig: four particle picture catalan number U(1)}
\end{figure}
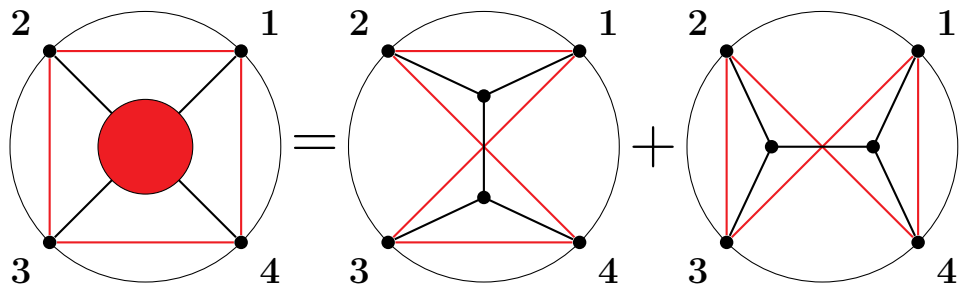

\begin{comment}
   To see this correspondence more explicitly, we can look at colour-ordered amplitudes of such as in Figure \ref{fig:5 point case - U(1) decoupling identity}, where there is a structure constants with two dummy indices being summed over, suggesting a specific structure to the diagram. 
\begin{figure}
\centering
\begin{center}
 \tikzpicture [scale=1.0]
 \draw [line width=0.30mm]  (2,0.5) -- (1.5,1)node[left]{$a_2$}  ;
 \draw [line width=0.30mm]  (2,0.5) -- (1.5,0)node[left]{$a_1$} ;
 %
 \draw [line width=0.30mm]  (2,0.5) -- (3,0.5) ;
\draw [ultra thick, red, line width=0.30mm]  (2.5,0.5) -- (2.5,1) node[above]{$a_5$};
%
 \draw [line width=0.30mm]  (3,0.5) -- (3.5,1)node[right]{$a_3$}  ;
 \draw [line width=0.30mm]  (3,0.5) -- (3.5,0)node[right]{$a_4$} ;
 %
\draw(5,0.5)node{$\longleftrightarrow$}; 
 %
\draw(6,0.5)node{$\Bigg\{$}; 
\draw(8.8,0.5)node{$\sum_{b,c=1}^{N^2}f^{a_1a_2b}f^{ba_5c}f^{ca_3a_4}$};
 \endtikzpicture
\end{center} 
    \caption{Colour Diagram Type Example for $n=5$.}
    \label{fig:5 point case - U(1) decoupling identity}
\end{figure} 
\end{comment}
To see this correspondence more explicitly, for $n=4$ particles, we have the following sum over planar Feynman diagrams and the relevant $U(1)$ decoupling identity
\begin{align}
    &P_4\;\;\;\;\;\;\; = \;\;\;\;\;\;\;\;\;\;P_3P_2\;\;\;\;\;\;\;+ \;\;\;\;\;\;\;\;P_2P_3 \\
    &\updownarrow \;\;\;\;\;\;\;\;\;\;\;\;\;\;\;\;\;\;\;\;\;\;\;\;\;\;\updownarrow\;\;\;\;\;\; \;\;\;\;\;\;\;\;\;\;\;\;\;\;\;\;\;\;\updownarrow \nonumber\\
    m(1234&|1234)\:\: =-m(1234|1243)-m(1234|1423), 
\end{align}
or pictorially as shown in Fig. \ref{fig: four particle picture catalan number U(1)}. 
The recursive splitting of $n$-point bi-adjoint diagrams prompts several natural combinatorial questions. How many distinct diagrammatic structures arise for a given $n$? Among these, how many feature a maximal sub-diagram involving only $p<n$ particles? Furthermore, is there an underlying numerical pattern governing the frequency of each diagram type produced by the splitting? This will be the focus of the next section, including an intriguing connection with the \textit{Narayana numbers}.

\subsection{Counting Planar Diagram Splitting - The Narayana Numbers}
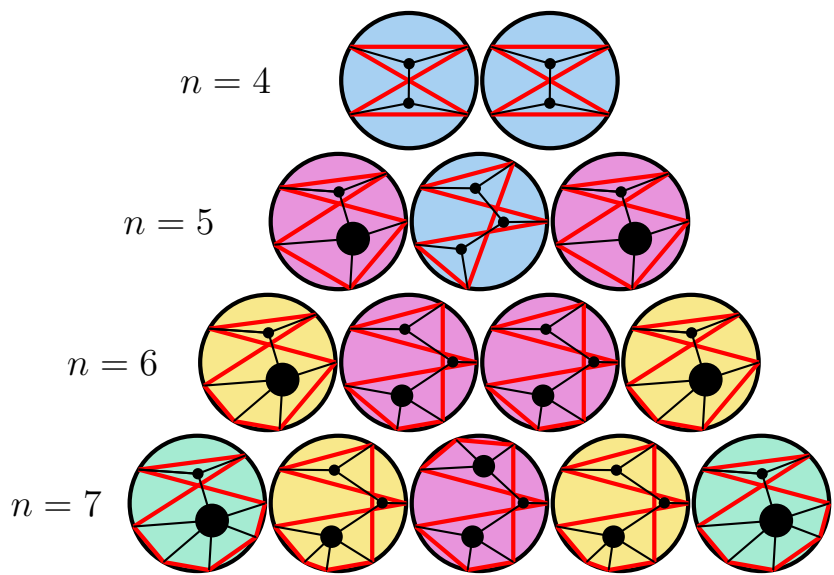
\begin{figure}[hbt!]
    \centering
\begin{tikzpicture}[scale=0.75]
\definecolor{myred}{RGB}{167, 208, 242}
\definecolor{mypink}{RGB}{232, 148, 220}
\definecolor{myyellow}{RGB}{248, 232, 139}
\definecolor{mygreen}{RGB}{165, 235, 210}

% Radius and spacing
\def\r{1.2}
\def\xgap{2.5}
\def\ygap{2.5}

% Y positions
\def\yFour{3*\ygap}
\def\yFive{2*\ygap}
\def\ySix{\ygap}
\def\ySeven{0}

% Row labels
\node at (-4.5,\yFour) {\textcolor{black}{\large $n = 4$}};
\node at (-5.5,\yFive) {\textcolor{black}{\large $n = 5$}};
\node at (-6.5,\ySix) {\textcolor{black}{\large $n = 6$}};
\node at (-7.5,\ySeven) {\textcolor{black}{\large $n = 7$}};

% n=4
\foreach \i/\c in {0/myred, 1/myred} {
    \pgfmathsetmacro{\x}{\i*\xgap - 0.5*\xgap}
    \pgfmathsetmacro{\y}{\yFour}

    % Draw the circle
    \draw[fill=\c, ultra thick] (\x,\y) circle (\r);

    % Triangle vertex coordinates
    \pgfmathsetmacro{\xA}{\x + \r*cos(30)}
    \pgfmathsetmacro{\yA}{\y + \r*sin(30)}
    \pgfmathsetmacro{\xB}{\x + \r*cos(150)}
    \pgfmathsetmacro{\yB}{\y + \r*sin(150)}
    \pgfmathsetmacro{\xC}{\x + \r*cos(210)}
    \pgfmathsetmacro{\yC}{\y + \r*sin(210)}
    \pgfmathsetmacro{\xD}{\x + \r*cos(330)}
    \pgfmathsetmacro{\yD}{\y + \r*sin(330)}

    % Draw red triangle edges
    \draw[red, ultra thick] (\xA,\yA) -- (\xB,\yB); % Top chord
    \draw[red, ultra thick] (\xC,\yC) -- (\xD,\yD); % Bottom chord
    \draw[red, ultra thick] (\xC,\yC) -- (\xA,\yA);
    \draw[red, ultra thick] (\xD,\yD) -- (\xB,\yB);
    
    % TOP triangle: A (30°), B (150°), D (330°)
    \pgfmathsetmacro{\xTop}{(\xA + \xB + \xD)/3-0.35}
    \pgfmathsetmacro{\yTop}{(\yA + \yB + \yD)/3-0.6}
    \fill[black] (\xTop,\yTop) circle (0.1);

    % BOTTOM triangle: A (30°), B (150°), C (210°)
    \pgfmathsetmacro{\xBot}{(\xA + \xB + \xC)/3+0.35}
    \pgfmathsetmacro{\yBot}{(\yA + \yB + \yC)/3+0.1}
    \fill[black] (\xBot,\yBot) circle (0.1);

    \draw[black, thick] (\xTop,\yTop) --  (\xBot,\yBot);
    \draw[black, thick] (\xA,\yA) --  (\xBot,\yBot);
    \draw[black, thick] (\xB,\yB) --  (\xBot,\yBot);
    \draw[black, thick] (\xC,\yC) --  (\xTop,\yTop);
    \draw[black, thick] (\xD,\yD) --  (\xTop,\yTop);
}

% n=5

\foreach \i/\c in {0/mypink, 1/myred, 2/mypink} {
    \pgfmathsetmacro{\x}{\i*\xgap - 1*\xgap}
    \pgfmathsetmacro{\y}{\yFive}
    \draw[fill=\c, ultra thick] (\x,\y) circle (\r);

    % Only apply for pink circles at positions 0 and 2
    \ifnum\i=0
        \def\doTriangle{1}
    \else
        \ifnum\i=2
            \def\doTriangle{1}
        \else
            \def\doTriangle{0}
        \fi
    \fi

    \ifnum\doTriangle=1
        % Triangle vertex coordinates
        \pgfmathsetmacro{\xA}{\x + \r*cos(45)}
        \pgfmathsetmacro{\yA}{\y + \r*sin(45)}
        \pgfmathsetmacro{\xB}{\x + \r*cos(150)}
        \pgfmathsetmacro{\yB}{\y + \r*sin(150)}
        \pgfmathsetmacro{\xC}{\x + \r*cos(200)}
        \pgfmathsetmacro{\yC}{\y + \r*sin(200)}
        \pgfmathsetmacro{\xD}{\x + \r*cos(280)}
        \pgfmathsetmacro{\yD}{\y + \r*sin(280)}
        \pgfmathsetmacro{\xE}{\x + \r*cos(360)}
        \pgfmathsetmacro{\yE}{\y + \r*sin(360)}

        % Draw red triangle edges
        \draw[red, ultra thick] (\xA,\yA) -- (\xB,\yB);
        \draw[red, ultra thick] (\xB,\yB) -- (\xE,\yE);
        \draw[red, ultra thick] (\xA,\yA) -- (\xC,\yC);
        \draw[red, ultra thick] (\xD,\yD) -- (\xC,\yC);
        \draw[red, ultra thick] (\xD,\yD) -- (\xE,\yE);

    \pgfmathsetmacro{\xBot}{(\xA + \xB + \xC)/3+0.45}
    \pgfmathsetmacro{\yBot}{(\yA + \yB + \yC)/3+0.18}
    \fill[black] (\xBot,\yBot) circle (0.1);
    \pgfmathsetmacro{\xsegment}{(\xA + \xB + \xC)/3+0.7}
    \pgfmathsetmacro{\ysegment}{(\yA + \yB + \yC)/3-0.65}
    \fill[black] (\xsegment,\ysegment) circle (0.3);

    \draw[black, thick] (\xA,\yA) -- (\xBot,\yBot);
    \draw[black, thick] (\xB,\yB) -- (\xBot,\yBot);
    \draw[black, thick] (\xB,\yB) -- (\xBot,\yBot);
    \draw[black, thick] (\xBot,\yBot) -- (\xsegment,\ysegment);
    \draw[black, thick] (\xC,\yC) -- (\xsegment,\ysegment);
    \draw[black, thick] (\xD,\yD) -- (\xsegment,\ysegment);
    \draw[black, thick] (\xE,\yE) -- (\xsegment,\ysegment);

    \fi
    % Add a section (wedge) to the red circle at i = 1
    \ifnum\i=1
       \pgfmathsetmacro{\xA}{\x + \r*cos(60)}
        \pgfmathsetmacro{\yA}{\y + \r*sin(60)}
        \pgfmathsetmacro{\xB}{\x + \r*cos(150)}
        \pgfmathsetmacro{\yB}{\y + \r*sin(150)}
        \pgfmathsetmacro{\xC}{\x + \r*cos(200)}
        \pgfmathsetmacro{\yC}{\y + \r*sin(200)}
        \pgfmathsetmacro{\xD}{\x + \r*cos(260)}
        \pgfmathsetmacro{\yD}{\y + \r*sin(260)}
        \pgfmathsetmacro{\xE}{\x + \r*cos(360)}
        \pgfmathsetmacro{\yE}{\y + \r*sin(360)}

        % Draw red triangle edges
        \draw[red, ultra thick] (\xA,\yA) -- (\xB,\yB);
        \draw[red, ultra thick] (\xB,\yB) -- (\xE,\yE);
        \draw[red, ultra thick] (\xE,\yE) -- (\xC,\yC);
        \draw[red, ultra thick] (\xD,\yD) -- (\xC,\yC);
        \draw[red, ultra thick] (\xA,\yA) -- (\xD,\yD);

    \pgfmathsetmacro{\xBot}{(\xA + \xB + \xC)/3+0.45}
    \pgfmathsetmacro{\yBot}{(\yA + \yB + \yC)/3+0.18}
    \fill[black] (\xBot,\yBot) circle (0.1);
    \pgfmathsetmacro{\xsegment}{(\xA + \xB + \xC)/3+0.2}
    \pgfmathsetmacro{\ysegment}{(\yA + \yB + \yC)/3-0.9}
    \fill[black] (\xsegment,\ysegment) circle (0.1);
    \pgfmathsetmacro{\xsegmenttwo}{(\xA + \xB + \xC)/3+0.95}
    \pgfmathsetmacro{\ysegmenttwo}{(\yA + \yB + \yC)/3-0.42}
    \fill[black] (\xsegmenttwo,\ysegmenttwo) circle (0.1);

    \draw[black, thick] (\xA,\yA) -- (\xBot,\yBot);
        \draw[black, thick] (\xB,\yB) -- (\xBot,\yBot);
        \draw[black, thick] (\xBot,\yBot) -- (\xsegmenttwo,\ysegmenttwo);
        \draw[black, thick] (\xsegmenttwo,\ysegmenttwo) -- (\xsegment,\ysegment);
        \draw[black, thick] (\xsegment,\ysegment) -- (\xD,\yD);
        \draw[black, thick] (\xsegment,\ysegment) -- (\xC,\yC);
        \draw[black, thick] (\xsegmenttwo,\ysegmenttwo) -- (\xE,\yE);

    \fi
}

% n=6
% n=6
\foreach \i/\c in {0/myyellow, 1/mypink, 2/mypink, 3/myyellow} {
    \pgfmathsetmacro{\x}{\i*\xgap - 1.5*\xgap}
    \pgfmathsetmacro{\y}{\ySix}
    \draw[fill=\c, ultra thick] (\x,\y) circle (\r);

    % === Custom for mypink circles at index 1 and 2 ===
    \ifnum\i=1
        \def\doPink{1}
    \else
        \ifnum\i=2
            \def\doPink{1}
        \else
            \def\doPink{0}
        \fi
    \fi

    \ifnum\doPink=1
       \pgfmathsetmacro{\xA}{\x + \r*cos(60)}
        \pgfmathsetmacro{\yA}{\y + \r*sin(60)}
        \pgfmathsetmacro{\xB}{\x + \r*cos(150)}
        \pgfmathsetmacro{\yB}{\y + \r*sin(150)}
        \pgfmathsetmacro{\xC}{\x + \r*cos(200)}
        \pgfmathsetmacro{\yC}{\y + \r*sin(200)}
        \pgfmathsetmacro{\xD}{\x + \r*cos(260)}
        \pgfmathsetmacro{\yD}{\y + \r*sin(260)}
        \pgfmathsetmacro{\xE}{\x + \r*cos(360)}
        \pgfmathsetmacro{\yE}{\y + \r*sin(360)}
        \pgfmathsetmacro{\xF}{\x + \r*cos(300)}
        \pgfmathsetmacro{\yF}{\y + \r*sin(300)}

        % Draw red triangle edges
        \draw[red, ultra thick] (\xA,\yA) -- (\xB,\yB);
        \draw[red, ultra thick] (\xB,\yB) -- (\xE,\yE);
        \draw[red, ultra thick] (\xE,\yE) -- (\xC,\yC);
        \draw[red, ultra thick] (\xD,\yD) -- (\xC,\yC);
        \draw[red, ultra thick] (\xF,\yF) -- (\xD,\yD);
        \draw[red, ultra thick] (\xF,\yF) -- (\xA,\yA);

    \pgfmathsetmacro{\xBot}{(\xA + \xB + \xC)/3+0.45}
    \pgfmathsetmacro{\yBot}{(\yA + \yB + \yC)/3+0.18}
    \fill[black] (\xBot,\yBot) circle (0.1);
    \pgfmathsetmacro{\xsegment}{(\xA + \xB + \xC)/3+0.4}
    \pgfmathsetmacro{\ysegment}{(\yA + \yB + \yC)/3-1.0}
    \fill[black] (\xsegment,\ysegment) circle (0.2);
    \pgfmathsetmacro{\xsegmenttwo}{(\xA + \xB + \xC)/3+1.3}
    \pgfmathsetmacro{\ysegmenttwo}{(\yA + \yB + \yC)/3-0.4}
    \fill[black] (\xsegmenttwo,\ysegmenttwo) circle (0.1);

    \draw[black, thick] (\xA,\yA) -- (\xBot,\yBot);
        \draw[black, thick] (\xB,\yB) -- (\xBot,\yBot);
        \draw[black, thick] (\xBot,\yBot) -- (\xsegmenttwo,\ysegmenttwo);
        \draw[black, thick] (\xsegmenttwo,\ysegmenttwo) -- (\xsegment,\ysegment);
        \draw[black, thick] (\xsegment,\ysegment) -- (\xD,\yD);
        \draw[black, thick] (\xsegment,\ysegment) -- (\xC,\yC);
        \draw[black, thick] (\xsegment,\ysegment) -- (\xF,\yF);
        \draw[black, thick] (\xsegmenttwo,\ysegmenttwo) -- (\xE,\yE);
        
    \fi

    % === Custom for myyellow circles at index 0 and 3 ===
    \ifnum\i=0
        \def\doYellow{1}
    \else
        \ifnum\i=3
            \def\doYellow{1}
        \else
            \def\doYellow{0}
        \fi
    \fi

    \ifnum\doYellow=1
        \pgfmathsetmacro{\xA}{\x + \r*cos(45)}
        \pgfmathsetmacro{\yA}{\y + \r*sin(45)}
        \pgfmathsetmacro{\xB}{\x + \r*cos(150)}
        \pgfmathsetmacro{\yB}{\y + \r*sin(150)}
        \pgfmathsetmacro{\xC}{\x + \r*cos(200)}
        \pgfmathsetmacro{\yC}{\y + \r*sin(200)}
        \pgfmathsetmacro{\xD}{\x + \r*cos(280)}
        \pgfmathsetmacro{\yD}{\y + \r*sin(280)}
        \pgfmathsetmacro{\xE}{\x + \r*cos(360)}
        \pgfmathsetmacro{\yE}{\y + \r*sin(360)}
        \pgfmathsetmacro{\xF}{\x + \r*cos(240)}
        \pgfmathsetmacro{\yF}{\y + \r*sin(240)}

        % Draw red triangle edges
        \draw[red, ultra thick] (\xA,\yA) -- (\xB,\yB);
        \draw[red, ultra thick] (\xB,\yB) -- (\xE,\yE);
        \draw[red, ultra thick] (\xA,\yA) -- (\xC,\yC);
        \draw[red, ultra thick] (\xF,\yF) -- (\xC,\yC);
        \draw[red, ultra thick] (\xD,\yD) -- (\xF,\yF);
        \draw[red, ultra thick] (\xD,\yD) -- (\xE,\yE);

    \pgfmathsetmacro{\xBot}{(\xA + \xB + \xC)/3+0.45}
    \pgfmathsetmacro{\yBot}{(\yA + \yB + \yC)/3+0.18}
    \fill[black] (\xBot,\yBot) circle (0.1);
    \pgfmathsetmacro{\xsegment}{(\xA + \xB + \xC)/3+0.7}
    \pgfmathsetmacro{\ysegment}{(\yA + \yB + \yC)/3-0.65}
    \fill[black] (\xsegment,\ysegment) circle (0.3);

    \draw[black, thick] (\xA,\yA) -- (\xBot,\yBot);
    \draw[black, thick] (\xB,\yB) -- (\xBot,\yBot);
    \draw[black, thick] (\xB,\yB) -- (\xBot,\yBot);
    \draw[black, thick] (\xBot,\yBot) -- (\xsegment,\ysegment);
    \draw[black, thick] (\xC,\yC) -- (\xsegment,\ysegment);
    \draw[black, thick] (\xD,\yD) -- (\xsegment,\ysegment);
    \draw[black, thick] (\xE,\yE) -- (\xsegment,\ysegment);
    \draw[black, thick] (\xF,\yF) -- (\xsegment,\ysegment);
    \fi
}

% n=7
% Define positions and colors for n = 7
\foreach \i/\c in {0/mygreen, 1/myyellow, 2/mypink, 3/myyellow, 4/mygreen} {
    \pgfmathsetmacro{\x}{\i*\xgap - 2*\xgap}
    \pgfmathsetmacro{\y}{\ySeven}
    \draw[fill=\c, ultra thick] (\x,\y) circle (\r);

    % === Green circle logic (vertical chord)
    \ifnum\i=0
        \def\doGreen{1}
    \else
        \ifnum\i=4
            \def\doGreen{1}
        \else
            \def\doGreen{0}
        \fi
    \fi
    \ifnum\doGreen=1
        \pgfmathsetmacro{\xA}{\x + \r*cos(45)}
        \pgfmathsetmacro{\yA}{\y + \r*sin(45)}
        \pgfmathsetmacro{\xB}{\x + \r*cos(150)}
        \pgfmathsetmacro{\yB}{\y + \r*sin(150)}
        \pgfmathsetmacro{\xC}{\x + \r*cos(200)}
        \pgfmathsetmacro{\yC}{\y + \r*sin(200)}
        \pgfmathsetmacro{\xD}{\x + \r*cos(280)}
        \pgfmathsetmacro{\yD}{\y + \r*sin(280)}
        \pgfmathsetmacro{\xE}{\x + \r*cos(360)}
        \pgfmathsetmacro{\yE}{\y + \r*sin(360)}
        \pgfmathsetmacro{\xF}{\x + \r*cos(240)}
        \pgfmathsetmacro{\yF}{\y + \r*sin(240)}
        \pgfmathsetmacro{\xG}{\x + \r*cos(330)}
        \pgfmathsetmacro{\yG}{\y + \r*sin(330)}

        % Draw red triangle edges
        \draw[red, ultra thick] (\xA,\yA) -- (\xB,\yB);
        \draw[red, ultra thick] (\xB,\yB) -- (\xE,\yE);
        \draw[red, ultra thick] (\xA,\yA) -- (\xC,\yC);
        \draw[red, ultra thick] (\xF,\yF) -- (\xC,\yC);
        \draw[red, ultra thick] (\xD,\yD) -- (\xF,\yF);
        \draw[red, ultra thick] (\xD,\yD) -- (\xG,\yG);
        \draw[red, ultra thick] (\xE,\yE) -- (\xG,\yG);

    \pgfmathsetmacro{\xBot}{(\xA + \xB + \xC)/3+0.45}
    \pgfmathsetmacro{\yBot}{(\yA + \yB + \yC)/3+0.18}
    \fill[black] (\xBot,\yBot) circle (0.1);
    \pgfmathsetmacro{\xsegment}{(\xA + \xB + \xC)/3+0.7}
    \pgfmathsetmacro{\ysegment}{(\yA + \yB + \yC)/3-0.65}
    \fill[black] (\xsegment,\ysegment) circle (0.3);

    \draw[black, thick] (\xA,\yA) -- (\xBot,\yBot);
    \draw[black, thick] (\xB,\yB) -- (\xBot,\yBot);
    \draw[black, thick] (\xB,\yB) -- (\xBot,\yBot);
    \draw[black, thick] (\xBot,\yBot) -- (\xsegment,\ysegment);
    \draw[black, thick] (\xC,\yC) -- (\xsegment,\ysegment);
    \draw[black, thick] (\xD,\yD) -- (\xsegment,\ysegment);
    \draw[black, thick] (\xE,\yE) -- (\xsegment,\ysegment);
    \draw[black, thick] (\xF,\yF) -- (\xsegment,\ysegment);
    \draw[black, thick] (\xG,\yG) -- (\xsegment,\ysegment);
    \fi

    % === Yellow circle logic (wedge sector)
    \ifnum\i=1
        \def\doYellow{1}
    \else
        \ifnum\i=3
            \def\doYellow{1}
        \else
            \def\doYellow{0}
        \fi
    \fi
    \ifnum\doYellow=1
        \pgfmathsetmacro{\xA}{\x + \r*cos(60)}
        \pgfmathsetmacro{\yA}{\y + \r*sin(60)}
        \pgfmathsetmacro{\xB}{\x + \r*cos(150)}
        \pgfmathsetmacro{\yB}{\y + \r*sin(150)}
        \pgfmathsetmacro{\xC}{\x + \r*cos(200)}
        \pgfmathsetmacro{\yC}{\y + \r*sin(200)}
        \pgfmathsetmacro{\xG}{\x + \r*cos(240)}
        \pgfmathsetmacro{\yG}{\y + \r*sin(240)}
        \pgfmathsetmacro{\xD}{\x + \r*cos(260)}
        \pgfmathsetmacro{\yD}{\y + \r*sin(260)}
        \pgfmathsetmacro{\xE}{\x + \r*cos(360)}
        \pgfmathsetmacro{\yE}{\y + \r*sin(360)}
        \pgfmathsetmacro{\xF}{\x + \r*cos(300)}
        \pgfmathsetmacro{\yF}{\y + \r*sin(300)}

        % Draw red triangle edges
        \draw[red, ultra thick] (\xA,\yA) -- (\xB,\yB);
        \draw[red, ultra thick] (\xB,\yB) -- (\xE,\yE);
        \draw[red, ultra thick] (\xE,\yE) -- (\xC,\yC);
        \draw[red, ultra thick] (\xG,\yG) -- (\xC,\yC);
        \draw[red, ultra thick] (\xD,\yD) -- (\xG,\yG);
        \draw[red, ultra thick] (\xD,\yD) -- (\xF,\yF);
        \draw[red, ultra thick] (\xF,\yF) -- (\xA,\yA);

    \pgfmathsetmacro{\xBot}{(\xA + \xB + \xC)/3+0.45}
    \pgfmathsetmacro{\yBot}{(\yA + \yB + \yC)/3+0.18}
    \fill[black] (\xBot,\yBot) circle (0.1);
    \pgfmathsetmacro{\xsegment}{(\xA + \xB + \xC)/3+0.4}
    \pgfmathsetmacro{\ysegment}{(\yA + \yB + \yC)/3-1.0}
    \fill[black] (\xsegment,\ysegment) circle (0.2);
    \pgfmathsetmacro{\xsegmenttwo}{(\xA + \xB + \xC)/3+1.3}
    \pgfmathsetmacro{\ysegmenttwo}{(\yA + \yB + \yC)/3-0.4}
    \fill[black] (\xsegmenttwo,\ysegmenttwo) circle (0.1);

    \draw[black, thick] (\xA,\yA) -- (\xBot,\yBot);
        \draw[black, thick] (\xB,\yB) -- (\xBot,\yBot);
        \draw[black, thick] (\xBot,\yBot) -- (\xsegmenttwo,\ysegmenttwo);
        \draw[black, thick] (\xsegmenttwo,\ysegmenttwo) -- (\xsegment,\ysegment);
        \draw[black, thick] (\xsegment,\ysegment) -- (\xD,\yD);
        \draw[black, thick] (\xsegment,\ysegment) -- (\xC,\yC);
        \draw[black, thick] (\xsegment,\ysegment) -- (\xF,\yF);
        \draw[black, thick] (\xsegmenttwo,\ysegmenttwo) -- (\xE,\yE);
        \draw[black, thick] (\xsegment,\ysegment) -- (\xG,\yG);
    \fi

    % === Pink circle logic (triangle)
    \ifnum\i=2
         \pgfmathsetmacro{\xA}{\x + \r*cos(60)}
        \pgfmathsetmacro{\yA}{\y + \r*sin(60)}
        \pgfmathsetmacro{\xG}{\x + \r*cos(110)}
        \pgfmathsetmacro{\yG}{\y + \r*sin(110)}
        \pgfmathsetmacro{\xB}{\x + \r*cos(150)}
        \pgfmathsetmacro{\yB}{\y + \r*sin(150)}
        \pgfmathsetmacro{\xC}{\x + \r*cos(200)}
        \pgfmathsetmacro{\yC}{\y + \r*sin(200)}
        \pgfmathsetmacro{\xD}{\x + \r*cos(260)}
        \pgfmathsetmacro{\yD}{\y + \r*sin(260)}
        \pgfmathsetmacro{\xE}{\x + \r*cos(360)}
        \pgfmathsetmacro{\yE}{\y + \r*sin(360)}
        \pgfmathsetmacro{\xF}{\x + \r*cos(300)}
        \pgfmathsetmacro{\yF}{\y + \r*sin(300)}

        % Draw red triangle edges
        \draw[red, ultra thick] (\xA,\yA) -- (\xG,\yG);
        \draw[red, ultra thick] (\xB,\yB) -- (\xG,\yG);
        \draw[red, ultra thick] (\xB,\yB) -- (\xE,\yE);
        \draw[red, ultra thick] (\xE,\yE) -- (\xC,\yC);
        \draw[red, ultra thick] (\xD,\yD) -- (\xC,\yC);
        \draw[red, ultra thick] (\xF,\yF) -- (\xD,\yD);
        \draw[red, ultra thick] (\xF,\yF) -- (\xA,\yA);

    \pgfmathsetmacro{\xBot}{(\xA + \xB + \xC)/3+0.6}
    \pgfmathsetmacro{\yBot}{(\yA + \yB + \yC)/3+0.25}
    \fill[black] (\xBot,\yBot) circle (0.2);
    \pgfmathsetmacro{\xsegment}{(\xA + \xB + \xC)/3+0.4}
    \pgfmathsetmacro{\ysegment}{(\yA + \yB + \yC)/3-1.0}
    \fill[black] (\xsegment,\ysegment) circle (0.2);
    \pgfmathsetmacro{\xsegmenttwo}{(\xA + \xB + \xC)/3+1.3}
    \pgfmathsetmacro{\ysegmenttwo}{(\yA + \yB + \yC)/3-0.4}
    \fill[black] (\xsegmenttwo,\ysegmenttwo) circle (0.1);

    \draw[black, thick] (\xA,\yA) -- (\xBot,\yBot);
        \draw[black, thick] (\xB,\yB) -- (\xBot,\yBot);
        \draw[black, thick] (\xG,\yG) -- (\xBot,\yBot);
        \draw[black, thick] (\xBot,\yBot) -- (\xsegmenttwo,\ysegmenttwo);
        
        \draw[black, thick] (\xsegmenttwo,\ysegmenttwo) -- (\xsegment,\ysegment);
        \draw[black, thick] (\xsegment,\ysegment) -- (\xD,\yD);
        \draw[black, thick] (\xsegment,\ysegment) -- (\xC,\yC);
        \draw[black, thick] (\xsegment,\ysegment) -- (\xF,\yF);
        \draw[black, thick] (\xsegmenttwo,\ysegmenttwo) -- (\xE,\yE);
    \fi
}
\end{tikzpicture}
    \caption{Block triangle structure of the diagram splitting due to Eq.~\eqref{eq:catalan recursion relation} up to $n=7$ particles.}
    \label{fig: Catalan diagram factorisation triangle}
\end{figure}

The pictorial manifestation of the amplitude splitting from the $U(1)$ decoupling relation is given by a block triangle structure such as in Fig. \ref{fig: Catalan diagram factorisation triangle}.

This triangle has several nice properties. Obviously the first one being that summing the number of diagrams in each row is just $C_{n-2}$, for $n$ being the number of particles, counting via the recursion relation of Eq.~\eqref{eq:catalan recursion relation}. 

The colour code of the triangle is no coincidence --- one can visually track the number of diagrams with $p-$particle sub-diagram, which is the maximal sub-diagram. Take for example diagrams in the pink circles, they all have a maximal sub-diagram with $p=4$ particles, which  stops at $n=7$ particle amplitudes, given that for $n=8$ particle amplitudes, there are no diagrams in which a $4-$particle sub-diagram is maximal. The number of such diagrams in general along these diagonals is given by the recursion relation
\begin{equation}
    C_{p-2}\left(\sum_{i=0}^{p-3}C_2C_i+C_{p-2}\right),
\end{equation}
the proof of which is given in Appendix \ref{sec:appendix1}.

Another nice feature about Fig. \ref{fig: Catalan diagram factorisation triangle} is how the diagram types are partitioned into blocks. Stopping at a particle particle number $n$ on the triangle, one can see that the blocks (visually represented as circles) of the triangle are partitioned as $\{n-2,n-3,...,3,2\}$, which is the same partition structure as in Fig. \ref{fig:ferrers} when counting the planar kinematic invariants. In combinatorics, two sets can often be compared when they admit Ferrers shapes of the same type~\cite{stanley_ec1}. This correspondence is natural: one counts planar kinematic invariants, the other, planar diagrams constructed from them. 
\begin{figure}[hbt!]
\centering

% Subfigure A: Narayana table
\begin{subfigure}[t]{0.47\textwidth}
\centering
\begin{tabular}{ccccccc}
\toprule
$\mathbf{m \backslash k}$ & \textbf{1} & \textbf{2} & \textbf{3} & \textbf{4} & \textbf{5} \\
\midrule
\textbf{1} & 1 \\
\textbf{2} & 1 & 1 \\
\textbf{3} & 1 & 3 & 1 \\
\textbf{4} & 1 & 6 & 6 & 1 \\
\textbf{5} & 1 & 10 & 20 & 10 & 1 \\
\bottomrule
\end{tabular}
\caption{Narayana numbers table $ N(m,k) $.}
\label{fig:narayana}
\end{subfigure}
\begin{subfigure}[t]{0.47\textwidth}
\centering
\begin{tabular}{ccccccc}
\toprule
$\mathbf{m \backslash k}$ & \textbf{1} & \textbf{2} & \textbf{3} & \textbf{4} & \textbf{5}  \\
\midrule
\textbf{1} & \colorbox{White}{1} \\
\textbf{2} & \colorbox{Cyan}{1}  & \colorbox{Cyan}{1}  \\
\textbf{3} & \colorbox{magenta}{2}  & \colorbox{Cyan}{1}  & \colorbox{magenta}{2}  \\
\textbf{4} & \colorbox{Goldenrod}{5}  & \colorbox{magenta}{2}  & \colorbox{magenta}{2}  & \colorbox{Goldenrod}{5}  \\
\textbf{5} & \colorbox{green}{14}  & \colorbox{Goldenrod}{5}  & \colorbox{magenta}{4}  & \colorbox{Goldenrod}{5}  & \colorbox{green}{14} \\
\bottomrule
\end{tabular}
\caption{A table counting the number of planar Feynman diagrams factorised as given by Fig. \ref{fig: Catalan diagram factorisation triangle}, with $n=m+2$ being the number of particles. }
\label{fig:second}
\end{subfigure}

\caption{Two sub-figures: (a) Narayana table, (b) Table of the number of diagrams for each block in Fig. \ref{fig: Catalan diagram factorisation triangle}.}
\label{fig:combined}
\end{figure}

Another interesting direction is to study the correspondence between the two sets based on how the blocks are filled. A connection can be found between the number of diagrams in each block (circle), categorised in Fig. \ref{fig:second}, and the Narayana numbers Fig. \ref{fig:narayana}. The Narayana numbers $N(m,k)$ are a refinement\footnote{A refinement in this case breaks the count into finer categories using an additional parameter, revealing deeper structure. See Eq.~\eqref{eq: catalan refinement narayana}} of the Catalan numbers \cite{petersen2016eulerian}, which similarly have recursive and combinatorial definitions. It can be seen that the row label $m$ is related to the number of particles and each column $k$ contains diagrams all of which have a sub-diagram(s) with $k+1$ external legs.  A key feature connecting the two tables is that the sum of their rows is equal to the same Catalan number:
\begin{align}
    C_m &= \sum_{k=1}^mN(m,k) \label{eq: catalan refinement narayana}\\
    C_m &=\sum_{k=1}^mC_{k-1}C_{m-k}.
\end{align}

Figure~\ref{fig:second} illustrates the correspondence between a Narayana number represented as a rooted tree and a planar Feynman diagram in the bi-adjoint $\phi^3$ theory. In the figure, the number of edges in the rooted tree matches the number of internal nodes in the corresponding Feynman diagram. Moreover, for extremal cases like $N(m,1)$ or $N(m,m)$, the number of nodes in the tree equals the number of external legs in the largest sub-diagram. This visual and combinatorial parallel reflects how planar $\phi^3$ diagrams encode recursive tree-like structures.
\begin{figure}[hbt!]
  \centering

  % Subfigure: Tree Diagram
  \begin{subfigure}[b]{0.4\textwidth}
    \centering
    \begin{tikzpicture}[scale=2, every node/.style={circle, draw, fill=white, inner sep=2pt}]
      \node (root) at (0,0) {};
      \node (a) at (-1,-1.5) {};
      \node (b) at (-0.5,-1.5) {};
      \node (c) at (0.5,-1.5) {};
      \node (d) at (1,-1.5) {};
      
      \draw (root) -- (a);
      \draw (root) -- (b);
      \draw (root) -- (c);
      \draw (root) -- (d);
    \end{tikzpicture}
    \caption{$N(4,4)$ as an unlabelled, ordered, rooted tree.}
    \label{subfig1: narayana number as a rooted tree}
  \end{subfigure}
  \hfill
  % Subfigure: Plain circle with filled nodes and red path
  \begin{subfigure}[b]{0.5\textwidth}
    \centering
    \begin{tikzpicture}[scale=2, every node/.style={inner sep=0pt}]
      % Draw the black circle
      \draw[black, thick] (0,0) circle(1);

      % Coordinates and node positions
      \foreach \i in {1,...,6} {
        \coordinate (P\i) at ({90 - (\i - 1)*60}:1);
        \filldraw[black] (P\i) circle (1pt);
        \node at ($(P\i) + ({90 - (\i - 1)*60}:0.2)$) {\i};
      }

      % red path: 1 → 6 → 2 → 3 → 4 → 5 → 1
      \draw[Blue, thick] (P1) -- (P6);
      \draw[Blue, thick] (P6) -- (P2);
      \draw[Blue, thick] (P2) -- (P3);
      \draw[Blue, thick] (P3) -- (P4);
      \draw[Blue, thick] (P4) -- (P5);
      \draw[Blue, thick] (P5) -- (P1);

      \draw[black, thick] (P5) -- (0.15,-0.25);
      \draw[black, thick] (P4) -- (0.15,-0.25);
      \draw[black, thick] (P3) -- (0.15,-0.25);
      \draw[black, thick] (P2) -- (0.15,-0.25);

      \draw[black, thick] (P1) -- (-0.375, 0.65);
      \draw[black, thick] (P6) -- (-0.375, 0.65);
      \draw[black, thick] (0.15,-0.25) -- (-0.375, 0.65);
      \filldraw[fill=Blue, draw=black] (0.15,-0.25) circle (0.3);
      \filldraw[fill=Blue, draw=black] (-0.375, 0.65) circle (0.04);
      
    \end{tikzpicture}
    \caption{Diagram with colour-structure $m_6(123456|162345)$.}
    \label{subfig1: planar feynman diagram}
  \end{subfigure}

  \caption{Graphical Comparison between the Narayana numbers and planar Feynman diagrams in the bi-adjoint $\phi^3$ theory.}
\end{figure}
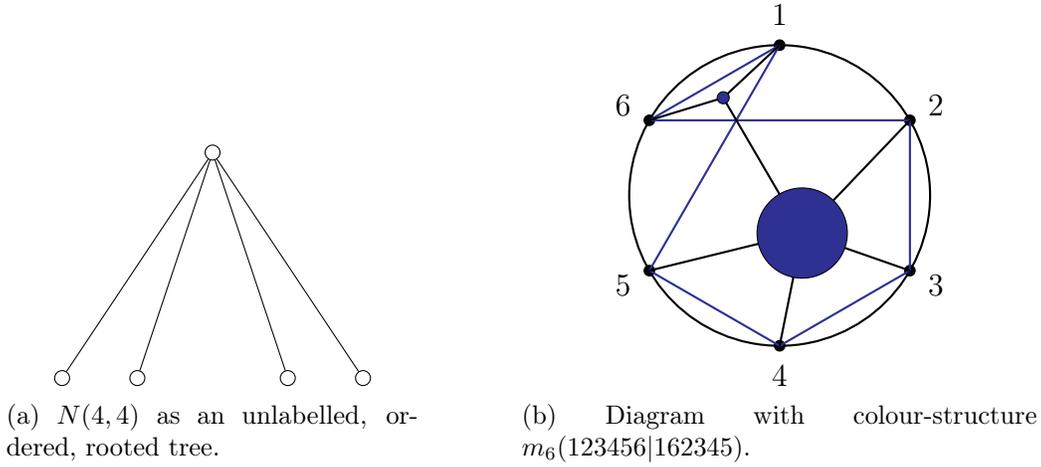

The full amplitude seems to decompose into a sum over channels, where a propagator splits the external legs into left and right subsets.
Each side computes a lower-point sub-amplitude, corresponding to a Catalan number. The Narayana numbers suggest a different partition entirely, potentially hinting at a deeper, more hidden structure within the diagrams. Further discussion and possible future directions are deferred to Section \ref{sec: conclusions}.

After exploring the $U(1)$ decoupling identity, we found a second Catalan recursion that carries a physical meaning — it governs how amplitudes can be built recursively through the splitting of soft factors. This is the subject of the next section.

\subsection{Building Diagrams: Another Catalan Recursion Relation}

Aside from Eq.~\eqref{eq:catalan recursion relation}, the Catalan numbers satisfy another set of recursion relations
\begin{equation}
    (n-1)C_{n-2} = 2(2n-5)C_{n-3},  \; \ \  C_0 = 1,
    \label{eq: U1 nonplanar catalan relations}
\end{equation}
with $C_{n-2}$ counting the number of $n-$point planar Feynman diagrams in the bi-adjoint $\phi^3$. It turns out, Eq.~\eqref{eq: U1 nonplanar catalan relations} could have a physical interpretation for building amplitudes in the bi-adjoint $\phi^3$ theory again related to the $U(1)$ decoupling identity, but this time in terms of building diagrams instead of factorising them. Let's look at the counting argument to explore more of this in detail.

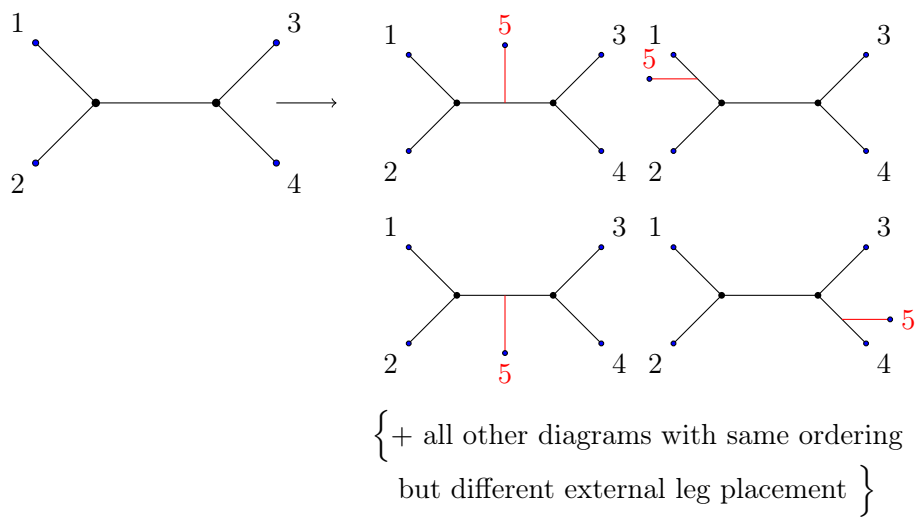
\begin{figure}[hbt!]
\begin{center}
\begin{tikzpicture}[scale=0.8, every node/.style={font=\small},
  squiggle/.style={decorate, decoration={snake, amplitude=0.4mm, segment length=2mm}, draw=blue}
]

  % Offset for centering the top diagram above the 2x2 grid
  \begin{scope}[shift={(-2.5,0)}]  % shift right by half of (5.5/2), the total width of the grid
    % Base diagram (4-point)
    \coordinate (v1) at (0,0);
    \coordinate (v2) at (2,0);
    \coordinate (a1) at (-1,1);
    \coordinate (a2) at (-1,-1);
    \coordinate (a3) at (3,1);
    \coordinate (a4) at (3,-1);

    \draw[-] (a1) -- (v1);
    \draw[-] (a2) -- (v1);
    \draw[-] (v1) -- (v2);
    \draw[-] (a3) -- (v2);
    \draw[-] (a4) -- (v2);
    \fill (v1) circle (2pt);
    \fill (v2) circle (2pt);
    \filldraw[fill=blue, draw=black] (a1) circle (0.05);
    \node[anchor=south east] at (a1) {1};
    \filldraw[fill=blue, draw=black] (a2) circle (0.05);
    \node[anchor=north east] at (a2) {2};
    \filldraw[fill=blue, draw=black] (a3) circle (0.05);
    \node[anchor=south west] at (a3) {3};
    \filldraw[fill=blue, draw=black] (a4) circle (0.05);
    \node[anchor=north west] at (a4) {4};

    % Centered downward arrow
    \draw[->] (3,0) -- (4,0);
  \end{scope}

  % Group of four smaller diagrams centered below
  \begin{scope}[shift={(3.5,0)}, scale=0.8]

    % First 5-point: 12345
    \begin{scope}[shift={(0,0)}]
      \coordinate (v1) at (0,0);
      \coordinate (v2) at (2,0);
      \coordinate (a1) at (-1,1);
      \coordinate (a2) at (-1,-1);
      \coordinate (a3) at (3,1);
      \coordinate (a4) at (3,-1);
      \coordinate (a5) at (1,1.2);
      \draw[-] (a1) -- (v1);
      \draw[-] (a2) -- (v1);
      \draw[-] (v1) -- (v2);
      \draw[-] (a3) -- (v2);
      \draw[-] (a4) -- (v2);
      \draw[-,red] (a5) -- (1,0);
      \fill (v1) circle (2pt);
      \fill (v2) circle (2pt);
      \foreach \pt in {a1,a2,a3,a4,a5} {
        \filldraw[fill=blue, draw=black] (\pt) circle (0.05);
      }
      \node[anchor=south east] at (a1) {1};
      \node[anchor=north east] at (a2) {2};
      \node[anchor=south west] at (a3) {3};
      \node[anchor=north west] at (a4) {4};
      \node[red, anchor=south] at (a5) {5};
    \end{scope}

    % Second 5-point: 15234
    \begin{scope}[shift={(5.5,0)}]
      \coordinate (v1) at (0,0);
      \coordinate (v2) at (2,0);
      \coordinate (a1) at (-1,1);
      \coordinate (a2) at (-1,-1);
      \coordinate (a3) at (3,1);
      \coordinate (a4) at (3,-1);
      \coordinate (a5) at (-1.5,0.5);
      \draw[-] (a1) -- (v1);
      \draw[-] (a2) -- (v1);
      \draw[-] (v1) -- (v2);
      \draw[-] (a3) -- (v2);
      \draw[-] (a4) -- (v2);
      \draw[-,red] (a5) -- (-0.5,0.5);
      \fill (v1) circle (2pt);
      \fill (v2) circle (2pt);
      \foreach \pt in {a1,a2,a3,a4,a5} {
        \filldraw[fill=blue, draw=black] (\pt) circle (0.05);
      }
      \node[anchor=south east] at (a1) {1};
      \node[anchor=north east] at (a2) {2};
      \node[anchor=south west] at (a3) {3};
      \node[anchor=north west] at (a4) {4};
      \node[red, anchor=south] at (a5) {5};
    \end{scope}

    % Third 5-point: 12534
    \begin{scope}[shift={(0,-4)}]
      \coordinate (v1) at (0,0);
      \coordinate (v2) at (2,0);
      \coordinate (a1) at (-1,1);
      \coordinate (a2) at (-1,-1);
      \coordinate (a3) at (3,1);
      \coordinate (a4) at (3,-1);
      \coordinate (a5) at (1,-1.2);
      \draw[-] (a1) -- (v1);
      \draw[-] (a2) -- (v1);
      \draw[-] (v1) -- (v2);
      \draw[-] (a3) -- (v2);
      \draw[-] (a4) -- (v2);
      \draw[-,red] (a5) -- (1,0);
      \fill (v1) circle (2pt);
      \fill (v2) circle (2pt);
      \foreach \pt in {a1,a2,a3,a4,a5} {
        \filldraw[fill=blue, draw=black] (\pt) circle (0.05);
      }
      \node[anchor=south east] at (a1) {1};
      \node[anchor=north east] at (a2) {2};
      \node[anchor=south west] at (a3) {3};
      \node[anchor=north west] at (a4) {4};
      \node[red, anchor=north] at (a5) {5};
    \end{scope}

    % Fourth 5-point: 12354
    \begin{scope}[shift={(5.5,-4)}]
      \coordinate (v1) at (0,0);
      \coordinate (v2) at (2,0);
      \coordinate (a1) at (-1,1);
      \coordinate (a2) at (-1,-1);
      \coordinate (a3) at (3,1);
      \coordinate (a4) at (3,-1);
      \coordinate (a5) at (3.5,-0.5);
      \draw[-] (a1) -- (v1);
      \draw[-] (a2) -- (v1);
      \draw[-] (v1) -- (v2);
      \draw[-] (a3) -- (v2);
      \draw[-] (a4) -- (v2);
      \draw[-,red] (a5) -- (2.5,-0.5);
      \fill (v1) circle (2pt);
      \fill (v2) circle (2pt);
      \foreach \pt in {a1,a2,a3,a4,a5} {
        \filldraw[fill=blue, draw=black] (\pt) circle (0.05);
      }
      \node[anchor=south east] at (a1) {1};
      \node[anchor=north east] at (a2) {2};
      \node[anchor=south west] at (a3) {3};
      \node[anchor=north west] at (a4) {4};
      \node[red, anchor=west] at (a5) {5};
    \end{scope}

  \end{scope}
\node[align=center] at (6.5, -6)
  {$\Big\{ +$ all other diagrams
   with same ordering \\ but
   different  external leg placement $\Big\}$};
\end{tikzpicture}
\end{center}
\caption{Constructing $5$-point diagrams by inserting leg $5$ into a $4$-point diagram, generating the $4$ diagrams of the $ U(1)$ decoupling identity under shuffles of $5$.}
\label{fig:soft particle insertion diagram}
\end{figure}
We start with a Feynman diagram with $n-1$ particles, thus $C_{n-3}$ diagrams in total. We can insert the $n^{th}$ particle either on an external leg \emph{or} on an internal propagator. We can insert the $n^{th}$ particle on one of the $n-1$ legs or on one of the $(n-1)-3$ internal propagators. But we also have two ways to insert the $n^{th}$ particle at each spot, i.e., on either side of an edge in the diagram. Hence in total we have
\begin{equation}
    2(n-1+(n-1)-3) = 2(2n-5)
\end{equation}
ways to insert the $n^{th}$ particle into our $n-1$ particle diagram. 

Each diagram resulting from the insertion on the right-hand side of Eq.\eqref{eq: the U(1) decoupling relation}) can be mapped to the left-hand-side. Specifically, the $U(1)$ decoupling identity yields $(n-1)$ distinct relations, and each of these gives rise to $C_{n-2}$ diagrams when there are $n$ external particles. This establishes the bijection.

Having uncovered the role of the $U(1)$ decoupling identity in revealing a physical connection between tree-level amplitudes and Catalan recursion relations---and having identified a hidden structure governed by the Narayana numbers---we now shift perspective. In the final lens through which we examine colour-ordered amplitudes, we turn to \emph{correlation functions}, using them to explore how the CHY formalism accommodates off-shell kinematics and encodes rich combinatorial structure.
\section{Emergent Correlation Functions from Colour-Ordered  Amplitudes}\label{sec: section4}
In their $1955$ paper, Lehmann, Symanzik and Zimmermann (LSZ) proved how to obtain the S-matrix from time-ordered Green functions \cite{Lehmann:1955zz}, making it possible to calculate the S-matrix from correlation functions. The proof of the so-called \textit{LSZ formula} does not depend on perturbation theory, relying instead on general properties such as locality, Lorentz invariance, and the existence of asymptotic states. As a result, it demonstrates that crossing symmetry is not merely an artifact of Feynman diagrams, but a more profound and general feature of quantum field theory itself \cite{Peskin:1995ev}.

The Fourier transform of the Green's function for $n$ scalar fields is simply
\begin{equation}
    \mathcal{G}(P_1,...,P_n) = \int \prod_{a=1}^n dx_a e^{iP_a\cdot x_a}\langle \phi(x_1)\cdots\phi(x_n)\rangle
    \label{eq: Fourier transformed Green's function},
\end{equation}
where $\mathcal{G}(P_1,...,P_n)$ is time-ordered and includes vacuum expectation values, with the $P_a$'s being completely arbitrary, $D-$dimensional spacetime vectors. More explicitly, we have not specified that they are momenta. However, translation invariance of the Green's function in Eq.~\eqref{eq: Fourier transformed Green's function} implies that the sum of these $n$, $D-$dimensional spacetime vectors $P_a^{\mu}$ must vanish, i.e., 
\begin{equation}
    \sum_{a=1}^n P_a=0.
    \label{eq: translational invariance of the Green's function}
\end{equation}

In the on-shell limit, Eq.~\eqref{eq: Fourier transformed Green's function} gives rise to the on-shell amplitude $\mathcal{A}_n(P_1,...,P_n)$ for $n-$particles. Namely,  
\begin{equation}
    \mathcal{G}(P_1,...,P_n) \rightarrow  \left(\prod_{a=1}^{n}\frac{1}{P_a^2}\right)\mathcal{A}_n(P_1,...,P_n),
    \label{eq: directional greens}
\end{equation}
with the simple poles $P_a^2=0$ corresponding to the physical on-shell particles that represent the asymptotic states. However, Eq.~\eqref{eq: directional greens} holds only in one direction: the full Green’s function \( \mathcal{G}(P_1, \dots, P_n) \) cannot be reconstructed from its right-hand side. While only pole terms contribute to the S-matrix, the full Green’s function also includes non-pole (off-shell) terms. The amplitude appears as the residue of the Green’s function at poles corresponding to external on-shell particles.

This raises an intriguing question: is it possible to reconstruct the full Green’s function—-not just its residues—-from scattering amplitudes, possibly by including higher-point processes or off-shell generalisations? This is precisely the question we will explore in this section\footnote{For a different approach see \cite{lam2016offshellchyamplitudes}}. We will first lay out the kinematic notation for our correlation functions and test it out on an even, colour-ordered amplitude. We will then test it out in the simplest case - analysing the solutions to the scattering equations and calculating the amplitude. We will show the generalised derivation of the reduced number of solutions to these revised scattering equations. Finally, we will derive and take a look at the off-shell scattering potential.

\subsection{Correlation Functions \& Colour-Ordering}
\begin{figure}[hbt!]
\centering
\begin{tikzpicture}[scale=1.3, thick]

    % Outer circle
    \draw[cyan] (0,0) circle(2);

    % Boundary points (1, 1', ..., 4')
    \foreach \i/\angle/\label in {
        1/67.5/$1$,
        1p/22.5/$1'$,
        2/337.5/$2$,
        2p/292.5/$2'$,
        3/247.5/$3$,
        3p/202.5/$3'$,
        4/157.5/$4$,
        4p/112.5/$4'$
    } {
        \coordinate (\i) at (\angle:2);
        \fill[cyan] (\i) circle(0.03);
        \node at (\angle:2.3) {\small \label};
    }

    % Square corners
    \coordinate (M1) at (0.78, 0.78);
    \coordinate (M2) at (0.78, -0.78);
    \coordinate (M3) at (-0.78, -0.78);
    \coordinate (M4) at (-0.78, 0.78);

    % Outer square
    \draw[black, thick] (M1) -- (M2) -- (M3) -- (M4) -- cycle;

    % Center green circle
    \filldraw[fill=blue!60, draw=black] (0,0) circle(0.4);

    % Lines between i and i'
    \draw[black] (1) -- (1p);
    \draw[black] (2) -- (2p);
    \draw[black] (3) -- (3p);
    \draw[black] (4) -- (4p);

    % Lines from each unprimed and primed point to square corners
    \draw[black] (1) -- (M1); \draw[black] (1p) -- (M1);
    \draw[black] (2) -- (M2); \draw[black] (2p) -- (M2);
    \draw[black] (3) -- (M3); \draw[black] (3p) -- (M3);
    \draw[black] (4) -- (M4); \draw[black] (4p) -- (M4);

    % Triangle centroids
    \coordinate (C1) at (1.0, 1.0);
    \coordinate (C2) at (-1.075, 1.05);
    \coordinate (C3) at (1.05, -1.05);
    \coordinate (C4) at (-1.05, -1.05);

    \foreach \C/\n in {C1/1, C2/2, C3/3, C4/4} {
        \fill[red] (\C) circle(0.04);
    }
    % Lines from each unprimed and primed point to square corners
    \draw[blue] (1) -- (C1); \draw[blue] (1p) -- (C1);
    \draw[blue] (2) -- (C3); \draw[blue] (2p) -- (C3);
    \draw[blue] (3) -- (C4); \draw[blue] (3p) -- (C4);
    \draw[blue] (4) -- (C2); \draw[blue] (4p) -- (C2);
\draw[blue, thick] (-0.29,-0.29) -- (C4) node[midway, sloped, above] {$\longrightarrow$} node[near end, sloped, below] {$P_3$}; 
\draw[blue] (-0.29,0.29) -- (C2) node[midway, sloped, above] {$\longrightarrow$} node[near end, sloped, below] {$P_4$};
\draw[blue] (0.29,-0.29) -- (C3) node[midway, sloped, above] {$\longleftarrow$} node[near end, sloped, below] {$P_2$}; 
\draw[blue] (0.29,0.29) -- (C1) node[midway, sloped, above] {$\longleftarrow$} node[near end, sloped, below] {$P_1$};

     \filldraw[fill=red!60, draw=red] (C1) circle(0.04);
      \filldraw[fill=red!60, draw=red] (C2) circle(0.04);
       \filldraw[fill=red!60, draw=red] (C3) circle(0.04);
        \filldraw[fill=red!60, draw=red] (C4) circle(0.04);
        
        \filldraw[fill=cyan!60, draw=cyan] (1) circle(0.04);
      \filldraw[fill=cyan!60, draw=cyan] (2) circle(0.04);
       \filldraw[fill=cyan!60, draw=cyan] (3) circle(0.04);
        \filldraw[fill=cyan!60, draw=cyan] (4) circle(0.04);
        \filldraw[fill=cyan!60, draw=cyan] (1p) circle(0.04);
      \filldraw[fill=cyan!60, draw=cyan] (2p) circle(0.04);
       \filldraw[fill=cyan!60, draw=cyan] (3p) circle(0.04);
        \filldraw[fill=cyan!60, draw=cyan] (4p) circle(0.04);
\end{tikzpicture}
\caption{Colour-ordered eight-point amplitude $m(11'22'33'44'|1'12'23'34'4)$, with the blue circle including all possible diagrams for four-particle amplitudes and $P_i^\mu$ are the off-shell amplitudes.}
\label{fig: 8 particle amplitude correlation}
\end{figure}
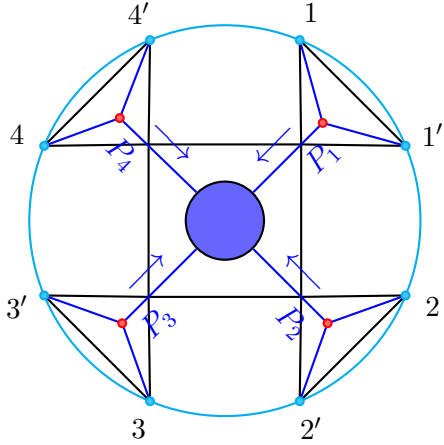

We start by expressing the off-shell momentum $P_a^\mu$ in terms of on shell momenta for massless particles $p_a^\mu$ and reference momentum $q^\mu$
\begin{equation}
    P_a^{\mu}=p_a^{\mu}+\tau_a q^{\mu},
    \label{eq: off shell momenta in terms of on-shell momenta}
\end{equation}
with $p_a^\mu p_{a,\mu}=0$ and $q^\mu q_{\mu}=0$. The numbers $\tau_a$ parameterise the ``off-shellness'' of the external particle labelled $a$ and has $1$ degree of freedom, given that $q^\mu$ has no degrees of freedom and $p_a^\mu$ has $D-1$ degrees of freedom\footnote{due to momentum conservation.}. The idea is to start with $2m$ particles $\{1,1',2,2',\cdots,m,m'\}$ in an amplitude with colour ordering $m_n(1,1',2,2',\cdots,m,m'|1',1,2',2,\cdots,m',m)$ (see Fig. \ref{fig: 8 particle amplitude correlation}), with momentum conservation
\begin{equation}
    \sum_{a=1}^n P_a^{\mu} = \sum_{a=1}^m k_a +\sum_{a=1}^m k_{a'}=0,
    \label{eq: off shell momentum conservation}
\end{equation}
for $k_a=p_a^\mu$ and $k_{a'}=\tau_a q^{\mu}$.
The observation here is that this $2m-$particle amplitude, under this construction, is evaluated on 
\begin{equation}
    s_{a'b'}=(k_{a'}+k_{b'})^2 = (\tau_{a}+\tau_{b})^2q^2 = 0,
    \label{eq: the sa'b' are zero}
\end{equation}
and the complicated sub-diagram of the interaction (such as the blue blob in Fig. \ref{fig: 8 particle amplitude correlation}) would actually be $\mathcal{G}(P_1,\cdots,P_n)$. Eq.~\eqref{eq: the sa'b' are zero} tells us that a portion of the kinematic invariants under this construction are zero. This infers there could be degenerate, or invalid solutions to the scattering equations \cite{Cachazo_2014, Cachazo_2014_scalars}. Let's test this construction for the simplest possible case, namely for $m=3$ or $2m = 6$ particles, i.e., for the amplitude $m(11'22'33'|1'12'23'3)$.

Now normally with the scattering equations with $2m$ particles, there are $(2m-3)!$ solutions. However, we find computationally that there is only one valid, non-degenerate solution to the scattering equations under this construction. Four solutions are found to have values at gauge fixed points, while the fifth solution is degenerate.  When gauge fixing $x_1=0, \; x_2=1$ and $x_3=-1$, the analytic result for the unique solution is simply
\begin{align}
    x_4 &= \frac{s_{12'}s_{1'2}}{2s_{11'}s_{22'}-s_{12'}s_{1'2}} \label{eq: x4 for correlation m=6} \\ 
    x_5 &= -\frac{s_{1'2}}{2s_{11'}+s_{1'2}} \label{eq: x5 for correlation m=6} \\
    x_6 &= \frac{s_{13'}s_{1'2}}{s_{1'2}(s_{12}+s_{12'}+s_{13})+s_{11'}(2s_{13}-s_{1'2}-2s_{22'})} \label{eq: x6 for correlation m=6}
\end{align}
The amplitude as a result of the solution to Eq.'s \eqref{eq: x4 for correlation m=6}-\eqref{eq: x6 for correlation m=6} is just simply 
\begin{equation}
    m(11'22'33'|1'12'23'3) = \frac{1}{s_{11'}s_{22'}s_{33'}},
    \label{eq: 3pt correlation function result from the scattering equation}
\end{equation}
which is \textit{exactly} the three point correlation given that 
\begin{equation*}
    \frac{1}{s_{11'}s_{22'}s_{33'}} = \frac{1}{(2\tau_1p_1\cdot q)(2\tau_2p_2\cdot q)(2\tau_3p_3\cdot q)} = \frac{1}{P_1^2P_2^2P_3^2},
\end{equation*}
since $P_a^2 = 2\tau_a p_a\cdot q$. 

One can do the same numerical evaluations and find that for $2m=8$ particles, there are only $16$ solutions in total. A pattern does emerge in counting valid, non-degenerate solutions for any $m$, and to uncover it, we turn to the positive kinematic counting methods of Ref.~\cite{Cachazo_2017}.
\subsection{Counting Non-Degenerate Solutions with Vanishing Invariants}\label{subsec: number of solutions}
In the spirit of Ref. \cite{Cachazo_2017}, we go back to the idea of the so-called scattering potential, which gives rise to the scattering equations when extremised. It acts as a potential for the $2m-3$ particles moving in the real interval $I=[0,1]$. The proof of reality of the solutions comes from the fact that all $(2m-3)!$ solutions to the scattering equations are of this form.

When none of the kinematic invariants are zero, then all $2m-3$ particles repel one another given that on the interval the $\log|x_i-x_j|$ terms will be negative, and the potential diverges when two such particles coalesce or if a particle approaches the potential barriers, synonymous with the repulsive nature of the electrostatic force. Hence, there is only one equilibrium point for each of the possible orderings of the particles, proving the result. A natural question arises: what is the effect of setting certain Mandelstam invariants to zero?

The physical set-up is as follows\footnote{For more details on the potential setup, see Ref.~\cite{Cachazo_2017}}. In our system, we have  $m$ primed and $m$ unprimed particles living on a real interval $I=[0,1]$, whereby the kinematic invariants $s_{ij}$ encode the strength of the interaction between particles $i$ and $j$. The primed particles do not interact with one another-they may pass through each other, approach arbitrarily close, or even coincide without any divergence in the interaction potential. In contrast, the unprimed particles interact among themselves, and each primed particle can also interact with the unprimed particles. So when counting the number of valid solutions, we need only need to consider the potential placements of the $m-3$ non-primed, non-gauge fixed particles, i.e, ${4,5,...,m}$, with respect to the primed particle placements ${1',2',3'...,m'}$, whose $s_{a'b'}=0$ and the original boundary conditions.

For $m = 3$ particles, we have one unique solution. With the unprimed particles fixed by gauge choice, we only need to place the three primed particles between the barriers, which do not interact. Combinatorially, this is equivalent to arranging all three unlabelled objects, yielding $ \binom{3}{3} = 1 $ valid solution.

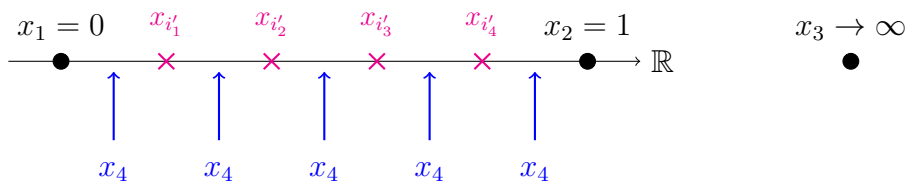
\begin{figure}[hbt!]
    \centering
    \begin{tikzpicture}[scale=7] % scale up for visibility

  % Draw the real line
  \draw[->] (-0.1,0) -- (1.1,0) node[right] {$\mathbb{R}$};

  % Mark the endpoints
  \filldraw (0,0) circle (0.015) node[above=4pt] {$x_1 = 0$};
  \filldraw (1,0) circle (0.015) node[above=4pt] {$x_2 = 1$};

  % Draw 4 evenly spaced X-shaped crosses between 0 and 1
  \foreach \i in {1,2,3,4} {
    \pgfmathsetmacro{\x}{\i/5}
    \draw[thick, magenta] 
      ({\x - 0.015},{-0.015}) -- ({\x + 0.015},{0.015});
    \draw[thick, magenta] 
      ({\x - 0.015},{0.015}) -- ({\x + 0.015},{-0.015});
      % Add label above the cross
  \node[above=3pt, scale=0.9, magenta] at (\x, 0.015) {$x_{i'_\i}$};
  }
    \filldraw (1.5,0) circle (0.015) node[above=4pt] {$x_3 \to \infty$};
    % Draw arrows to the 5 midpoint slots between the dots and crosses
  \foreach \i in {0,1,2,3,4} {
    \pgfmathsetmacro{\x}{(\i + 0.5)/5}
    \draw[->, thick, blue] (\x,-0.15) node[below=4pt] {$x_4$} -- (\x,-0.02);
  }

\end{tikzpicture}
    \caption{Particles on a line for $2m=8$ particles. The crosses in the figure are the positions of the primed particles arranged in some way.}
    \label{fig: Particles on a line for $2m=8$ particles}
\end{figure}
The counting becomes more intricate with more particles, but a symmetry emerges that leads to a general closed-form expression for any $m$. Let's now look at the $m=4$ example, illustrated in Fig. \ref{fig: Particles on a line for $2m=8$ particles}. In this configuration, we have five spots to place the only non-primed particle $x_4$. If we place $x_4$ in the slot with $k$ primed particles to its left (or equivalently $4-k$ to its right) where $k=0,1,2,3,4$, then we have $\binom{4}{k} = \binom{4}{4-k}$ choices for this configuration. Adding all of our choices together, we obtain 
\begin{equation*}
    \sum_{k=0}^4 \binom{4}{k}= 16
\end{equation*}
possible solutions, which is indeed correct. 

We can sense a pattern emerging, so much so that we can generalise this result: for the $2m-$particle off-shell amplitude of this colour-ordering, the number of non-degenerate, valid solutions to the scattering equations is 
\begin{equation*}
    (m-3)!(m-2)^m.
\end{equation*}
This result can be generalised even further to obtain the number of solutions $\mathcal{N}_{m,p}$ for $m+p$ particle amplitudes, where $m$ is the number of unprimed particles and $p$ is the number of primed particles. The result is 
\begin{equation}
    \mathcal{N}_{m,p} = (m-3)!(m-2)^p.
\end{equation}
To prove this, we look at the following counting exercise from our physical potential set-up. We know that $(m-3)!$ is the number of ways to permute $m-3$ distinguishable objects, in this case, our unprimed particles. So, let's just count the number of solutions for just one of these orderings, and then we multiply our answer by this number to obtain the total number of solutions.

Through the binomial theorem, we know we can rewrite $(m-2)^p$ in the form
\begin{equation}
    (m-2)^p = \sum_{k=0}^p \binom{p}{k}(m-3)^{p-k}1^k.
    \label{eq: binomial counting expression}
\end{equation}
Eq.~\eqref{eq: binomial counting expression} is equal to the number of ways to choose our unprimed labels from our set and arrange them in a way that is compatible with our set-up. We choose one primed particle, the $k^{th}$ primed particle, after which to place our $m-3$ unprimed particles. We then pick the $m-3$ unprimed labels to be placed around the $(p-k)$ primed slots available. For a given $k$, the number of ways to choose the $k^{th}$ primed slot out of $p$ primed particles is just $\binom{p}{k}$. To count properly, we must sum over all the possible values of $k$, giving us our result.

Now that we have determined the modified number of solutions to the scattering equations in this setup, we turn to deriving the scattering potential in terms of the off-shell momenta \( P_a^2 \), providing a clearer window into the dynamics of this construction.

\subsection{The Off-Shell Scattering Potential}
While the results thus far seem encouraging, to really probe the dynamics of off-shell scattering, we will now look at expressing Eq.~\eqref{eq: potential of the scattering equations} in terms of our new formalism.
The scattering potential (Eq.~\eqref{eq: potential of the scattering equations}) is expressed in terms of the kinematic invariants $s_{i,j}$ 
for $i,j\in \{1,2,\ldots,m,1',2',\ldots,m'\}$, so to reach our goal, we try and find ways to express our kinematic invariants in terms of the off-shell momenta. We know that $s_{a'b'}=0$ (Eq.~\eqref{eq: the sa'b' are zero}), and similarly we can check that
\begin{equation}
    s_{ab} = 2p_a\cdot p_b, \quad s_{a'b} = 2\tau_ap_b\cdot q, \quad s_{ab'} = 2\tau_bp_a\cdot q.
    \label{eq: the other s_{ab}'s correlation}
\end{equation}
Using Eq's~\eqref{eq: off shell momenta in terms of on-shell momenta} and \eqref{eq: the other s_{ab}'s correlation}, we find the following potential
\begin{align}
    \mathcal{S}(x) = \sum_{a \neq b} p_a \cdot p_b \log|x_a - x_b| 
+ \sum_{a',b} 2 \tau_a p_b\cdot q\log|x_{a'} - x_b|, \label{eq: scattering potential off shell regime}
\end{align}
where in the second line we used the fact that $s_{a'b'}=0$. In Appendix \ref{sec: SL2C invariance}, it is shown how this potential is $\mathrm{SL}(2,\mathbb{C})$ invariant thus proving the validity of the formalism. 

After checking for $SL(2,\mathbb{C})$ invariance under our new formulation, let's now express our scattering potential in terms of these off-shell momenta to gauge the behaviour of off-shell dynamics. Using Eq.'s~\eqref{eq: off shell momenta in terms of on-shell momenta} and \eqref{eq: the other s_{ab}'s correlation}, and knowing that $P_a^2 = 2\tau_ap_a\cdot q$, we find the following expressions for the kinematic invariants in terms of the off-shell momenta: 
\begin{align}
    s_{ab}+s_{a'b}+s_{ab'}&= 2P_a\cdot P_b, \label{eq: expression offshell 1}\\
    s_{ab'} &= \frac{\tau_b}{\tau_a}(P_a)^2. \label{eq: expression offshell 2}
\end{align}
We can now substitute Eq's~\eqref{eq: expression offshell 1}-\eqref{eq: expression offshell 2} into the scattering potential, utilising the fact that terms with $a \leftrightarrow b$ are equivalent, as we are summing over both $a$ and $b$. Presented in a symmetrised form, the modified CHY potential for off-shell scattering is
\begin{equation}
    \mathcal{S}(x) = \sum_{a\neq b}\left(P_a\cdot P_b-\frac{\tau_a}{\tau_b}(P_b)^2\right)\log{|x_a-x_b|}+\sum_{a, b}\frac{\tau_a}{\tau_b}(P_b)^2\log{|x_{a'}-x_b|}.
    \label{eq: scattering potential in terms of off-shell momenta}
\end{equation}
This expression for the potential, Eq.~\eqref{eq: scattering potential in terms of off-shell momenta}, completes our reformulation of the CHY scattering potential in the off-shell regime. The presence of the $\tau$ parameters-unavoidable in the off-shell formulation--emphasises their role in controlling deviations from the on-shell limit and encapsulating the kinematic structure of the extended theory. This sets the foundation for probing off-shell dynamics within the CHY framework and opens the door to further exploration of how off-shell corrections manifest in scattering amplitudes and their underlying geometric structures.
\section{Conclusions}\label{sec: conclusions}

This essay examined the structure of colour-ordered tree-level amplitudes in bi-adjoint $\phi^3$ theory from three perspectives: finding the physical intuition behind the constructed massive scattering equations, probing the physical interpretation of the Catalan recursion relations, and using the CHY formalism to derive correlators from amplitudes.

A compact, $\mathrm{SL}(2,\mathbb{C})$-invariant expression for the massive planar scattering potential was constructed using a planar kinematic basis, enabling the re-derivation of the Dolan-Goddard extension of the CHY scattering equations, grounded in a physical picture of planar scattering in bi-adjoint $\phi^3$ theory. Expressed neatly in terms of cross ratios, the potential is manifestly $\mathrm{SL}(2,\mathbb{C})$ invariant. The form of the potential also provides a more symmetric way to count the basis of kinematic invariants using staircase Ferrers diagrams.

The combinatorics of bi-adjoint $\phi^3$ diagrams revealed that the $U(1)$ decoupling identity provides physical insight into two Catalan recursion relations: one describing diagram factorisation and the other diagram construction. The recursion relation describing diagram factorisation is realised as a combinatorial array that mirrors the partition structure of the staircase Ferrer's shapes introduced earlier. This provides a combinatorial connection between planar kinematic invariants and planar diagrams. The combinatorial array also allowed the derivation of a closed-form expression for the number of diagrams containing maximal sub-diagrams involving $p \geq 3$ particles. Narayana numbers emerged naturally from diagram-splitting processes associated with the $U(1)$ identity through this combinatorial array. 

Analysing $2m$-particle amplitudes under a specific colour ordering showed that correlation functions with $m$ off-shell momenta arise naturally from these amplitudes in the CHY formalism. The off-shell system reduces the number of solutions to the scattering equations, with closed-form expressions obtained by analysing the system of particles interacting on the interval $[0,1]$ via an off-shell scattering potential. Such a potential was derived explicitly, generalising the on-shell formula and offering a framework for understanding correlation functions in $\phi^3$ theory from a CHY perspective.

Exploiting colour ordering in the bi-adjoint $\phi^3$ theory uncovers a unifying mathematical structure across the three domains studied. However, several open directions remain.

A natural next step is to extend the massive CHY formalism beyond the planar regime. Preliminary calculations suggest a non-planar extension using Strebel differentials is within reach, but the full construction of the non-planar CHY amplitude will be left for future work.

Further questions arise from the combinatorial analysis, particularly regarding whether other Catalan-like recursion relations capture deeper amplitude properties such as soft limits or Kleiss-Kuijf relations. Interestingly, similar combinatorial structures emerge in the study of the positive Grassmannian $ \mathrm{Gr}_{k,n}^{\geq 0} $, particularly in the so-called \textit{BCFW cells}. These cells admit a recursive description via plabic, or \emph{on-shell}, graphs and their enumeration matches the entries in Table~\ref{fig:second}, which, as we know, is connected to the Narayana numbers of Table~\ref{fig:narayana}. Despite the numerical match, the external particle counts differ in each factorisation process, suggesting a non-trivial correspondence.

It remains an open question whether these similarities point to a deeper combinatorial connection between the two factorisation processes. Ongoing work seeks to understand whether the appearance of Narayana numbers in both contexts signals a more universal structure underlying tree-level amplitudes in the bi-adjoint $ \phi^3 $ theory.

Additional directions arise from the CHY analysis. The symmetric structure in the solution counting of Section~\ref{subsec: number of solutions} raises the question of whether there exists a connection to Galois theory that could simplify the decoupled scattering equations. This could be especially beneficial in managing the increasing computational complexity for larger numbers of particles.

Finally, the off-shell scattering potential itself suggests further investigation. It remains to be seen how the associated scattering equations behave off-shell and what implications this has for amplitudes and correlators. It would also be interesting to explore whether particular factorisation channels emerge in specific limits - for example, as the $\tau$ parameters approach critical values -potentially revealing new structural insights into the underlying dynamics.

\section{Acknowledgements}
I would like to thank my supervisor, Dr Freddy Cachazo, for his guidance, insight, and support throughout this project. I am also grateful to all my PSI classmates who gave me their thoughtful feedback on the presentation of this essay, help proofreading earlier drafts, and for their encouragement throughout. A special thanks to both Emilia and Besi for many helpful discussions on amplitudes, both during group meetings and in informal settings.

%%%%%%%%%%%%%%%%%%%%%%%%%%%%%%%%%%%%%%%
%% THE REFERENCES - EDIT THESE FILES %%
%%%%%%%%%%%%%%%%%%%%%%%%%%%%%%%%%%%%%%%
\bibliographystyle{style_files/utphys}
\bibliography{references}

%%%%%%%%%%%%%%%%%%%%%%%%%%%%%%%%%%%%%%%
%% THE APPENDICES - EDIT THESE FILES %%
%%%%%%%%%%%%%%%%%%%%%%%%%%%%%%%%%%%%%%%
\begin{subappendices} % DO NOT EDIT THIS LINE
\renewcommand\thesection{\Alph{section}} % DO NOT EDIT THIS LINE
\section{\texorpdfstring{$SL(2,\mathbb{C})$ Redundancy in CHY \& Calculating CHY Amplitudes}{SL(2,C) Redundancy in CHY \& Calculating CHY Amplitudes}}\label{sec: chy subtleties}
Here, we provide a detailed account of several subtleties encountered in the CHY formalism, with particular focus on gauge fixing, momentum conservation, and the verification of collinear limits. Additionally, it provides a guide to explicitly calculating CHY amplitudes, by introducing an example for $n=4$ particles in the bi-adjoint $\phi^3$ theory. The content presented here stems from ongoing discussions during both individual group meetings conducted throughout the year, as part of a broader effort to understand the CHY framework.

\subsection{\texorpdfstring{Möbius Invariance and the Geometry of $\mathbb{CP}^1$}{Möbius Invariance and the Geometry of CP1}}\label{subsec: cp1 and chy}

As evident in Eq.~\eqref{eq:genCHY}, a $\mathrm{SL}(2,\mathbb{C})$ redundancy exists in the CHY amplitude. This arises because the complex integrals are defined over punctures on the \emph{Riemann sphere} $\mathbb{CP}^1$. These spaces are central in theoretical physics as a natural two-dimensional complex manifolds on which quantum field theory (QFT) - and particularly string theory - may be formulated~\cite{Schwarz:2013uza}.

The $\mathbb{CP}^1$ is a complex manifold: a topological 2-sphere equipped with a complex structure. It extends the complex plane by adding a point at infinity and can be covered by two stereographic coordinate charts whose transition functions are holomorphic~\cite{Polchinski:1998rq}.

Understanding a mathematical object is often facilitated by studying its group of automorphisms that preserve its structure. For $\mathbb{CP}^1$, the automorphisms are precisely the invertible conformal (biholomorphic \footnote{A bijective holomorphic map with a holomorphic inverse.}) maps from the $\mathbb{CP}^1$ to itself~\cite{Polchinski:1998rq}. These are the Möbius transformations, which take the form
\begin{equation}
z' = \frac{Az + B}{Cz + D}, \qquad
\begin{pmatrix}
A & B \\
C & D
\end{pmatrix} \in \mathrm{SL}(2,\mathbb{C}).
\label{eq: moebius transformation}
\end{equation}
Given an overall scaling of the matrix leaves the transformation unchanged, the true symmetry group of $\mathbb{CP}^1$ is the projective special linear group $\mathrm{PSL}(2,\mathbb{C}) = \mathrm{SL}(2,\mathbb{C})/\mathbb{Z}_2$~\cite{Polchinski:1998rq}. This group acts transitively on $\mathbb{CP}^1$ and reflects its conformal invariance. Two coordinate systems $z$ and $z'$ related by such a transformation describe the same Riemann surface up to biholomorphic equivalence~\cite{Polchinski:1998rq}.

In scattering theory, massless particles correspond to null momenta, which define points on the celestial sphere \cite{Pasterski:2021rjz}. Noting the isomorphism $S^2 \simeq \mathbb{CP}^1$, the conformal group acting on these directions is precisely $\mathrm{SL}(2,\mathbb{C})$. Thus, the CHY framework naturally incorporates this symmetry.

\begin{figure}[hbt!]
\begin{center}
\begin{tikzpicture}
% Draw Riemann sphere
\draw[purple] (0,0) circle (2);
\filldraw[black] (-1,1) circle (2pt) node[anchor=south] {$x_1$};
\filldraw[black] (1,1) circle (2pt) node[anchor=south] {$x_2$};
\filldraw[black] (-1,-1) circle (2pt) node[anchor=north] {$x_3$};
\filldraw[black] (1,-1) circle (2pt) node[anchor=north] {$x_4$};
\draw[thick, purple] (-2,0) arc (180:360:2 and 0.5);
\draw[thick, purple, dashed] (-2,0) arc (180:0:2 and 0.5);
\end{tikzpicture}
\end{center}
    \caption{Punctures on a Riemann sphere for $n=4$ particles, corresponding to the location of the particles in the scattering process.}
    \label{fig: punctures on a Riemann sphere for n=4}
\end{figure}
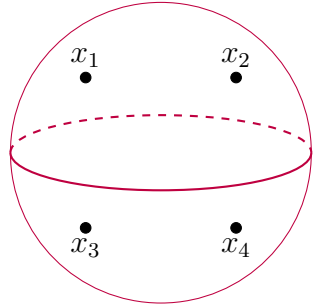

The CHY formulation expresses scattering amplitudes as integrals over the moduli space of punctured Riemann spheres, which serves as the domain of integration in Eq.~\eqref{eq:genCHY}. The punctures represent particle insertions, see Fig. \ref{fig: punctures on a Riemann sphere for n=4}, and configurations related by Möbius transformations are modded out.

A natural question is how solutions to the scattering equations behave as the kinematic data varies. For instance, as a kinematic invariant $s_{ab} = (k_a + k_b)^2$ tends to zero, the corresponding punctures $x_a$ and $x_b$ coalesce:
\begin{equation}
    s_{ab} \to 0 \quad \Rightarrow \quad |x_a - x_b| \to 0.
\end{equation}
This limit, the \emph{collinear} limit, corresponds physically to an intermediate particle going on-shell, and the amplitude factorising accordingly, potentially forming an intermediate on-shell state \cite{Dixon:1996wi, Bern_2007, Geyer_2016}. 
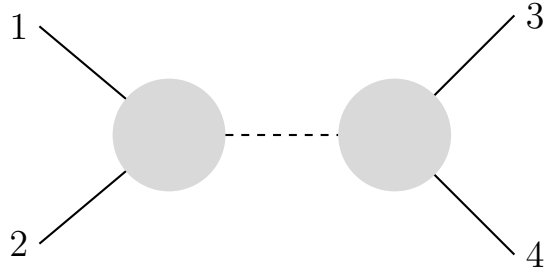
\begin{figure}[hbt!]
    \centering
    \begin{tikzpicture}[scale=1.5]
    \draw[thick, black] (0.5,0) -- ++(220:1.5cm) node[left, black] {\large$2$};
\draw[thick, black] (0.5,0) -- ++(140:1.5cm) node[left, black] {\large$1$};
% Gluon line into S
% Right side: S factor and A_{n-1}
% S factor
\fill[fill=gray!30] (0.5,0) circle (0.5cm);
\draw[thick, dashed, black] (1,0) -- (2,0);

\draw[thick, black] (2.5,0) -- ++(45:1.5cm) node[right, black] {\large$3$};
\draw[thick, black] (2.5,0) -- ++(315:1.5cm) node[right, black] {\large$4$};
\fill[fill=gray!30] (2.5,0) circle (0.5cm);

\end{tikzpicture}
    \caption{Crossing symmetry and collinear factorisation into two parts. The dotted line is the intermediate particle going on shell.}
    \label{fig:n4 collinear limit}
\end{figure}

Geometrically, this is reflected in the degeneration of the moduli space: the Riemann sphere degenerates into two spheres connected at a node -- a process that geometrically realises the factorisation of the amplitude into two lower--point amplitudes, just like in Fig.\ref{fig:n4 collinear limit} \cite{Dolan_2014, Geyer_2016}. The interaction becomes simpler and the system behaves as if it is in two parts, which interact separately, leading to a \emph{soft limit}.

Consider the case of four massless external momenta $k_i$ satisfying
\begin{equation}
    k_1 + k_2 + k_3 + k_4 = 0, \quad k_i^2 = 0, 
\end{equation}
and Mandelstam invariants~\cite{peskin2019concepts, Thomson_2013}
\begin{equation}
    s = (k_1 + k_2)^2 = 2k_1 \cdot k_2, \quad
t = (k_1 + k_3)^2 = 2k_1 \cdot k_3, \quad
u = (k_1 + k_4)^2 = 2k_1 \cdot k_4,
\end{equation}
with the massless condition for the invariants given by $s + t + u = 0$. After gauge fixing three of the punctures using $SL(2,\mathbb{C})$, the scattering equations reduce to a single equation for the remaining puncture as will be seen in the example detailed in Section \ref{sec: calculating a 4point chy amplitude}.

Using a common gauge-fixing choice, such as
\begin{equation}
    x_1 = 0, \quad x_2 = 1, \quad x_3 = \infty,
\end{equation}
$x_4$ is left as the unfixed coordinate. Therefore, the behaviour of $x_4$ in various kinematic limits encodes the singularity structure of the amplitude.

But what happens if we use another gauge choice? Say we let $x_1 = 0$, $x_3 = 1$, and $x_4 = \infty$. Well, one finds $x_2 \to 0$ as $s_{12} \to 0$. These two gauge-fixed configurations are related by Möbius transformations, confirming the CHY prescription’s co-ordinate invariance under this check.

Permuting external particles corresponds to permuting the associated punctures on the sphere. The analytic structure of the CHY integrand ensures that such permutations yield different kinematic channels (e.g., $s$, $t$, $u$) via analytic continuation. For example, interchanging $x_2$ and $x_3$ maps the $s$-channel into the $t$-channel~\cite{Schwartz_2013}. This reflects the crossing symmetry intrinsic to quantum field theory amplitudes \cite{Peskin:1995ev, Schwartz_2013}.

While the degenerating behaviour of punctures is physically intuitive, a complete understanding requires detailed analysis of the scattering equations. At higher $n$, one must classify how various subsets of punctures cluster and identify the corresponding factorisation channels. 

Because the moduli space $\mathcal{M}_{0,n}$ involves a quotient by $\mathrm{SL}(2,\mathbb{C})$, which is a 3-complex-dimensional group, this gauge redundancy allows us to fix the positions of any three marked points on $\mathbb{CP}^1$ without loss of generality. This means that among the $n$ puncture locations $\{x_i\}$, only $n-3$ are physically independent\cite{Cachazo_2013}. As a result, although there are $n$ scattering equations-one for each external particle-only $n-3$ of them are linearly independent. The remaining three are redundant due to the Möbius invariance of the formulation~\cite{Cachazo_2014}. This reduction is crucial in the CHY formalism, as it ensures the correct dimensionality of the integration over $\mathcal{M}_{0,n}$, and it allows one to localise the amplitude onto the solutions of just $n-3$ independent scattering equations.

\subsection{\texorpdfstring{$SL(2,\mathbb{C})$ Invariance of the Scattering Potential Implies Momentum Conservation}{SL(2,C) Invariance of the Scattering Potential Implies Momentum Conservation}}\label{subsec: invariance of the scattering potential implies momentum conservation}
Using the definition of the scattering potential in Eq.~\eqref{eq: potential of the scattering equations}, we study its behaviour under the Möbius transformation defined in Eq.~\eqref{eq: moebius transformation}. Under this transformation, we have:
\begin{align}
\ln |\tilde{x}_a - \tilde{x}_b| &= \ln \left| \frac{(AD - BC)(x_a - x_b)}{(C x_a + D)(C x_b + D)} \right| \nonumber \\
&= \ln |x_a - x_b| - \ln |(C x_a + D)(C x_b + D)|,
\end{align}
where we used the fact that $AD-BC=1$ in the second line.

Substituting into Eq.~\eqref{eq: potential of the scattering equations}, we find:
\begin{equation}
\mathcal{S}(\tilde{x}_1, \dots, \tilde{x}_n) = \mathcal{S}(x_1, \dots, x_n)
- \sum_{a < b} s_{ab} \ln |(C x_a + D)(C x_b + D)|.
\end{equation}

The third term simplifies to:
\begin{equation}
\sum_{a < b} s_{ab} \left( \ln |C x_a + D| + \ln |C x_b + D| \right)
= \sum_{a} \left( \sum_{b \ne a} s_{ab} \right) \ln |C x_a + D|,
\end{equation}
where we used the fact that $\sum_{a<b} = \frac{1}{2}\sum_{b\neq a}$. 

For massless kinematics, we have $s_{ab} = s_{ba}$, $s_{aa} = 0$, and $\sum_{b \ne a} s_{ab} = 0$, so the term above vanishes, and we conclude:
\begin{equation}
\mathcal{S}(\tilde{x}_1, \dots, \tilde{x}_n) = \mathcal{S}(x_1, \dots, x_n).
\end{equation}
Thus, the potential is invariant under global $\text{SL}(2,\mathbb{C})$ transformations if momentum is conserved.

\subsection{\texorpdfstring{Calculating a $4-$Point CHY Amplitude}{Calculating a 4-Point CHY Amplitude}}\label{sec: calculating a 4point chy amplitude}
Let's use the CHY recipe to calculate the partial amplitude $m_4(1234|1234)$ in the bi-adjoint $\phi^3$ theory. Using Eq.~\eqref{eq:genCHY} and Eq.~\eqref{eq:genintegrands}, the partial amplitude is simply 
\begin{equation}
A^{\rm massless}_{4} = \frac{1}{\mathrm{Vol}(\mathrm{SL}(2,\mathbb{C}))} \int \prod_{a=1}^n dx_a \prod_{a=1}^n \delta\left( E_a \right)\, \mathrm{PT}(1234)\mathrm{PT(1234)}\,,
\label{eq: chy n4 example}
\end{equation}
where
\begin{equation}
    E_a\equiv\sum_{\substack{b=1 \\ b\neq a}}^n \frac{s_{ab}}{x_a - x_b}.
\end{equation}
Now, clearly as discussed in Section \ref{subsec: cp1 and chy}, there is an $SL(2,\mathbb{C})$ redundancy we need to take care of. So, we need to choose any three punctures and gauge fix them to three different values. In this case, let's choose the following gauge
\begin{equation}
    \label{eq: gauge fixing for example}
    x_2=0, \quad x_3=1 \quad \& \quad x_4=\infty.
\end{equation}
To do this, we outline the following Fadeev-Popov gauge fixing procedure, as outlined in Ref.~\cite{Cachazo_2014_short}. We remove the delta functions corresponding to our gauge fixed variables, which in our case are $\delta(E_2),\delta(E_3)$ and $\delta(E_4)$. The final stage of our gauge fixing procedure is to compensate for the above by adding to our integral the factor
\begin{equation}
    (x_2-x_3)^2(x_3-x_4)^2(x_2-x_4)^2,
\end{equation}
such that now our $dx_a$'s and $\delta(E_a)$ are fully gauge invariant.

With our gauge fixing in place, Eq.~\eqref{eq: chy n4 example} becomes
\begin{equation}
    A_4 = \int dx_1 \delta(E_1)\frac{(x_2-x_3)^2(x_3-x_4)^2(x_2-x_4)^2}{(x_1-x_2)^2(x_2-x_3)^2(x_3-x_4)^2(x_4-x_1)^2},
\end{equation}
which when noting that $x_4=\infty$ and $x_2=0$, simplifies to 
\begin{equation}
    A_4 = \int dx_1 \delta(E_1)\frac{1}{x_1^2}.
    \label{eq: nearly last step}
\end{equation}
To fully evaluate this amplitude, we must solve the scattering equations $E_1$ to determine the puncture location $x_1$. Given Eq.~\eqref{eq: gauge fixing for example}, $E_1$ simply becomes
\begin{equation}
    E_1=\frac{s_{12}}{x_1}+\frac{s_{13}}{x_1-1}=0.
\end{equation}
\begin{figure}[hbt!]
    \centering
    \begin{center}
\begin{tikzpicture}[thick, scale=1.2]

% s-channel
\draw[<-] (-2,1) -- (-1,0) node[midway, above left=18pt] {$p_1$};
\draw[<-] (-2,-1) -- (-1,0) node[midway, below left=18pt] {$p_2$};
\draw[-] (-1,0) -- (1,0) node[midway, above=2pt] {$s_{12}= p_1 + p_2$};
\draw[->] (1,0) -- (2,1) node[midway, above right=18pt] {$p_3$};
\draw[->] (1,0) -- (2,-1) node[midway, below right=18pt] {$p_4$};

\node at (0,-1.8) {\textbf{s-channel}};

\end{tikzpicture}
\hspace{2.5cm}
\begin{tikzpicture}[thick, scale=1.2]

% t-channel
\draw[<-] (-2,1) -- (-0.5,0.5) node[midway, above left=4pt] {$p_1$};
\draw[<-] (-2,-1) -- (-0.5,-0.5) node[midway, below left=4pt] {$p_2$};
\draw[-] (-0.5,0.5) -- (-0.5,-0.5) node[midway, left] {$s_{14} = p_1 + p_4$};
\draw[->] (-0.5,0.5) -- (1,1) node[midway, above right=4pt] {$p_4$};
\draw[->] (-0.5,-0.5) -- (1,-1) node[midway, below right=4pt] {$p_3$};

\node at (-0.5,-1.8) {\textbf{t-channel}};

\end{tikzpicture}
\end{center}
    \caption{S-channel and t-channel diagrams for the amplitude under consideration.}
    \label{fig: s and t channel}
\end{figure}
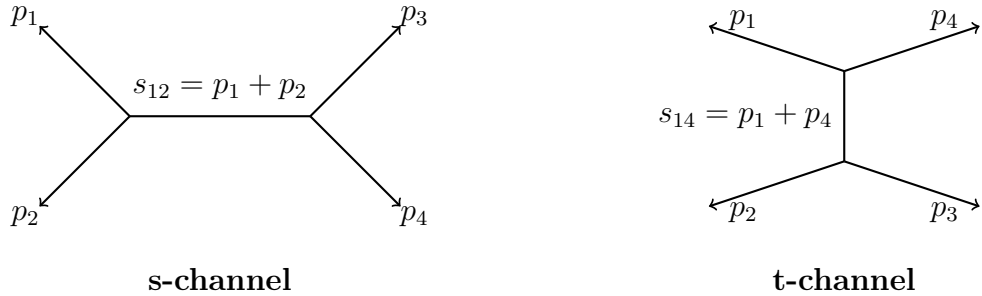

Plugging this into Eq.~\eqref{eq: nearly last step} and using the standard delta-function identities, we obtain our amplitude
\begin{equation}
    A_4 = \int dx_1 \frac{x_1(x_1-1)}{s_{12}+s_{13}}\: \delta\left(x_1-\frac{s_{12}}{s_{12}+s_{13}}\right) = \frac{s_{13}}{s_{12}(s_{12}+s_{13})}.
\end{equation}
If we set $s_{12}=s$, $s_{13}=u$ and $s_{14}=t$ and use the massless identity for Mandelstam invariants
\begin{equation}
    s+t+u=0,
\end{equation}
our amplitude simplifies to the expected result of an s-channel and t-channel diagram: 
\begin{equation}
    A_4 = -\frac{u}{st} = \frac{1}{s}+\frac{1}{t}.
\end{equation}

\section{The CHY Building Blocks}\label{sec: chy building blocks}
As suggested in Section \ref{sec:introduction}, the two main mathematical building blocks for constructing CHY integrands thus far are \emph{Parke--Taylor} factors and \emph{Pfaffians}. The following analysis is based of personal notes taken from group PSI essay meetings with the aim of learning more about the CHY formalism. Checks behind multi-linearity in the polarisation vectors, $\mathrm{SL}(2,\mathbb{C})$ redundancy and mass dimension of the Pfaffian mentioned in Ref.~\cite{Cachazo_2014_short} will be covered in these notes.

From our study of the bi-adjoint amplitude, we are quite comfortable with Parke-Taylor factors, and how they encode the necessary flavour degrees of freedom of a scattering theory. But what about these Pfaffians? What are they, and how are they viable candidates as a CHY building block?

For any even dimensional\footnote{If $M$ is of odd dimensionality, then its Pfaffian is zero.}, complex $2n\times2n$ anti-symmetric matrix $M$, the Pfaffian of $M$, denoted by $\mathrm{Pf}M$ is given by
\begin{equation}
\mathrm{Pf}M\equiv \frac{1}{2^n n!}\epsilon_{i_1j_1i_2j_2\cdots i_nj_n}M_{i_1j_1}M_{i_2j_2}\cdots M_{i_nj_n},
\end{equation}
where $M_{i_aj_a}$ is the element of $M$ located in the $i_a^{th}$ row and $j_a^{th}$ column~\cite{mehta2004random}. However, the Pfaffian of a complex, anti-symmetric matrix is actually related to the determinant of that matrix:
\begin{equation}
    \label{eq: determinant and Pfaffian}
    \mathrm{det}\ M = (\mathrm{Pf}M)^2.
\end{equation}
Rich in mathematical structure, the Pfaffian can be seen as more fundamental than the determinant of antisymmetric matrices, as it forms their core building block.

To see how the CHY amplitudes are constructed, let's do some dimensional analysis to determine how Parke-Taylor factors and Pfaffians feature in these tree--level amplitudes. 

So, how do we find the mass dimension? Well, we know that these amplitudes are sums over the final diagrams, and it's impossible to add two Feynman diagrams with two different mass dimensions, so it is completely okay to just look at one Feynmann diagram in our dimensional analysis. 

We know that momentum has mass dimension one~\cite{Schwartz_2013}, i.e., 
\begin{equation*}
    [k^\mu] = m,
\end{equation*}
and propagators have mass dimension minus two, i.e., 
\begin{equation}
    \left[\frac{1}{k^2}\right] =m^{-2}.
    \label{eq: scalar dimension}
\end{equation}
In an $n-$point, planar tree--level amplitudes, there are $n-3$ propagators in total~\cite{Elvang:2013cua}.  External legs typically do not affect a diagram’s mass dimension, as Feynman rules assign factors only to internal lines and vertices~\cite{Schwartz_2013,Peskin:1995ev}-except when external polarisation vectors are involved. In Yang-Mills theory and Einstein gravity, these polarisation vectors appear explicitly, with one and two copies respectively, reflecting the spin of the particles: $\mathbf{s}=1$ for vector bosons and $\mathbf{s}=2$ for gravitons. 

For partial amplitudes $m_n(\alpha|\beta)$ in the bi-adjoint $\phi^3$ scalar theory, the only contributors to the mass dimension in the tree--level amplitude are the propagators, which for $n-$external particles in the interaction, there are $n-3$ as stated above, thus the mass dimension is given by 
\begin{equation}
    [m_n] = m^{-2(n-3)} = m^{-2n+6}.
\end{equation}
For Yang--Mills amplitudes $A_n$ and Einstein gravity amplitudes $M_n$, we need to work only slightly harder, as these theories involve external particles carrying polarisation. Since there are $n - 2$ internal vertices in a tree-level diagram~\cite{Elvang:2013cua}, and each vertex involves polarisation degrees of freedom, the overall contribution from polarisation to the amplitude's mass dimension is $\mathbf{s}\cdot(n - 2)$. Therefore, the mass dimension of the amplitudes are
\begin{align}
    [A_n] &= m^{-2(n - 3) + (n - 2)} = m^{-n + 4}, \label{eq: YM dimension} \\
    [M_n] &= m^{-2(n - 3) + 2(n - 2)} = m^{2}. \label{eq: Einstein dimension}
\end{align}
From this simple dimensional analysis, we observe the relation:
\begin{equation}
    [M_n][m_n] = [A_n]^2,
\end{equation}
which is the statement that gravity times $\phi^3$ is Yang-Mills squared~\cite{Cachazo_2014_scalars}. This observation was made by Hodges, independent of CHY, by using twistor diagrams~\cite{hodges2011newexpressionsgravitationalscattering}. Moreover, the fact that the mass dimension in Eq.~\eqref{eq: Einstein dimension} is independent of the number of external particles suggests that gravitational amplitudes are, in a certain sense, the simplest among them all.

We can now use this dimensional analysis to convince us of the CHY amplitudes for these three theories, as presented in Eq.~\eqref{eq:genCHY} and Section \ref{subsec: introduction to CHY}. We begin by analysing each part of the CHY integrand. The integration variables themselves are dimensionless, but the delta functions contribute to the overall mass dimension due to their dependence on the scattering equations. Each delta function $\delta(E_a)$ carries mass dimension:
\begin{equation}
    [\delta(E_a)] = \left[\delta(s_{ab})\right]^{n - 3} = \left[\frac{1}{s_{ab}}\right]^{n - 3} = m^{-2(n - 3)},
\end{equation}
where in the first step we used the fact that the puncture coordinates $x_a$ are dimensionless, and that there are $n - 3$ independent scattering equations. The mass dimension follows from the Jacobian associated with the delta function.

Already, this matches the mass dimension of the bi-adjoint $\phi^3$ scalar theory amplitudes. This confirms that the building blocks of that theory can indeed be constructed from two Parke--Taylor factors alone. As we have seen, the Parke--Taylor factors enforce the necessary cyclic ordering dictated by the global $U(N) \times U(\tilde{N})$ symmetry.

However, for theories with polarisation degrees of freedom, such as Yang--Mills and Einstein gravity, we must go further: Parke--Taylor factors alone are insufficient to build the correct amplitudes. What kind of object, then, can be used to construct the integrand? Such an object must encode the polarisation data of the external particles, be symmetric under particle relabelling, and-crucially-have the correct mass dimension and gauge invariance properties. Understanding the required properties of this building block is essential to fathom the structure of amplitudes in these gauge and gravity theories. 

First, the expression must be multilinear in the polarisation vectors $\epsilon^\mu$. For a type $(0,n)$ tensor, the whole object should be a Lorentz scalar, i.e., 
    \begin{equation}
     \epsilon^{\mu_1}_1\epsilon^{\mu_2}_2\cdots\epsilon^{\mu_n}_nT_{\mu_1\cdots \mu_n}(x,k).
    \end{equation}
Physically, this means that the expression includes exactly one polarization vector for each external particle.

Secondly, the expression must also be \emph{gauge invariant}, meaning that under the transformation
\begin{equation}
    \epsilon_{a}^{\mu} \to \epsilon_{a}^{\mu}+\alpha k^{\mu}_a,
\end{equation}
then the answer shouldn't change~\cite{Cachazo_2014_scalars, Cachazo_2014_short}. Polarisation vectors alone are not sufficient to describe gluons or gravitons; they yield correct physical results only when embedded within expressions that ensure gauge invariance. This is equivalent to requiring that if $\epsilon_a^\mu$ is replaced by $k_a^\mu$, then the expression vanishes, i.e., 
\begin{equation}
    k_a^{\mu_a}T_{\mu_1\cdots\mu_a\cdots\mu_n}(x,k)=0,
\end{equation}
which is a manifestation of the Ward identity. This condition guarantees gauge invariance required. 

Luckily for us, the Pfaffian, $\mathrm{Pf}\Psi$ satisfies such conditions, with gauge invariance occuring via a simply linear algebra property of the Pfaffians~\cite{jashi2024scaffoldingresiduesyangmillsscalara}. The proposed $\Psi$ is a $2n\times 2n$ matrix, defined as~\cite{Cachazo_2014_short}
\begin{equation}
\label{eq: Psi matrix}
\Psi_{2n} = \begin{bmatrix}
A & -C^T \\
C & B
\end{bmatrix}.
\end{equation}
The $n\times n$ block matrices $A,B$ and $C$ serve also as physical building block in the amplitudes. To exploit the convenient properties of the Pfaffian for antisymmetric matrices, the structure in Eq.~\eqref{eq: Psi matrix} is chosen to be antisymmetric by taking $A$ and $B$ to be antisymmetric matrices with off-diagonal components defined in terms of the punctures $x_a$ and the momentum $k_a^\mu$, and polarisation vectors $\epsilon_a^\mu$ assigned to each external particle
\begin{equation}
 A_{ab}:= \frac{k_a\cdot k_b}{x_a-x_b} \quad {\rm and} \quad B_{ab}:= \frac{\epsilon_a\cdot \epsilon_b}{x_a-x_b}.   
\end{equation}
The matrix $C$ was constructed to make gauge invariance manifest~\cite{jashi2024scaffoldingresiduesyangmillsscalara}, with components
\begin{equation}
    C_{aa}:=-\sum_{\substack{b=1\\b\neq a}}^n C_{ab} \quad {\rm and}\quad C_{ab}:= \frac{\epsilon_a\cdot k_b}{x_a-x_b}.
\end{equation}

However, by considering the Pfaffian $\mathrm{Pf}\Psi$ alone for a Yang--Mills amplitude, we run into some problems with dimensional analysis and the fact that $\mathrm{Pf}\Psi=0$. The latter observation follows from the fact that the first $n$ rows (or columns) are linearly dependent due to momentum conservation. The former follows from the definition of the Pfaffian (Eq.~\eqref{eq: determinant and Pfaffian}) and Eq.~\eqref{eq: Psi matrix}. From Eq.~\eqref{eq: determinant and Pfaffian} and Eq.~\eqref{eq: Psi matrix}, the mass dimension of the Pfaffian is 
\begin{equation}
    \left[\mathrm{Pf}\Psi\right] = [k_1^{\mu_1}\cdots k_n^{\mu_n}] = m^n,
\end{equation}
but this does clearly not give the necessary contribution to the mass-dimension to satisfy Eq. \eqref{eq: YM dimension}. Both these observations lead us to consider the  \emph{reduced} Pfaffian $\mathrm{Pf'}\Psi$ as a candidate for a mathematical building block of the CHY formalism~\cite{Cachazo_2014_short}. 

The reduced Pfaffian $\mathrm{Pf'}\Psi$ is defined as \cite{Cachazo_2014_short}
\begin{equation}
    \mathrm{Pf'}\Psi \equiv 2\frac{(-1)^{a+b}}{(x_a-x_{b})}\mathrm{Pf}(\Psi^{ab}_{ab}),
    \label{eq: the reduced Pfaffian}
\end{equation}
where $\Psi^{ab}_{ab}$ is the matrix $\Psi$ with both rows and columns $a$ and $b$ removed for $1\leq a<b\leq n$. Eq.~\eqref{eq: the reduced Pfaffian} is non-zero and independent of the choice of $a$ and $b$. Given this elimination, the mass dimension of Eq.~\eqref{eq: the reduced Pfaffian} is exactly 
\begin{equation}
    [\mathrm{Pf'}\Psi] = m^{n-2},
\end{equation}
as needed for our amplitude. It was proved in Ref. \cite{Cachazo_2014_short} that the reduced Pfaffian is also invariant under permutations of particle labels. So, truly the reduced Pfaffian is a correct candidate for our amplitudes. This explains why Yang--Mills amplitudes under the CHY construction are built with one copy of $\mathrm{Pf'}\Psi$ and a Parke-Taylor factor (to induce flavour ordering), while Einstein-Gravity amplitudes need two copies of $\mathrm{Pf'}\Psi$.

\section{Colour Decomposition of Tree--Level Amplitudes}\label{sec: colour ordered amplitudes}

To better understand the analytic structure of tree-level amplitudes, it is useful to separate their colour and kinematic components. To gain some intuition, a natural starting point for this is Yang--Mills theory, where colour degrees of freedom of the gluons are most commonly known. The dynamics are governed by the Lagrangian

\begin{equation}
\mathcal{L} = -\frac{1}{4} \sum_{a_1} F^{a_1}_{\mu\nu} F^{{a_1}\mu\nu}, \quad F^{a_1}_{\mu\nu} = \partial_\mu A^{a_1}_\nu - \partial_\nu A^{a_1}_\mu + g f^{a_1a_2a_3} A^{a_2}_\mu A^{a_3}_\nu,
\label{eq: yang mills lagrangian}
\end{equation}
where $g$ measures the strength of the interaction. 

The indices $ a_1, a_2 \; \& \; a_3 $ label elements of the Lie algebra, with structure constants $f^{a_1a_2a_3}$. These are often informally referred to as \emph{colour indices}, since gluons in Yang--Mills theory carry a so-called \emph{colour charge}~\cite{peskin2019concepts,Thomson_2013}. The non-Abelian nature of the theory manifests in the field strength tensor through a self-interaction term proportional to the structure constants $f^{a_1a_2a_3}$ of the gauge group $SU(N)$~\cite{pirsa_PIRSA:18040035, peskin2019concepts}.

To avoid confusion, it is worth noting that this colour decomposition can be constructed for the bi-adjoint $\phi^3$ theory, which carries a \emph{global} (not a gauge) $U(N)\times U(\tilde{N})$ symmetry. In this case, the structure constants arise from the theory's interaction term~\cite{Cachazo_2014_scalars}
\begin{equation}
    \mathcal{L}_{\mathrm{int.}}^{\mathrm{BA}-\phi^3} = -f_{abc}\tilde{f}_{\tilde{a}\tilde{b}\tilde{c}}\phi^{a\tilde{a}}\phi^{b\tilde{b}}\phi^{c\tilde{c}}
\end{equation}
instead of a gauge connection like in Yang-Mills theory. In this case, the structure constants $f_{abc}$ and $\tilde{f}_{\tilde{a}\tilde{b}\tilde{c}}$ are from the Lie algebra corresponding to the global symmetry groups of the theory $U(N)$ and $U(\tilde{N})$ respectively.  Although the theory has two global symmetry groups, we will focus on the colour decomposition with respect to one group at a time, since the amplitude sums linearly over independent colour structures from each group. 

Our symmetry group $U(N)$, the group of unitary $N\times N$ matrices, form a vector space consisting of $N^2-1$ generators. Any unitary matrix $\mathcal{U}\in U(N)$ can be expressed as 
\begin{equation}
    \mathcal{U}=\exp(-i\alpha^{a_1}T^{a_1}),
\end{equation}
with the adjoint index $a_1$ running over the $N^2$ elements of $U(N)$. Analogous to an exponential expansion, the vector space decomposes into the identity matrix (generating a $U(1)$ subgroup) and traceless matrices generating the $SU(N)$ subgroup~\cite{pirsa_PIRSA:18040035, peskin2019concepts}. Hence, we have
\begin{equation}
    U(N) =SU(N)\times U(1).
\end{equation}
The structure constants are defined via the commutation relations of the generators $T^{a_1}$~\cite{Georgi:1999wka}:
\begin{equation}
    i[T^{a_1}, T^{a_2}] = \sum_{a_3} f^{a_1a_2a_3} T^{a_3}.
\end{equation}

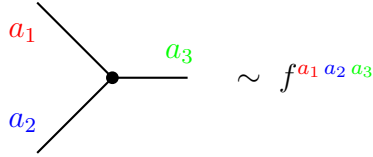
\begin{figure}[hbt!]
    \centering
\begin{tikzpicture}[thick, scale=1, baseline=(v)]
  % Vertex
  \coordinate (v) at (0,0);

  % External lines
  \coordinate (a) at (-1,1);
  \coordinate (b) at (-1,-1);
  \coordinate (c) at (1,0);

  % Draw lines with labels placed outside
  \draw (a) -- (v) node[pos=0.4, left=7pt] {\textcolor{red}{$a_1$}};
  \draw (b) -- (v) node[pos=0.4, left=7pt] {\textcolor{blue}{$a_2$}};
  \draw (v) -- (c) node[midway, above right=2pt] {\textcolor{green}{$a_3$}};

  % Vertex dot
  \filldraw[black] (v) circle (2pt);

  % Equals sign and structure constant
  \node[right=1.5cm of v] (label) {$\sim\; f^{\textcolor{red}{a_1}\,\textcolor{blue}{a_2}\,\textcolor{green}{a_3}}$};
\end{tikzpicture}
      \caption{How the structure constants $f^{a_1a_2a_3}$ of the $U(N)$ group appear in a $3-$point vertex.}
    \label{fig: structure constant at three point vertex}
\end{figure}

\begin{figure}
    \centering
    \begin{tikzpicture}[scale = 2]
    \begin{scope}[shift ={(0,1.5)}]
        \node[scale=1.5] at (-2,0) {$(T^a)_{ij}$};
    \node[scale=1.5] at (-1.1,0) {$\longrightarrow$};
\draw[-] (1,0.5) arc (90:270:0.5);
  \draw[->] (0.5,0) arc (180:135:0.5);
  \draw[->] (1,-0.5) arc (270:225:0.5);
    \node[left] at (0.6,0.35) {$j$};
    \node[left] at (0.6,-0.35) {$i$};

  % Straight line connecting the two arcs
  \draw[-] (-0.5,0) -- (0.5,0);
    \node[above] at (0,0) {$a$};
    \end{scope}
    \begin{scope}[shift ={(0,0)}]
        \node[scale=1.5] at (-2,0) {$\mathrm{Tr}(T^aT^b) = (T^a)_{ij}(T^b)_{ji} = $};
        \draw[-] (0.0,0) -- (1,0); 
        \node[above] at (0.5,0) {$a$};
        \draw[->] (1,0) arc (180:90:0.5);
    \draw[>-] (1.5,-0.5) arc (270:180:0.5);
    \draw[-] (1.5,0.5) arc (90:-90:0.5);
    \draw[-] (2,0) -- (3,0);
    \node[above] at (2.5,0) {$b$};
    \node[above] at (1.6,0.5) {$j$};
    \node[below] at (1.6,-0.45) {$i$};
    \node[scale=1.5] at (-0.33,-1) {$=$};
    \draw[-] (0.0,-1) -- (1,-1);
    \node[above] at (0.2,-1) {$a$};
    \draw[-] (1,-1) -- (2,-1);
    \node[above] at (1.8,-1) {$b$};

    \node[scale=1.5] at (0.05,-1.5) {$= \quad \delta_{ab}$};
    \end{scope}
    \end{tikzpicture}
    \caption{Diagrammatic Representation of the generators of $U(N)$ and their traces \cite{pirsa_PIRSA:18040035}.}
    \label{fig: generators and traces }
\end{figure}
The three-point vertex is decorated with a factor of $f^{a_1a_2a_3}$ according to the Feynman rules, encoding the colour structure of the interaction. To streamline the colour structure, one can employ a colour-ordering strategy that rewrites the structure constants in terms of traces over generators\cite{https://doi.org/10.5170/cern-2014-008.31, pirsa_PIRSA:18040035}:
\begin{equation}
     f^{a_1a_2a_3} = \text{Tr}(T^{a_1} T^{a_2} T^{a_3}) - \text{Tr}(T^{a_1} T^{a_3} T^{a_2}),
    \label{eq: structure constants as sum over traces}
\end{equation}
where the following normalisation identity of the generators was used
\begin{equation}
    \mathrm{Tr}(T^{a_1}T^{a_2})=\frac{1}{2}\delta^{a_1a_2}.
\end{equation}

Following the graphical notation of Fig.~\ref{fig: generators and traces }, this corresponds to replacing the vertex with a sum over loops, as shown in Fig.\ref{fig: diagrammatic rep of structure constants as traces}. This reformulation allows us to define colour-ordered amplitudes, where colour factors are stripped off and organized into sums of single trace terms\cite{https://doi.org/10.5170/cern-2014-008.31}, allowing the remaining kinematic part to be studied in isolation, which in the end is what we really want to study.
\begin{figure}[hbt!]
\begin{center}
\begin{tikzpicture}[thick, scale=0.8]
  % Vertex
  \coordinate (v) at (0,0);

  % External lines
  \coordinate (a) at (-1,1);
  \coordinate (b) at (-1,-1);
  \coordinate (c) at (1,0);

  % Draw lines with labels placed outside
  \draw (a) -- (v) node[pos=0.4, left=7pt] {\textcolor{red}{$a_1$}};
  \draw (b) -- (v) node[pos=0.4, left=7pt] {\textcolor{blue}{$a_2$}};
  \draw (v) -- (c) node[midway, above right=2pt] {\textcolor{green}{$a_3$}};

  % Vertex dot
  \filldraw[black] (v) circle (2pt);
\end{tikzpicture}
\hspace{1em}
\raisebox{2em}{$=$}
\hspace{1em}
\begin{tikzpicture}[scale=0.8]
  \coordinate (v) at (0,0);

  \draw (120:0.7) -- +(120:1.2) node[pos=0.4, left=7pt] 
 {\textcolor{red}{$a_1$}};
  \draw (360:0.7) -- +(360:1.2) node[midway, above right=2pt] {\textcolor{green}{$a_3$}};
  \draw (220:0.7) -- +(220:1.2) node[pos=0.4, left=7pt] {\textcolor{blue}{$a_2$}};

  \draw[fermion] (-0.7,0) arc (180:-180:0.7);
\end{tikzpicture}
\hspace{1em}
\raisebox{2em}{$-$}
\hspace{1em}
\begin{tikzpicture}[scale=0.8]
  \coordinate (v) at (0,0);
  \draw[thick, ->] (0.7,0) arc (0:360:0.7);

  \draw (120:0.7) -- +(120:1.2) node[pos=0.4, left=7pt] 
 {\textcolor{red}{$a_1$}};
  \draw (360:0.7) -- +(360:1.2) node[midway, above right=2pt] {\textcolor{green}{$a_3$}};
  \draw (220:0.7) -- +(220:1.2) node[pos=0.4, left=7pt] {\textcolor{blue}{$a_2$}};
\end{tikzpicture}
\end{center}
    \caption{Diagrammatic Representation of Eq.~\eqref{eq: structure constants as sum over traces} \cite{pirsa_PIRSA:18040035,https://doi.org/10.5170/cern-2014-008.31}.}
    \label{fig: diagrammatic rep of structure constants as traces}
\end{figure}

\begin{figure}[hbt!]
\begin{center}
\begin{tikzpicture}[baseline=(current bounding box.center), scale=1.8]
  % Tree-level again (leftmost)
  \coordinate (a1) at (0,0);
  \coordinate (a2) at (0,2);
  \coordinate (a3) at (2,2);
  \coordinate (a4) at (2,0);
  \coordinate (leftmid) at (0.5,1);
  \coordinate (rightmid) at (1.5,1);

  \draw (a1) -- (leftmid) -- (a2);
  \draw (a3) -- (rightmid) -- (a4);
  \draw (leftmid) -- (rightmid);

  \node at ($(a1) + (-0.2,-0.1)$) {$a_1$};
  \node at ($(a2) + (-0.2,0.1)$) {$a_2$};
  \node at ($(a3) + (0.2,0.1)$) {$a_3$};
  \node at ($(a4) + (0.2,-0.1)$) {$a_4$};

  \node at (2.5,1) {$=$};
  \filldraw (leftmid) circle (0.05);
\filldraw (rightmid) circle (0.05);

  % Double bubble with cut
  \begin{scope}[xshift=3cm]
    \coordinate (a1) at (0,0);
    \coordinate (a2) at (0,2);
    \coordinate (a3) at (2,2);
    \coordinate (a4) at (2,0);
    \coordinate (leftmid) at (0.8,1);
    \coordinate (rightmid) at (1.2,1);

    % Internal loops
    \draw[fermion] (0.2,1) arc (180:-180:0.3);
    \draw[fermion] (1.2,1) arc (180:-180:0.3);
    \draw[postaction={decorate}, decoration={markings, mark=at position 0.55 with {\arrow{>}}}] (0.2,1) arc (180:-180:0.3);
    \draw[postaction={decorate}, decoration={markings, mark=at position 0.55 with {\arrow{>}}}] (1.2,1) arc (180:-180:0.3);

    % External legs
    \draw (a1) -- (0.5,0.7);
    \draw (a2) -- (0.5,1.3);
    \draw (a3) -- (1.5,1.3);
    \draw (a4) -- (1.5,0.7);
    \draw (leftmid) -- (rightmid);

    % Labels
    \node at ($(a1) + (-0.2,-0.1)$) {$a_1$};
    \node at ($(a2) + (-0.2,0.1)$) {$a_2$};
    \node at ($(a3) + (0.2,0.1)$) {$a_3$};
    \node at ($(a4) + (0.2,-0.1)$) {$a_4$};

  \end{scope}

  \node at (6,1) {$\pm \:\:\:\text{(3 other terms)}$};
\end{tikzpicture}
\end{center}
    \caption{Diagrammatic representation of breaking down a $4-$point amplitude into trace structures \cite{pirsa_PIRSA:18040035}.}
    \label{fig:first colour ordered amplitude QCD appendix}
\end{figure}
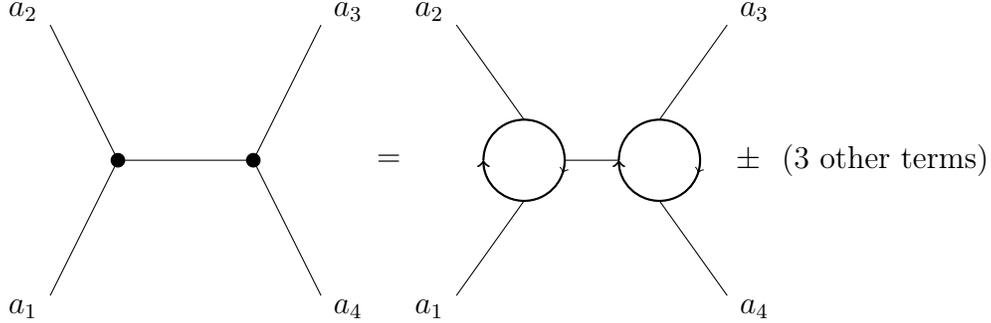

Before generalizing to the $n$-point case, we examine the 4-point tree-level amplitude. In the same style as Fig.\ref{fig: diagrammatic rep of structure constants as traces}, the colour-labelled diagram on the left of Fig.\ref{fig:first colour ordered amplitude QCD appendix} can be expanded using Eq.~\eqref{eq: structure constants as sum over traces} into four terms, each traceable at the internal vertex points as shown on the right.

\begin{figure}
    \centering
    \begin{tikzpicture}[thick,scale=2]

  % Incoming semicircle arrow (left)
  \draw[-] (-0.5,0.5) arc (90:-90:0.5);
  \draw[->] (-0.5,0.5) arc (90:45:0.5);
  \draw[->] (0,0) arc (0:-45:0.5);
    \node[right] at (-0.15,0.35) {$i$};
    \node[right] at (-0.15,-0.35) {$j$};

  % Outgoing semicircle arrow (right)
  \draw[-] (2.5,0.5) arc (90:270:0.5);
  \draw[->] (2,0) arc (180:135:0.5);
  \draw[->] (2.5,-0.5) arc (270:225:0.5);
    \node[left] at (2.1,0.35) {$l$};
    \node[left] at (2.1,-0.35) {$k$};

  % Straight line connecting the two arcs
  \draw[->] (0,0) -- (1,0);
  \node[above] at (1,0) {$a$};
  \draw[-] (1,0) -- (2,0);
  \node at (3,0.0) {$=$};
    \draw[->] (3.5,0.5) -- (4.5,0.5);
    \draw[-] (3.5,-0.5) -- (4.5,-0.5);
    \draw[-] (4.5,0.5) -- (5.5,0.5);
    \draw[<-] (4.5,-0.5) -- (5.5,-0.5);
     \node[above] at (3.5,0.5) {$i$};
    \node[above] at (5.5,0.5) {$l$};
     \node[below] at (3.5,-0.5) {$j$};
    \node[below] at (5.5,-0.5) {$k$};

\end{tikzpicture}
\caption{Graphical representation of the $U(1)$ completeness relation \cite{pirsa_PIRSA:18040035}.}
\label{fig: u1 completeness relation}
\end{figure}
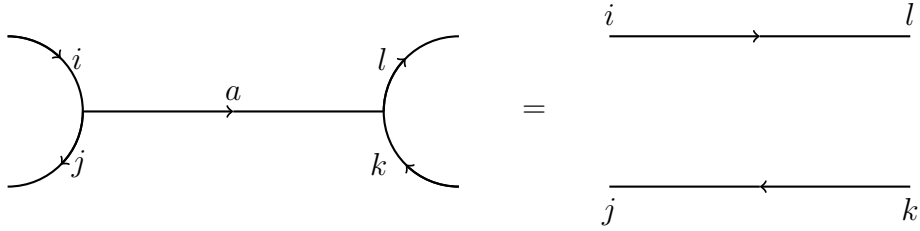

Now, you might be wondering why we have chosen to use $U(N)$, rather than the more commonly used $SU(N)$ in gauge theories. One motivation is that $U(N)$ naturally incorporates both the non-abelian $SU(N)$ and the abelian $U(1)$ in its basis decomposition, which allows the inclusion of photons and other particles transforming under $U(1)$. More importantly for us, the $U(N)$ algebra provides us with a single-term completeness relation for generators $T^{a_1}$ in the fundamental representation
\begin{equation}
    \label{eq: completness relation for U(N)}
    \sum_{a_1=1}^{N^2}(T^{a_1})_{ij}(T^{a_1})_{kl} = \delta_{il}\delta_{jk},
\end{equation}
which will bring the trace decomposition to life (see Fig.~\ref{fig: u1 completeness relation}). In contrast, the completeness relation for $\mathrm{SU}(N)$ is more complicated: the $N^2-1$ generators of $\mathrm{SU}(N)$ are \emph{traceless}, so we would require an extra term to subtract the trace part. 

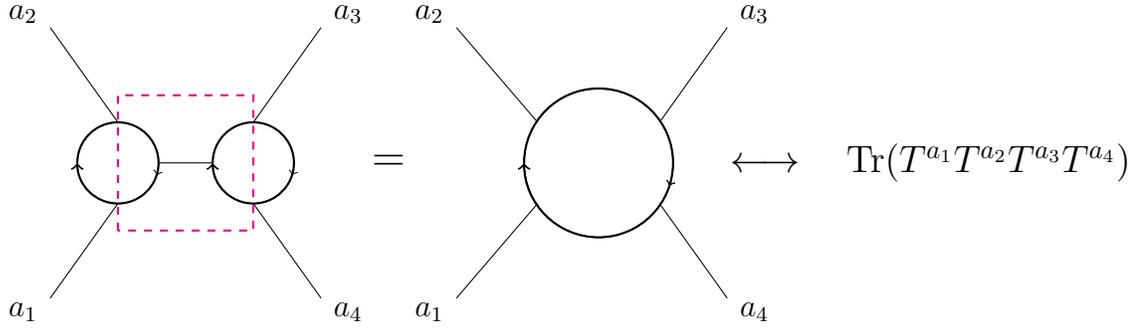
\begin{figure}[hbt!]
    \begin{center}
\begin{tikzpicture}[baseline=(current bounding box.center), scale=1.8]
  \usetikzlibrary{decorations.markings}
  
  \coordinate (a1) at (0,0);
  \coordinate (a2) at (0,2);
  \coordinate (a3) at (2,2);
  \coordinate (a4) at (2,0);
  \coordinate (leftmid) at (0.8,1);
  \coordinate (rightmid) at (1.2,1);

  % Internal loops with arrows
  \draw[fermion] (0.2,1) arc (180:-180:0.3);
  \draw[fermion] (1.2,1) arc (180:-180:0.3);
  \draw[postaction={decorate}, decoration={markings, mark=at position 0.55 with {\arrow{>}}}] 
    (0.2,1) arc (180:-180:0.3);
  \draw[postaction={decorate}, decoration={markings, mark=at position 0.55 with {\arrow{>}}}] 
    (1.2,1) arc (180:-180:0.3);

  % External legs
  \draw (a1) -- (0.5,0.7);
  \draw (a2) -- (0.5,1.3);
  \draw (a3) -- (1.5,1.3);
  \draw (a4) -- (1.5,0.7);
  \draw (leftmid) -- (rightmid);

  % Labels
  \node at ($(a1) + (-0.2,-0.1)$) {$a_1$};
  \node at ($(a2) + (-0.2,0.1)$) {$a_2$};
  \node at ($(a3) + (0.2,0.1)$) {$a_3$};
  \node at ($(a4) + (0.2,-0.1)$) {$a_4$};

  % Dashed pink box intersecting centers of the loops
  \draw[dashed, thick, magenta] (0.5,0.5) rectangle (1.5,1.5);
  \node at (2.5,1) {\Large$=$};
  
  \begin{scope}[xshift=3cm]
    \coordinate (a1) at (0,0);
  \coordinate (a2) at (0,2);
  \coordinate (a3) at (2,2);
  \coordinate (a4) at (2,0);

  % Internal loops with arrows
  \draw[fermion] (0.5,1) arc (180:-180:0.55);
  \draw[postaction={decorate}, decoration={markings, mark=at position 0.55 with {\arrow{>}}}] 
    (0.5,1) arc (180:-180:0.55);
  \draw[postaction={decorate}, decoration={markings, mark=at position 0.55 with {\arrow{>}}}] 
    (0.5,1) arc (180:-180:0.55);

  % External legs
  \draw (a1) -- (0.6,0.7);
  \draw (a2) -- (0.6,1.3);
  \draw (a3) -- (1.5,1.3);
  \draw (a4) -- (1.5,0.7);

  % Labels
  \node at ($(a1) + (-0.2,-0.1)$) {$a_1$};
  \node at ($(a2) + (-0.2,0.1)$) {$a_2$};
  \node at ($(a3) + (0.2,0.1)$) {$a_3$};
  \node at ($(a4) + (0.2,-0.1)$) {$a_4$};
  \end{scope}
  \node at (6.5,1) {\large$\longleftrightarrow \quad \mathrm{Tr}(T^{a_1}T^{a_2}T^{a_3}T^{a_4})$};
\end{tikzpicture}
\end{center}
    \caption{Single-trace color structure for a $4$-point colour-ordered amplitude \cite{pirsa_PIRSA:18040035}.}
    \label{fig: final breakdown of amplitude}
\end{figure}

Each term in Eq.~\eqref{eq: completness relation for U(N)} corresponds to a single trace over generators, representing a specific colour ordering (see Fig.\ref{fig: final breakdown of amplitude}).

Hence, the $4-$point colour-dressed amplitude for the $\beta$ ordering is given by
\begin{align}
    \mathcal{M}_4(\alpha) =  &\mathrm{Tr}(T^{a_1}T^{a_2}T^{a_3}T^{a_4})A(\alpha|1234)+(T^{a_1}T^{a_3}T^{a_4}T^{a_2})A(\alpha|1342) \nonumber \\
    &+(T^{a_1}T^{a_2}T^{a_4}T^{a_3})A(\alpha|1243)+(T^{a_1}T^{a_3}T^{a_2}T^{a_4})A(\alpha|1324). \label{eq: beta ordering}
\end{align}
The result of this decomposition over one symmetry group at tree level for the $n-$external particle is the expression for the amplitude \cite{Dixon:1996wi, https://doi.org/10.5170/cern-2014-008.31}:
\begin{equation}
    \mathcal{M}_n^{\text{tree}}(\alpha) = \sum_{\beta\in S_n/\mathbb{Z}_n} \text{Tr}(T^{\tilde{a}_{\beta(1)}} \cdots T^{\tilde{a}_{\beta(n)}})\, m_n(\alpha| \beta(1), \dots, \beta(n)),
    \label{eq: yang-mills amplitude}
\end{equation}
where $m_n$ are the colour-ordered partial amplitudes in the bi-adjoint $\phi^3$ theory with the first ordering fixed, since we decided to decompose one flavour group at a time. Again as in Section \ref{subsec:22}, the sum is over the set $S_n/\mathbb{Z}_n$, begin careful to keep only the cyclically inequivalent orderings. Once fully decomposed into colour and kinematics (Eq.~\eqref{eq: biadjoint phi3 colour ordered formal expression}), the corresponding partial amplitudes capture all the kinematic dependence, while the trace over generators encodes the colour structure. This highlights the significance of planar Feynman diagrams at tree level, as only they contribute to specific colour orderings~\cite{pirsa_PIRSA:18040035}.

\section{Counting Mandelstam Invariants via Maximal Length \& Ferrers Partitions}\label{sec: appendix counting kinematic invariants}

As noted in Section \ref{subsec: Planar Feynman Diagrams and Planar Kinematic Invariants}, an alternative method for counting the dimension of the basis of planar kinematic invariants was suggested. However, this method requires separate consideration of the number of particles $n$ depending on whether $n$ is odd or even. The counting argument will centre around the maximal length a planar kinematic invariant can have for an odd or even number of particles after applying momentum conservation maximally.

Starting with the case for odd $n$, the maximum length a kinematic invariant can have is $\mathcal{L}_{max} = \frac{n-1}{2}$. For each kinematic invariant of length $2 \leq j \leq \mathcal{L}_{max}$, there are $n$ kinematic invariants of that length. For example, for length $2$, we have $s_{12}, s_{23}, \ldots, s_{n1}$, which gives $n$ terms in total. Thus, the total number of independent kinematic invariants is given by multiplying $(\mathcal{L}_{max} - 1)$ by $n$, since there is no kinematic invariant of length $1$. This yields
\begin{equation}
    n\left(\mathcal{L}_{max} -1\right) = n\left(\frac{n-1}{2} -1\right) = \frac{n(n-3)}{2}
\end{equation}
independent kinematic invariants in total. 

For the case of even $n$, the maximum length a kinematic invariant can have is $\mathcal{L}_{max} = \frac{n}{2}$. For each kinematic invariant of length $2 \leq j \leq \mathcal{L}_{max}-1$, there are $n$ kinematic invariants of that length. However, due to momentum conservation\footnote{Double check the case for $n=6$ in Section \ref{subsec: Planar Feynman Diagrams and Planar Kinematic Invariants} if you do not believe me!}, we only have $\frac{n}{2}$ kinematic invariants for kinematic invariants of length $\mathcal{L}_{max}$. Therefore, in a similar manner to the odd-particle case, we subtract $\frac{n}{2}$ from the total count for momentum conservation, yielding
\begin{equation}
    n\left(\mathcal{L}_{max}-1\right) - \frac{n}{2} = n\left(\frac{n}{2}-1\right) - \frac{n}{2} = \frac{n(n-3)}{2}.
\end{equation}

This counting method can be effectively visualized using Ferrers shapes. While Ferrers shapes were introduced in Section \ref{subsec: Counting Planar Mandelstam Invariants via Maximally Symmetric Ferrers Shapes.}, their definition extends beyond maximally symmetric partitions.
\begin{figure}[hbt!]
    \centering
    \ytableausetup{boxsize=2em} % Set larger box size
    \newcommand{\inv}[1]{\scriptstyle #1} % Shrink text inside boxes
    \begin{tabular}{c @{\hspace{3em}} c}% Create two columns (left and right)
        % n=5 Ferrers Diagram (Left Side)
        \begin{ytableau}
            s_{12} \\
            s_{23} \\
            s_{34}\\
            s_{45}         \\
            s_{51}
        \end{ytableau}
        &
        % n=7 Ferrers Diagram (Right Side)
        \begin{ytableau}
            s_{12} & s_{123}\\
            s_{23} & s_{234} \\
            s_{34} & s_{345} \\
            s_{45} & s_{456} \\
            s_{56} & s_{567}\\
            s_{67}& s_{671}\\
            s_{71}& s_{712}
        \end{ytableau}
    \end{tabular}  
    \caption{Counting planar kinematic invariants visualised as a Ferrer's diagram for an odd number of particles. Examples in this figure include $n=5$ (left) and $n=7$  (right) particles, with the planar kinematic invariants ordered in the blocks.}
    \label{fig:ferrers odd}
\end{figure}

\begin{figure}[hbt!]
    \centering
    \ytableausetup{boxsize=2em} % Set larger box size
    \newcommand{\inv}[1]{\scriptstyle #1} % Shrink text inside boxes
    \begin{tabular}{c @{\hspace{3em}} c}% Create two columns (left and right)
        % n=4 Ferrers Diagram (Left Side)
        \begin{ytableau}
            s_{12} \\
            s_{23} \\
        \end{ytableau}
        &
        % n=6 Ferrers Diagram (Right Side)
        \begin{ytableau}
            s_{12} & s_{123}\\
            s_{23} & s_{234}\\
            s_{34} & s_{345} \\
            s_{45}  \\
            s_{56}       \\
            s_{61}       \\
        \end{ytableau}
    \end{tabular}  
    \caption{Counting planar kinematic invariants visualised as a Ferrer's diagram for an even number of particles. Examples in this figure include $n=4$ (left) and $n=6$  (right) particles, with the planar kinematic invariants ordered in the blocks.}
    \label{fig:ferrers even}
\end{figure}
As shown in Figure \ref{fig:ferrers odd}, the Ferrer's shape is still of size $n(n-3)/2$ but of partition $(n,n,\ldots,n)$ with $\mathcal{L}_{max}-1$ columns.
Since our counting differs slightly for the case of an even number of particles, the Ferrers shape partitioning will also be different. As shown in Figure \ref{fig:ferrers even}, the Ferrers shape still has size $\frac{n(n-3)}{2}$, but the partition is given by $(n, n, \ldots, n, \frac{n}{2})$, with $\mathcal{L}_{max}-1$ columns. The first $\mathcal{L}_{\max} - 2$ columns have $n$ blocks each, with the final column containing $\frac{n}{2}$ blocks.

\section{Combinatorial Connection to the Polynomial Form of the Scattering Equations}\label{appendix: Combinatorial Connection to the Polynomial Form of the Scattering Equations}

Interestingly, counting planar kinematic invariants in this way also reveals a combinatorial connection to the homogeneous polynomial equations equivalent to the scattering equations, as presented by Dolan and Goddard in Ref. \cite{Dolan_2014_poly}.

A principal result of Ref. \cite{Dolan_2014_poly} is that the scattering equations for massless particles are equivalent to the homogenous polynomial equations
\begin{equation}
    \sum_{\substack{S\subset A \\ |S|=m}}k_S^2z_S = 0, \quad 2 \leq m \leq n-2,
\end{equation}
with $A=\{1,..., n\}$, $k_S = \sum_{b\in S}k_b$, i.e., the Mandelstam invariants, and $z_S =\prod_{a\in S}z_a$ for $S\subset A$. Now the sum is over subsets $S \subset A$ with $m$ elements. Notice that the length of the subsets in the sum are partitioned into $(2,...,n-2)$ sets, exactly like the partitioning of our Ferrer's shapes in the previous section. So the question is, is there a correspondence between counting the number of planar kinematic invariants and counting the number of subsets partitioned in sizes of $2\leq m \leq n-2$?

To count subsets to obtain a double sum as presented in Eq.~\eqref{eq: counting structure for planar kinematic invariants}, we count the number of subsets for each $m$, that do not have a distinct ordering and all elements are different within each subset\footnote{Physically, this relates to the fact that Mandelstam invariants have distinct index labels.}, and then add them altogether. For $m=2$, we order our elements $\{1,2,....,n\}$ and then count the number of subsets that contain each element, but excluding subsets containing any of the previous elements. For example, we have $n-1$ subsets with element (labelled) $1$ in it, then $n-2$ subsets with the element $2$, but no element $1$, since such subsets are already included in the previous $n-1$ subsets. We follow in this fashion until we get to the $(n-1)^{th}$ element, whereby there is only $1$ such subset left of size $2$ that hasn't been included. The number is simply 
\begin{equation}
   (n-1)+(n-2)+\cdots+1 = \sum_{k=1}^{n-1}\binom{k}{1}.
\end{equation}
Similarly for $m=3$, we obtain
\begin{equation}
   \sum_{k=2}^{n-1}\binom{k}{2},
\end{equation}
which can be thought of as summing over the number of subsets obtained by having $1,2,\ldots,n-2$, as the first element and then choosing two element from the list of elements which do not include any of the elements in the previously written down sets. One can do this in a similar fashion for $2\leq m\leq n-2$, and then the sum of all such subsets can be reduced to the double sum
\begin{equation}
   \sum_{k=1}^{n-1}\binom{k}{1}+\sum_{k=2}^{n-1}\binom{k}{2}+\ldots+ \sum_{k=n-3}^{n-1}\binom{k}{n-3}=\sum_{m=2}^{n-2}\sum_{k=m-1}^{n-1}\binom{k}{m-1}.
   \label{eq: double sum subset counting}
\end{equation}
We obtain a double sum, just as before. Also, by the way we counted the subsets, it gives us some intuition as to why counting the kinematic invariants of Section \ref{subsec: Counting Planar Mandelstam Invariants via Maximally Symmetric Ferrers Shapes.} by stopping when the outside index of the invariant was $(n-1)$. To mimic the sum structure of Eq.~\eqref{eq: counting structure for planar kinematic invariants}, Eq.~\eqref{eq: double sum subset counting} can be re-written as
\begin{equation}
    \sum^{n-2}_{m=2}\sum^{n-m+1}_{i=1}\binom{i+(m-1)-1}{m-1} = \sum^{n-2}_{m=2}\sum^{k}_{i=1}\binom{i+(n-k-1)}{i-1} = 2^n-2,
    \label{eq: number of subsets given the same partitioning}
\end{equation}
which is just the number of non-empty, proper subsets of an $n-$element set. Evaluating the inner sum of Eq.~\eqref{eq: number of subsets given the same partitioning}, one finds
\begin{align}
    \sum^{n-2}_{m=2}\ \sum^{k}_{i=1}\binom{i+(n-k-1)}{i-1} = \sum_{k=3}^{n-1} \binom{n}{k} &= \sum_{k=2}^{n-2} \binom{n}{k} - \binom{n}{2}+\binom{n}{n-1}\nonumber \\
    &=\sum_{k=2}^{n-2} \binom{n}{k} - \frac{n(n-3)}{2},
\end{align}
with the number of independent kinematic invariants appearing in the second term in the last line. Re-arranging, we can deconstruct the counting of the kinematic invariants much more clearly now:
\begin{equation}
    \sum_{k=2}^{n-2} \ \sum_{i=1}^{k}s_{i,i+1,\cdots, i+n-k-1} = \sum_{k=2}^{n-2} \binom{n}{k} -  \sum_{k=3}^{n-1}\ \sum^{k}_{i=1}\binom{i+(n-k-1)}{i-1}.
    \label{eq: breaking down the counting structure}
\end{equation}
The first term on the right hand side of Eq.~\eqref{eq: breaking down the counting structure} counts the number of non-empty subsets of sizes $2\leq k \leq n-2$ with unique elements. The second term is actually counting the sum over $i=\{1,\ldots,k\}$ of the number of integer compositions of $i$ into $n-3$ parts or fewer. 

We can now make the combinatorial structure behind the polynomial form of the scattering equations more transparent. In Eq.~\eqref{eq: breaking down the counting structure}, the first term, 
\begin{equation}
\sum_{k=2}^{n-2} \binom{n}{k},
\end{equation}
counts the number of subsets of the $ n $ external labels of size $ k $, where $ 2 \leq k \leq n-2 $. Each subset corresponds to a possible kinematic invariant $ s_A $ formed from the momenta of particles in set $ A $.

The second term,
\begin{equation}
\sum_{k=3}^{n-1} \sum_{i=1}^{k} \binom{i+n-k-1}{i-1},
\end{equation}
reveals a finer combinatorial refinement. For fixed $k$, the inner sum over $i$ accounts for different ways of distributing ``momentum weight'' across labels. Specifically, the binomial coefficient $\binom{i+n-k-1}{i-1}$ counts the number of weak compositions of $i$ into $n-k-1$ non-negative parts, where the order matters~\cite{bona2016walk}. In other words, it enumerates ways to distribute $i$ units among $n-k-1$ gaps between momenta of the external legs — directly generating the monomials in the polynomial form of the scattering equations.

It looks like each such composition corresponds to a distinct monomial. For example, terms like $(x_{i} - x_{j})^{-1}$ from the original scattering equations, when expanded, become part of a monomial structure reflecting these compositions. Each monomial encodes how residues distribute across the marked points $\sigma_i$ on the Riemann sphere, respecting momentum conservation.

This would mean that the second term captures the combinatorial structure underlying all monomials in the polynomial form. By subtracting this term from the naïve counting of subsets, we isolate the true number of independent kinematic constraints. In this way, the combinatorics of subsets and integer compositions intertwines with the algebraic geometry of the scattering map.

The appearance of $n-k-1$ refers to the diminishing number of options you have as you increase $k$: at each step, it counts the available labels beyond a fixed starting point, determining the rank of the basis of kinematic invariants. More precisely, the counting mirrors constrained integer partitions — selecting $k$ particles while preserving cyclic ordering resembles partitioning an interval into adjacent subsets. Each increment in $k$ corresponds to finer subdivisions of the external legs, directly governing the monomial structure. The factor $n-k-1$ thus tracks the freedom remaining at each stage, tightly binding the combinatorics of labellings to the scattering map\footnote{Here, the scattering map refers to the scattering equations themselves — how they map the moduli space of punctures on $\mathbb{CP}^1$ to the kinematic data.}.

\section{Proof of Diagonal Recursion Relation}\label{sec:appendix1}
We begin by considering diagrams with a $(n-1)$-particle sub-diagram for $n$ particles. The colour-ordered amplitude $m(12 \cdots n{-}1\, n\, |\beta)$ factorises into sub-diagrams of $n-1$ and $3$ particles, with the total number of diagrams given by $2 C_{n-3}$ due to the orderings
\begin{equation}
    \{1\,n\, 2 \cdots n{-}1\}, \quad \{1\,2 \cdots n\, n{-}1\}.
\end{equation}
Thus, the total number of diagrams is 
\begin{equation}
C_2 C_{n-3} C_0,
\end{equation}
where $C_2=2$ (from the two orderings) and $C_0=1$ (the factor from the smaller sub-diagram).

For $n+1$ particles, the diagrams factorise into sub-diagrams of $n-1$, $3$, and $3$ particles, yielding
\begin{equation}
C_2 C_{n-3} C_1 C_0,
\end{equation}
with $C_1$ counting the diagrams in the $(n-1)$-particle sub-diagram.

For general $n+i$ with $i \in \{2, \dots, n-4\}$, the number of diagrams is
\begin{equation}
C_2 C_{n-3} C_i,
\end{equation}
where $C_i$ counts the diagrams in the smaller sub-diagram.

For $i = n-3$, where the largest sub-diagram contains $(n-1)$-particles, the number of diagrams is
\begin{equation}
C_{n-3} \cdot C_{n-3}.
\end{equation}
This case arises when two $n-1$-particle sub-diagrams are joined by an additional particle, leading to $2n-3$ particles in total.

Summing over all diagrams, we obtain
\begin{equation}
\sum_{i=0}^{n-4} C_2 C_{n-3} C_i C_0 + C_{n-3}^2 = C_{n-3} \left( C_{n-3} + \sum_{i=0}^{n-4} C_2 C_i \right).
\end{equation}

\section{\texorpdfstring{$\mathrm{SL}(2,\mathbb{C})$ Invariance of Off-Shell Potential}{SL(2,C) Invariance of Off-Shell Potential}}\label{sec: SL2C invariance}
Under an $\mathrm{SL}(2,\mathbb{C})$ transformation, $|x_a-x_b|$ transforms as follows:
\begin{equation}
    |x_a-x_b|\to \frac{|x_a-x_b|}{|Cx_a+D|\cdot|Cx_b+D|}.
\end{equation}
To simplify the derivation, the following notation $\Gamma_a \equiv|Cx_a+D|$ will be utilised. Also, the following observation will be useful in determining $\mathrm{SL}(2,\mathbb{C})$ invariance. Namely that from Eq.~\eqref{eq: translational invariance of the Green's function}, we have 
\begin{equation}
  \sum_{a=1}^m P_a^\mu = 0 
    \Rightarrow \sum_{a=1}^m p_a^\mu = -\sum_{a=1}^m \tau_a q^\mu ,
\end{equation}
and 
\begin{equation}
    \quad \sum_{a=1}^m p_a \cdot q = 0,
    \label{eq: sl2c helper Equation}
\end{equation}
since $q^2 = 0$.

Now looking at the variation of the scattering potential under an $\mathrm{SL}(2,\mathbb{C})$ transformation, we have the following:

\begin{align}
    - \delta \mathcal{S} &= \sum_{a \neq b}  p_a \cdot p_b \log{\Gamma_a\Gamma_b} 
+ \sum_{a,b} 2 \tau_a \, p_b \cdot q (\log \Gamma_{a'} + \log \Gamma_b)\nonumber \\
&= \sum_{a \neq b}  2p_a \cdot p_b \log{\Gamma_b} 
+ \sum_{a,b} 2 \tau_a \, p_b \cdot q (\log \Gamma_{a'} + \log \Gamma_b),
\end{align}
where in the second line we exploited the symmetric form of the first term.  Moving on, we split up the sum over $a$ and $b$ in the second term to find
\begin{align}
 - \delta \mathcal{S}&= \sum_{a \neq b} 2 (p_a + \tau_a q) \cdot k_b \log \Gamma_b
+ \sum_{b=1}^{m} 2 \tau_b \, q \cdot k_b \log \Gamma_b 
+ \sum_{a,b} 2 \tau_a \, k_b \cdot q \log \Gamma_{a'}  \nonumber \\
 &=\sum_{a,b} 2 \tau_a \, k_b \cdot q \log \Gamma_{a'},
\end{align}
where in the second line, we used the fact that $\sum_{b=1}^mP_b^{\mu} = -\sum_{a=1}^mP_a^{\mu}$ and Eq.~\eqref{eq: sl2c helper Equation}. This last term also vanishes by noting that
\begin{equation}
\sum_{a,b} 2 \tau_a \, k_b \cdot q \log \Gamma_{a'} 
= 2 \sum_a \tau_a \log \Gamma_{a'} \left( \sum_b k_b \cdot q \right) = 0.    
\end{equation}
Thus $\mathrm{SL}(2,\mathbb{C})$ invariance has been established.

\end{subappendices} % DO NOT EDIT THIS LINE

% \cleardoublepage % makes sure there is an even number of pages for the book compilation (ignore this)

\end{document}